\documentclass[preprint,showpacs,showkeys,aps,prd,amsfonts,eqsecnum,nofootinbib]
{revtex4-1}
\usepackage{graphicx,epsfig}
\usepackage[dvips]{color}
\usepackage{rotate}
\usepackage{rotating}
\usepackage{multirow}
\usepackage{longtable}
\usepackage[T1]{fontenc}
\def\gsim{\lower0.5ex\hbox{$\:\buildrel >\over\sim\:$}}
\def\lsim{\lower0.5ex\hbox{$\:\buildrel <\over\sim\:$}}
\begin{document}
\preprint{CUMQ/HEP 156}

\title{\Large B Decays in an Asymmetric Left-Right  Model}
\author{Mariana Frank$^{a}$}
\author{Alper Hayreter$^{a}$} 
\author{Ismail Turan$^{b}$ }

\affiliation{$^{a}$Department of Physics, Concordia University, 7141 Sherbrooke St. 
West, Montreal, Quebec, CANADA H4B 1R6,}
 \affiliation{$^{b}$Ottawa-Carleton 
Institute of Physics,
Carleton University,
1125 Colonel By Drive
Ottawa, Ontario, Canada, K1S 5B6.}

\date{\today}

\begin{abstract}
Motivated by recently observed disagreements with the SM predictions in $B$ decays, we study $b \to d, s$ transitions in an asymmetric class of 
$SU(2)_L \times SU(2)_R \times U(1)_{B-L}$ models, with a simple one-parameter structure of the right handed mixing matrix for the quarks, which obeys 
 the  constraints from kaon physics. We use experimental constraints on the branching ratios of $b \to s \gamma$, 
$b \to c e {\bar \nu}_e$,  and $B_{d,s}^0 -\bar{B}^0_{d,s}$ mixing to restrict the parameters of the model: $\displaystyle {g_R}/{g_L}, M_{W_2}, M_{H^\pm}, 
\tan \beta$ as well as the elements of the right-handed quark mixing matrix $V^R_{CKM}$. We present a comparison with the more commonly used (manifest) left-right symmetric model.  Our analysis exposes the parameters most sensitive to $b$ 
transitions and reveals a large parameter space where left- and right-handed quarks mix differently, opening the possibility of observing marked differences 
in behaviour between the standard model and the left-right model.
\end{abstract}
\maketitle
 
\section{Introduction}
\label{sec:intro}
Within the next decade, significant progress is expected in experimental high energy physics. Most of the hope rests on  LHC, expected  to probe the standard model (SM) of electroweak 
interactions and models beyond it. The experimental explorations would complement efforts made by theorists over the last  decades. 
The common wisdom held that while the SM left some fundamental questions unanswered (such as stability of the Higgs mass, the origin of CP violation, the baryon asymmetry, or the presence of dark matter 
in the universe), it was experimentally sound. Several precision measurements have recently questioned the latter. First and foremost, there was evidence for the existence of 
neutrino masses and mixing,  inconsistent with the SM predictions, where neutrinos are assumed massless. 
Some of recent experimental results which might prove (at least) difficult to explain within the SM, and provide some hints of deviations from its predictions come mostly from $B$ physics. The values of the angle $\phi_1$ measured in some penguin process $b \to  s q{\bar q}$ and the precisely measured value in $B \to J/\psi K_S^0$ differ by two to three standard deviations ($B_0 \to \pi^0 \pi^0 K_S^0 , B_0 \to K^+K^-K^0$, \cite{Barberio:2008fa, a:2007xd, chao}) and may suggest the existence of a new CP phase in this penguin-dominated process; the lepton forward-backward asymmetry in $B \to K^*l^+l^-$  is measured to be around two standard deviations higher than the SM prediction \cite{adachi1}; direct CP asymmetries in $B_0 \to K^+\pi^- $ and $B^+ \to K^+\pi^0$ differ significantly from each other, although naively one would expect them to be the same \cite{lin}; the branching fraction for $B^+ \to \tau \nu $ is up to two standard deviations higher than expected, depending on the theoretical input chosen \cite{Barberio:2008fa, adachi2}; in purely leptonic $D_s^+ \to \mu \nu$ and $D_s^+ \to \tau \nu$ decays the deviation of the branching ratios is even larger  \cite{widhalm, rosner} if one uses the recent lattice QCD calculations of the meson decay constant; the measured production cross-section for $c {\bar c}$ states is higher than the calculated one \cite{abe}. A careful analysis combining all the experimental data on $B_s$ mixing \cite{Aubert:2007py} finds that the phase of the mixing amplitude deviates by about $3 \sigma$ 
from the SM prediction (or slightly less, if one does not use Gaussian error distributions{\footnote{ We thank Alexander Lenz for this observation.}}) \cite{Bona:2008jn}. 

Additionally, the CDF and D\O \, experiments have determined 
a sizable forward-backward asymmetry in top anti-top events, in which one top decays semileptonically, a measurement that is more than a $2 \sigma$ 
deviation from the SM prediction \cite{CDF}. 
 
Taken together, these  indicate that flavor and CP physics are highly non-trivial and that they may be governed by a new paradigm beyond the single 
CKM matrix  of the SM. Possibilities for non-SM flavor violation are present in the $b \to d, s $ non-leptonic decays. This justifies looking at rare B decays  in New Physics scenarios. 
  
Perhaps the simplest such scenario of models beyond the SM is the left-right symmetric model (LRSM) \cite{Pati:1974yy}. Motivated originally by the desire 
to understand parity violation in weak interactions \cite{Senjanovic:1978ev}, it gathered some more support due to its simplicity. It appears to be a 
natural extension of the SM, as it treats both left- and right-handed fermions as doublets. Additionally the model gauges the $B-L$ quantum number, left 
ungauged in the SM, and it provides an elegant explanation of neutrino masses through the see-saw mechanism \cite{Mohapatra:1979ia}.
 
The LRSM, based on the gauge group $SU(2)_L \times SU(2)_R \times U(1)_{B-L}$, has some immediate implications on the role played by the right-handed 
fermions in charged current interactions, both for flavor-changing and flavor-conserving, while leaving open how much, and with what strength.  Most authors assumed 
that LRSM is invariant under a discrete left-right symmetry, where the left- and right-handed fermions can be interchanged and the couplings of the two gauge groups, 
$g_L$ and $g_R$, are equal. If the discrete LR symmetry breaks down  at low (TeV scale) energy, then $g_L \ne g_R$. Furthermore, most previous works have assumed 
a relationship between quark flavor mixing in the left and right sectors, either that the Cabibbo-Kobayashi-Maskawa (CKM) matrices in the two sectors are equal, 
$V_{CKM}^R=V_{CKM}^L$, as in manifest left-right symmetry, or that they are related by   diagonal phase matrices $K^u$ and $K^d$,  $V_{CKM}^R=K^u\, V_{CKM}^{L\,\star}\,K^{d\,\star}$ (pseudo-manifest left-right models). The first scenario \cite{Senjanovic:1978ev} assumes CP violation to be produced by complex Yukawa couplings, and fermion masses to be generated by real vacuum expectation values of the Higgs fields. The second model \cite{Harari:1983gq} assumes that both parity (P) and charge parity (CP) are broken spontaneously, thus that the Yukawa couplings are real. Both of these scenarios have difficulties in accounting for the baryon asymmetry of the universe, and lead to cosmological domain-wall problems \cite{Widyan:1999bs}.
 
A notable exception to the above formulations of LRSM is the model proposed by Langacker and Sankar \cite{Langacker:1989xa}. The authors assume the left-right symmetry to be fundamental, 
superseding the Higgs, Yukawa or fermion structure, and analyze constraints on the charged gauge boson masses and mixings including a variety of constraints, 
coming from the kaon system, the $B^0_d-{\bar B}^0_d$ mixing, $b \to X \nu_e e$, universality, muon decays and neutrinoless double beta decays. They consider 
several neutrino masses scenarios (Dirac or Majorana, light, intermediate or heavy) and allow for $g_L \ne g_R$ as well as $V_{CKM}^L \ne V_{CKM}^R$. The form chosen 
for the $V_{CKM}^R$ is not arbitrary, nor is it the most general form for a $3 \times 3$ mixing matrix one could write down. The choice for right-handed  quark mixings is particularly attractive, as it is motivated by the 
$K^0-\bar {K}^0$ mass difference, which is strongly affected by the right-handed quark mixing matrix, and it depends on one parameter only, making it highly predictive. Their requirement is that $M_{W_R}$ be as general as 
possible, and the form of $V_{CKM}^R$ not be excessively fine-tuned. An additional reason to revisit this parametrization is that a recent analysis of CP violation in Pati-Salam type  left-right models \cite{Gielen:2010nv}  
concludes that manifest/pseudo-manifest  left-right models are disfavored, unless they  include an unnaturally large CP violating phase. In Langacker and Sankar parametrization, there are two possibilities for the right-handed CKM matrix,  known as 
$(A)$ and $(B)$, with
\begin{eqnarray}
\label{scenAB}
V^R_{(A)} = \left(\begin{array}{ccc}
1&0&0\\
0&c_{\alpha}&\pm s_{\alpha}\\
0& s_{\alpha}& \mp c_{\alpha}
\end{array}\right) ,~~~
V_{(B)}^R  = \left(\begin{array}{ccc}
0&1&0\\
c_{\alpha} &0&\pm s_{\alpha}\\
s_{\alpha} &0&\mp c_{\alpha}
\end{array}\right),
\end{eqnarray}
where $c_{\alpha} \equiv \cos\alpha$ and $s_{\alpha} \equiv \sin\alpha$, with $\alpha$ an arbitrary angle $\displaystyle (-\pi/2 \le \alpha \le \pi/2)$. The mixing between the first two families is trivial, removing the strict bounds on the new charged gauge boson mass required by $K^0-{\bar K}^0$ mixing. The two parametrizations allow for arbitrary mixing between the second and third, or first and third right-handed quark families, with an arbitrary parameter $\alpha$. Thus, although the ansatz seems specific, it is fairly general while fulfilling the constraints of kaon physics.

The aim of this work is to investigate the consequences of these parametrizations, referred from here on as the Asymmetric Left Right Model (ALRM)  on $b \to d,s $ transitions, concentrating at first on the CP-conserving, 
flavor violating processes $b \to s \gamma$ ($\Delta B=1$) and $B_{d,s}^0- {\bar B}_{d,s}^0$ mixing ($\Delta B=2$). Although the experimental data for these agrees with the predictions of the SM, we use the analysis  to establish consistency of the model parameters. These enter consideration of CP violating effects, which  
will be left for further work.

Our motivation is two-fold. First, flavor and CP violation in B decays have received a lot of theoretical and experimental interest recently, and careful 
analyses, as outlined before, show deviations from the SM predictions. Agreement with the branching ratio  for $b\to s\gamma$ is the cornerstone of any model beyond 
the SM. LHCb will uncover many new exciting results in B physics and may rule out certain models, as might a new (under discussion) Super KEKB factory. Second, strong flavor violation (which could come from the 
right-handed quarks in ALRM) has implications for new particles and interactions at the LHC, notable for new charged gauge bosons, which have received less 
attention than their neutral counterparts. We investigate this possibility in a forthcoming paper.

The analysis presented here follows several previous analyses of B decays in left-right models \cite{Rizzo:1994aj}. Although many discussions of the manifest or pseudo-manifest model  exist, very few are available for more general  left-right models. Our numerical analysis is more detailed and comprehensive than in previously works and clearly separates regions for all parameters of left-right models that are ruled out by existing measurements. As we were unable to find equally extensive discussions of manifest or pseudo-manifest left-right symmetric models, we include a comparison  with these models as well, and give the relevant values in the SM. Additionally, we have performed the analysis using well-established publicly available software, which allows exact numerical evaluations without using additional assumptions. As we had to modify the software to include evaluation of the box diagrams, we explain the modification in Appendix \ref{PVint} and give the relevant formulas.

Our paper is organized as follows. In Section \ref {sec:model} we present a succinct description of the  ALRM with the $(A)$ and $(B)$ parametrization for $V^R_{CKM}$, that is a summary of the model presented 
in \cite{Langacker:1989xa}. We then use the results to consider rare B decays in Section  \ref{sec:BDec}, in particular we investigate the process $ b \to s \gamma $ (including a short discussion of $b \to d \gamma$) in Section \ref{sec:bsg} and 
$B_{d,s}^0- {\bar B}_{d,s}^0$ mixing in Section \ref{sec:B0B0bar}, allowing for a large parameter space consistent with kaon physics constraints. We summarize our results and conclude in Section \ref{sec:conclude}. Some of our basic analytic expressions are included in the paper, and we delegate some details to the Appendices.

\section{Left-Right Symmetric Models}
\label{sec:model}
The left-right models of weak interactions are based on the gauge group $SU(2)_L \times SU(2)_R \times U(1)_{B-L}$.  Under the group symmetry, fermions (quarks and leptons) 
are assigned the following quantum numbers
\begin{eqnarray}
&&Q_L: \left ( \frac12, 0, \frac13  \right);~~~Q_R: \left ( 0, \frac12, \frac13  \right)  \nonumber\\
&&L_L: \left ( \frac12, 0, -1  \right);~~~L_R: \left ( 0, \frac12, -1  \right).
\end{eqnarray}
Interactions are mediated by three neutral gauge bosons $\gamma, Z_1, Z_2$ and four charged bosons $W_1^{\pm}, W_2^{\pm}$, which are mixtures of the fundamental gauge bosons of the three gauge groups. The electric charge formula is given by
$$Q=I_{3L}+I_{3R}+\frac{B-L}{2}.$$
The  parity symmetry is broken first, resulting in $g_L \ne g_R$ at the right-handed scale. The gauge symmetry is also broken, at the same or lower scale. 
The Higgs multiplets required for symmetry breaking are chosen so they are bilinears in the basic fermion multiplets. A bidoublet is needed to break LR symmetry
\begin{eqnarray}
\displaystyle
\Phi&&= \left( \begin{array}{cc} \phi^0_{1} &\phi^+_{2} \\ \phi^-_{1} & \phi^0_{2} \end{array} \right) \sim \left (2,2,0 \right).
\end{eqnarray}
Additional Higgs multiplets are needed to break the symmetry to the SM and to generate a large $M_{W_R} \gg M_{W_L}$. One has the option of introducing doublet Higgs representations
\begin{eqnarray}
\delta_L&=&\left( \begin{array}{c} \delta_L^+ \\ \delta_L^0 \end{array} \right)  \sim \left ( 2, 1,1 \right ), 
~~~\delta_R=\left( \begin{array}{c} \delta_R^+ \\ \delta_R^0 \end{array} \right)  \sim \left ( 1, 2,1 \right ) 
\end{eqnarray}
or Higgs triplets, a popular alternative as it can generate a small Majorana mass for the left-handed neutrinos and large masses for the right-handed neutrinos and $W_R$ bosons:
\begin{eqnarray}
\Delta_{L} = \left(\begin{array}{cc}
\frac {\Delta_L^-}{\sqrt{2}}&\Delta_L^0\\
\Delta_{L}^{--}&-\frac{\Delta_L^-}{\sqrt{2}}
\end{array}\right) \sim (3,1,2),~~~
\Delta_{R}  =
\left(\begin{array}{cc}
\frac {\Delta_R^-}{\sqrt{2}}&\Delta_R^0\\
\Delta_{R}^{--}&-\frac{\Delta_R^-}{\sqrt{2}}
\end{array}\right) \sim (1,3,2).
\end{eqnarray}

The Higgs develop vacuum expectation values (vevs)
\begin{eqnarray}
\displaystyle
\langle\Phi\rangle&&= \left( \begin{array}{cc} v_u &0 \\ 0 & v_d\end{array} \right) ~~,~~
\langle\delta_{L,R}\rangle=\left(\begin{array}{c} 0 \\ v_{\delta_{L,R}} \end{array} \right) ~~,~~
\langle\Delta_{L,R}\rangle=\left(\begin{array}{cc} 0 & v_{\Delta_{L,R}} \\ 0 & 0 \end{array} \right).
\end{eqnarray}
The Higgs triplet vev $v_{\Delta_R}$ can produce a large $M_{W_R}$ mass and generate a large Majorana mass for the right-handed neutrino.
If $ v_{\delta_R}\gg (v_u, v_d, v_{\delta_L})$, the Higgs doublet vev can generate a large $M_{W_R}$ and a large right-handed Dirac neutrino mass \cite{Gunion:1989in}. The Higgs doublets or triplets do not couple to quarks because of their $B-L$ quantum number assignments, and although they mix with the bidoublet Higgs bosons, only the eigenvectors corresponding to the bidoublet contribute to B decays. As the choice of doublet or triplet Higgs  does not play an essential role in our considerations, we will treat both possibilities together, and denote $v_L=v_{\Delta_L}, v_{\delta_L}$  and $v_R=v_{\Delta_R}, v_{\delta_R}$.

At the first stage of symmetry breaking, $W^{\pm}_{R}$ will pick up the mass $\displaystyle M_{W_R}=\frac{g_R v_R}{\sqrt{2} }$. The second stage of breaking is controlled by the $\langle \Phi  \rangle$. This contributes to the $Z_L, W_L$ masses, but since $\Phi$ transforms non-trivially under both $SU(2)_L$ and $SU(2)_R$, 
it mixes the charged gauge bosons with the following mass-squared matrix  
\begin{eqnarray}
{\cal{M}}^2=\left(
\begin{array}{cc}
\frac{g_L^2}{2}(v_L^2 + v_u^2 + v_d^2) & -g_L g_R v_u v_d \\ 
-g_L g_R v_u v_d & \frac{g_R^2}{2}(v_R^2 + v_u^2 + v_d^2)
\end{array}
\right) 
\end{eqnarray}
in which the two mass eigenstates mix with an orthogonal rotation matrix to construct physical W gauge bosons
\begin{eqnarray}
\label{Wmix}
W_1&=&c_\xi W_L +e^{-i \omega}s_\xi W_R,  
\nonumber \\
W_2&=&(-s_\xi W_L + e^{-i\omega}c_\xi W_R)
\end{eqnarray}
where $\omega$ is a CP violating phase \cite{Langacker:1984dp}, and $c_\xi \equiv \cos\xi$ , $s_\xi \equiv \sin\xi$ with $\xi$ a mixing angle which is severely restricted to be $\xi \le 3 \times 10^{-3}$ from $K^0-{\bar K}^0$ mixing \cite{Wolfenstein:1984ay}.
Since the electroweak analysis leads to the constraint $v_L \lesssim 10$ GeV and the see-saw mechanism 
for small left-handed neutrino masses requires $v_L \lesssim$ a few MeV, we will work in the limit $v_L \rightarrow 0$. Therefore the mixing angle and two mass eigenstates in this limit are defined
\begin{eqnarray}
 t_{2\xi} &=& \frac{2 \left(\frac{g_R}{g_L}\right) v_u v_d}{\left(\frac{g_R}{g_L}\right)^2 v_R^2 + \left[\left(\frac{g_R}{g_L}\right)^2 - 1\right]v^2}, 
\end{eqnarray}
\begin{eqnarray}
M_{W_1}^2 &=& \frac{g_L^2}{2} \left[ v^2c^2_{\xi} - 2 \left(\frac{g_R}{g_L}\right) v_u v_d s_{2\xi} + \left(\frac{g_R}{g_L}\right)^2(v_R^2 + v^2)s^2_{\xi}\right] , \nonumber \\
M_{W_2}^2 &=& \frac{g_L^2}{2} \left[ v^2s^2_{\xi} + 2 \left(\frac{g_R}{g_L}\right) v_u v_d s_{2\xi} + \left(\frac{g_R}{g_L}\right)^2(v_R^2 + v^2)c^2_{\xi}\right]
\end{eqnarray}
where we have introduced the shorthand notation $v^2 = v_u^2 + v_d^2$. Notice that, in the case of no mixing ($\xi \rightarrow 0 $) the mass eigenstates will exactly be  $M_{W_1}=M_{W_L}$ and $M_{W_2}=M_{W_R}$.
 The most common forms of left-right symmetric models are the manifest and the pseudo-manifest left-right models.

The manifest left-right symmetric model assumes that weak interactions enjoy a left-right symmetry in the Lagrangian (that is, the Lagrangian is invariant under $SU(2)_L \times SU(2)_R \times U(1)_{B-L}$ gauge symmetry), and that parity violation stems from the spontaneous breakdown of this symmetry \cite{Senjanovic:1978ev}. Manifest here indicates that the physical left-handed and right-handed currents have identical properties in flavor space and that 
$$V_{CKM}^R=V_{CKM}^L .$$
This model has complex Yukawa couplings  and real expectation values for the Higgs fields. 

In the pseudo-manifest left-right symmetric model \cite{Harari:1983gq}, the Lagrangian of the model is invariant under $SU(2)_L \times SU(2)_R \times U(1)_{B-L}$ gauge symmetry, but both  parity and charge conjugation  are broken spontaneously (unlike in the manifest case where charge conjugation is broken explicitly). In this model, the left- and right-handed quark mixing matrices are most generally related by
$$V_{CKM}^R=K^u V_{CKM}^{L\, \star} K^{ d\, \star}$$
with $K^u, K^d$ diagonal phase matrices, defined as  $K^u={\rm Diag} (e^{i\phi_u}, e^{i\phi_c}, e^{i\phi_t})$ and $K^d={\rm Diag}( e^{i\phi_d}, e^{i\phi_s}, e^{i\phi_b})$. Thus this model contains an additional set of CP violating phases. The pseudo-manifest model has real Yukawa couplings and complex vacuum expectation values for the Higgs fields. 

In the asymmetric left-right model, which we study here, left-right symmetry of the Lagrangian is seen as more  fundamental than the Higgs, Yukawa or fermion structure. The left- and right-handed quark mixing are independent of each other, and are fixed by experimental constraints from low energy physics. The mixing matrix for left-handed quarks is the known CKM matrix, while for right-handed quarks the mixing matrix is chosen to satisfy the kaon ($K^0-{\bar K}^0$ mixing, $\epsilon_K$) meson constraints. This fixes the mixing between the first two families (to be either minimal or maximal), allowing for arbitrary mixing between the second and third, or the first and third families, parametrized as 
$V^R_{(A)}$ and $V^R_{(B)}$ as in (\ref{scenAB}). The consequences of the asymmetric left-right model have received less attention \cite{Langacker:1989xa}, and we propose to investigate them here in $b \to s, d$ transitions.

 \section{B Decays}
 \label{sec:BDec}
 
Left-right  models are best constrained at low energies by flavor changing mixings and decays, as well as by the CP violating observables. 
In what follows, we will work with the $V_{(A)}^R$ and  $V_{(B)}^R$ parametrizations (denoted simply by $V^R$) and compare our results with the manifest left-right model where possible. The restrictions on these parametrizations in the $K_L-K_S$ mixing have been thoroughly examined \cite{Langacker:1989xa, Hou:1985ur}, and the experimental limits imply
\begin{equation}
\left (\frac{g_R M_1}{g_L M_2}\right)^2 \le 0.075, ~~{\rm or} ~~\frac{g_L}{g_R}M_2 \ge 300 ~{\rm GeV}
\end{equation}
with $M_1, M_2$ the masses of the charged gauge bosons in (\ref{Wmix}). These restrictions still hold, as the experimental data on kaon physics did not change significantly over the years.  However, we need to carefully re-examine the constraints on the model parameters coming from B physics, in light of the new measurements.
We proceed first with the analysis of the $\Delta B=1$ flavor changing decays, and follow in the next subsection with $\Delta B=2$ processes. Both $\Delta B=1$ and $\Delta B=2$ processes are generated by the same  Lagrangian, which is responsible for flavor changing. 
The charged current interactions for general B decays are, for the $W_{1,2}$  bosons
\begin{eqnarray}
{\cal L}^W_{cc}
&=&-\frac{1}{\sqrt{2}}{\bar u}_i\gamma^{\mu}\left[g_L c_\xi V^L_{ij} P_L + g_R e^{-i\omega} s_\xi V^R_{ij} P_R\right] d_j W_{1\mu}^+ \nonumber \\
&& +\frac{1}{\sqrt{2}}{\bar u}_i\gamma^{\mu}\left[g_L e^{i\omega} s_\xi V^L_{ij} P_L - g_R c_\xi V^R_{ij} P_R\right] d_j W_{2\mu}^+
\end{eqnarray}
and for the charged Higgs fields
\begin{eqnarray}
{\cal L}^H_{cc}&=&-\frac{\sin 2\beta}{\cos 2\beta}N_{H^+}{\bar u}_i\left [ m_{u_i}V^L_{ij}P_L-  m_{d_j}V^L_{ij}P_R\right] d_j H^+ \nonumber \\
&& -\frac{1}{\cos 2\beta}N_{H^+}{\bar u}_i\left [m_{u_i}V^R_{ij}P_R-m_{d_j}V^R_{ij}P_L\right] d_j H^+
\end{eqnarray}
with 
\begin{equation}
N_{H^+}=\left [ v_u^2+v_d^2+\frac{(v_u^2-v_d^2)^2}{2v_R^2}\right ]^{-\frac12}
\end{equation}
and $\displaystyle \tan \beta=\frac{v_u}{v_d}$. Note that there is a neutral Higgs boson which can violate flavor.  This   Higgs boson   
must be heavy to obey Flavor Changing Neutral Currents (FCNC) bounds (of order of $30-50$ TeV or heavier \cite{Pospelov:1996fq}, so we will {\it a priori} neglect its contribution here).  
Finally the interactions corresponding to the charged Goldstone bosons $G_{1,2}$ are:
\begin{eqnarray}
{\cal L}^G_{cc}=&-&\frac{1}{\sqrt{2}m_{W_1}}{\bar u}_i \left [ \left( g_L c_\xi m_{u_i} V^L_{ij} - g_R s_\xi m_{d_i} V^R_{ij}\right) P_L - \left( g_L 
c_\xi m_{d_i} V^L_{ij} - g_R s_\xi m_{u_i} V^R_{ij}\right) P_R\right] d_jG_{1}^+ \nonumber \\
&+&\frac{1 }{\sqrt{2 }m_{W_2}}{\bar u}_i \left [ \left( g_L s_\xi m_{u_i} V^L_{ij} + g_R c_\xi m_{d_i} V^R_{ij}\right) P_L - \left( g_L 
s_\xi m_{d_i} V^L_{ij} + g_R c_\xi m_{u_i} V^R_{ij}\right) P_R\right] d_jG_{2}^+. \nonumber\\ 
\end{eqnarray}
In all the above formulas, $u_i(d_i)$ denotes up(down)-type quarks, $m_{u_i(d_i)}$ are their respective masses and $P_{L,R}=(1\mp \gamma^5)/2$ are the left and right handed 
projection operators.
\subsection{$b \to s\gamma$ decay}
\label{sec:bsg}
The inclusive rate $B\to X_s \gamma$ has been measured precisely to 10\% \cite{aubert, poppenburg}
$$BR_{\rm Exp}(B\to X_s \gamma)=(3.55 \pm0.23)\times 10^{-4}.$$
The rate has been calculated in SM to ${\cal O}(\alpha_s^2)$ with the remaining uncertainty  7\% \cite{misiak}
 $$BR_{\rm SM}(B\to X_s \gamma)=(3.15 \pm0.23)\times 10^{-4}.$$
 While the difference is not too large, the window between the measurement and the SM can be used to severely constrain new physics.

The decay $b \to s \gamma$ has been considered by numerous authors in the context of manifest or pseudo-manifest left-right models \cite{Rizzo:1994aj}. Basically, this is a 
one-loop flavor-changing neutral current process, proceeding through an electromagnetic penguin diagram, with up-type quarks and charged bosons in the loop.
The low-energy effective Hamiltonian for $ b \to s \gamma$ is written as
\begin{eqnarray}
{\cal H}^{(\Delta B=1)}_{eff}=\frac{4G_F}{\sqrt{2}}\,\left[\left(V^L_{jb}V^{\star L}_{js}\right) C^7_L O^7_L + 
\frac{g_R^2}{g_L^2}\left(V^R_{jb}V^{\star R}_{js}\right) C^7_R O^7_R \right]
\end{eqnarray}
where the operators are
\begin{equation}
O^7_L = \frac{e\,m_b}{16\pi^2}\left(\bar{s}\,\sigma^{\mu\nu}P_R\,b\right)\,F_{\mu\nu}~~,~~
O^7_R = \frac{e\,m_b}{16\pi^2}\left(\bar{s}\,\sigma^{\mu\nu}P_L\,b\right)\,F_{\mu\nu} 
\end{equation}
with $F_{\mu \nu}$ the electromagnetic field tensor. 
We used {\tt FeynArts} \cite{Hahn:2000jm} for generating  the amplitudes, then {\tt FormCalc} and 
{\tt LoopTools} \cite{Hahn:2000jm} packages to evaluate the loop contributions $C^7_L$ and $C^7_R$ numerically. 
 The dominant contribution to $\Gamma(b \rightarrow s\,\gamma)$ comes from  
the top-quark in the loop, so below  we give the analytical expressions for the top-quark contribution.

The coefficients of pure left, pure right and LR interference are encoded in $C^7_L$ and $C^7_R$ ;
\begin{eqnarray}
C^7_L&=& c_{\xi}^2 A_{SM}(x_1) + s_{\xi}^2 A_{SM}(x_2) + s_{2\xi}\, \frac{g_R}{g_L}\frac{m_t}{m_b}\frac{V^R_{tb}}{V^L_{tb}}\sum_{i=1}^2 A_{LR}(x_i) \nonumber \\
&+& \frac{s_{2\beta}}{c_{2\beta}^2}\frac{m_t}{m_b}\frac{V^R_{tb}}{V^L_{tb}} A_{H^+}^1(y) + t^2_{2\beta} A_{H^+}^2(y),\\
C^7_R&=& s_{\xi}^2 \left(\frac{g_R}{g_L}\right)^2 A_{RH}(x_1) + c_{\xi}^2 \left(\frac{g_R}{g_L}\right)^2 A_{RH}(x_2) + s_{2\xi}\, \frac{g_R}{g_L} 
\frac{m_t}{m_b}\frac{V^L_{tb}}{V^R_{tb}}\sum_{i=1}^2 A_{LR}(x_i)\nonumber \\ 
&+& \frac{s_{2\beta}}{c_{2\beta}^2}\frac{m_t}{m_b}\frac{V^L_{tb}}{V^R_{tb}} A_{H^+}^1(y) + \frac{1}{c_{2\beta}^2}  A_{H^+}^2(y)
\end{eqnarray}
where the arguments of the functions are $x_i=\left (m_t/M_{W_i}\right)^2$ , $y=\left ( m_t/M_{H^\pm}\right)^2$. The loop integrals $A_{SM},\, A_{RH},\, A_{LR}$ and $A_{H^+}^{1,2}$ are calculated numerically in terms of 
scalar and tensor coefficient functions. The QCD corrections arising from the evolution of effective Hamiltonian down to $\mu = m_b$ scale are
\begin{eqnarray}
 C_L^{7(eff)}&=&\eta^{-16/23}\left[ C^7_L + \frac{3}{10}X(\eta^{10/23}-1) + \frac{3}{28}X(\eta^{28/23}-1)\right],  \nonumber \\
 C_R^{7(eff)}&=&\eta^{-16/23} C^7_R
\end{eqnarray}
with $X=\frac{208}{81}$ and $\eta=\frac{\alpha_s(m_b)}{\alpha_s(M_{W_1})}\simeq 1.8$. In the calculation of the branching ratio we have followed the traditional 
method of scaling   the decay width $\Gamma(b\rightarrow s \gamma)$  with  the semileptonic decay width $\Gamma(b\rightarrow c\, e\, \bar{\nu})$ 
\cite{Bertolini:1986th}
\begin{eqnarray}
BR(b\rightarrow s \gamma) = \frac{\Gamma(b\rightarrow s \gamma)}{\Gamma(b\rightarrow c\, e\, \bar{\nu})} \times BR(b\rightarrow c\, e\, \bar{\nu})
\end{eqnarray}
where we calculated  the width $\Gamma(b\rightarrow c\, e\, \bar{\nu})$ in our model and  for the branching ratio we used the  well-established value $BR(b\rightarrow c\, e\, \bar{\nu}) \simeq 11\%$ \cite{Amsler:2008zz}.
  
\begin{figure}[htb]
\vskip -0.3in
\begin{center}$
	\begin{array}{ccc}
\hspace*{-0.7cm}
	\includegraphics[width=2.2in,height=2.2in]{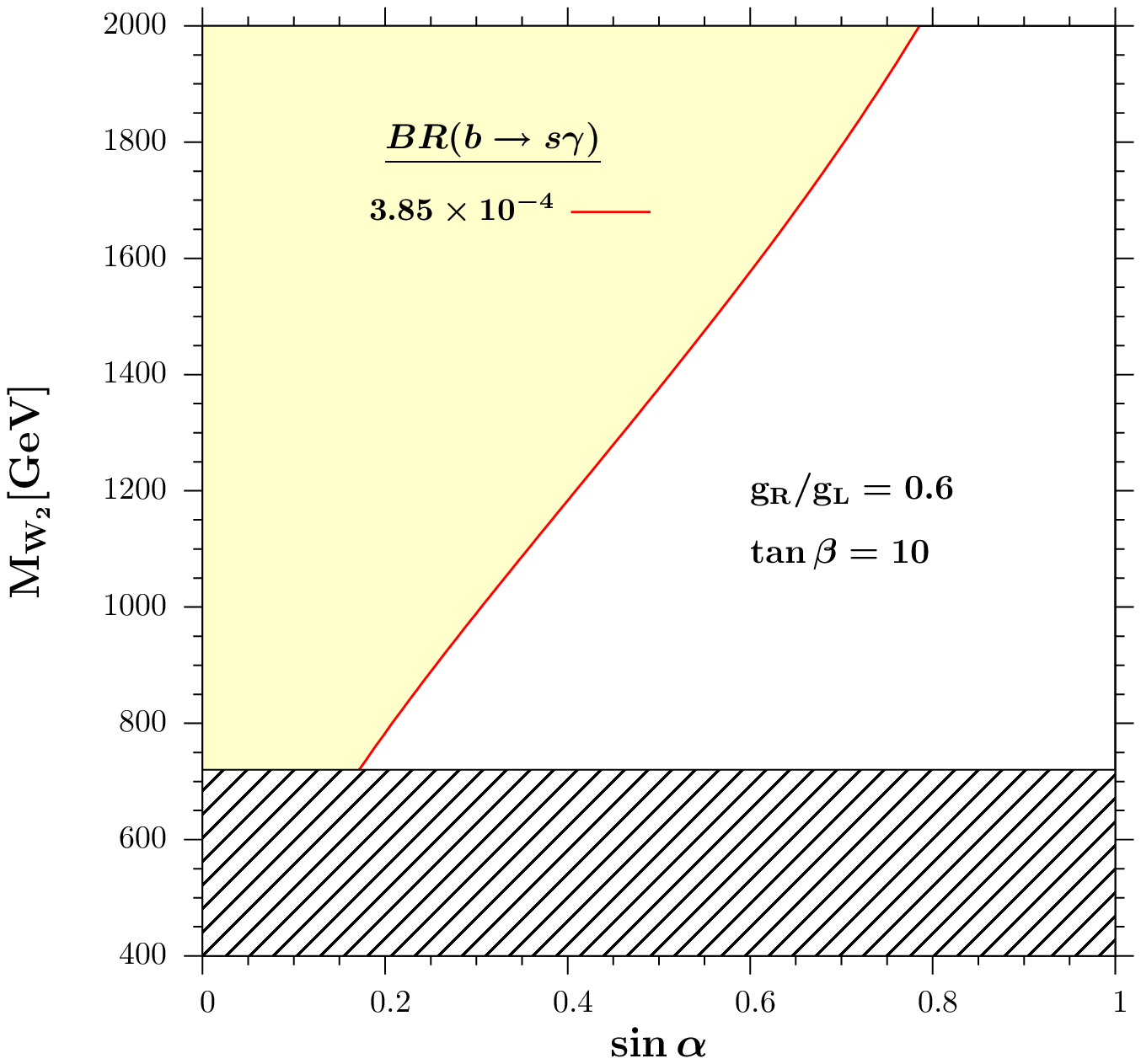} &\hspace*{-0.2cm}
	\includegraphics[width=2.2in,height=2.2in]{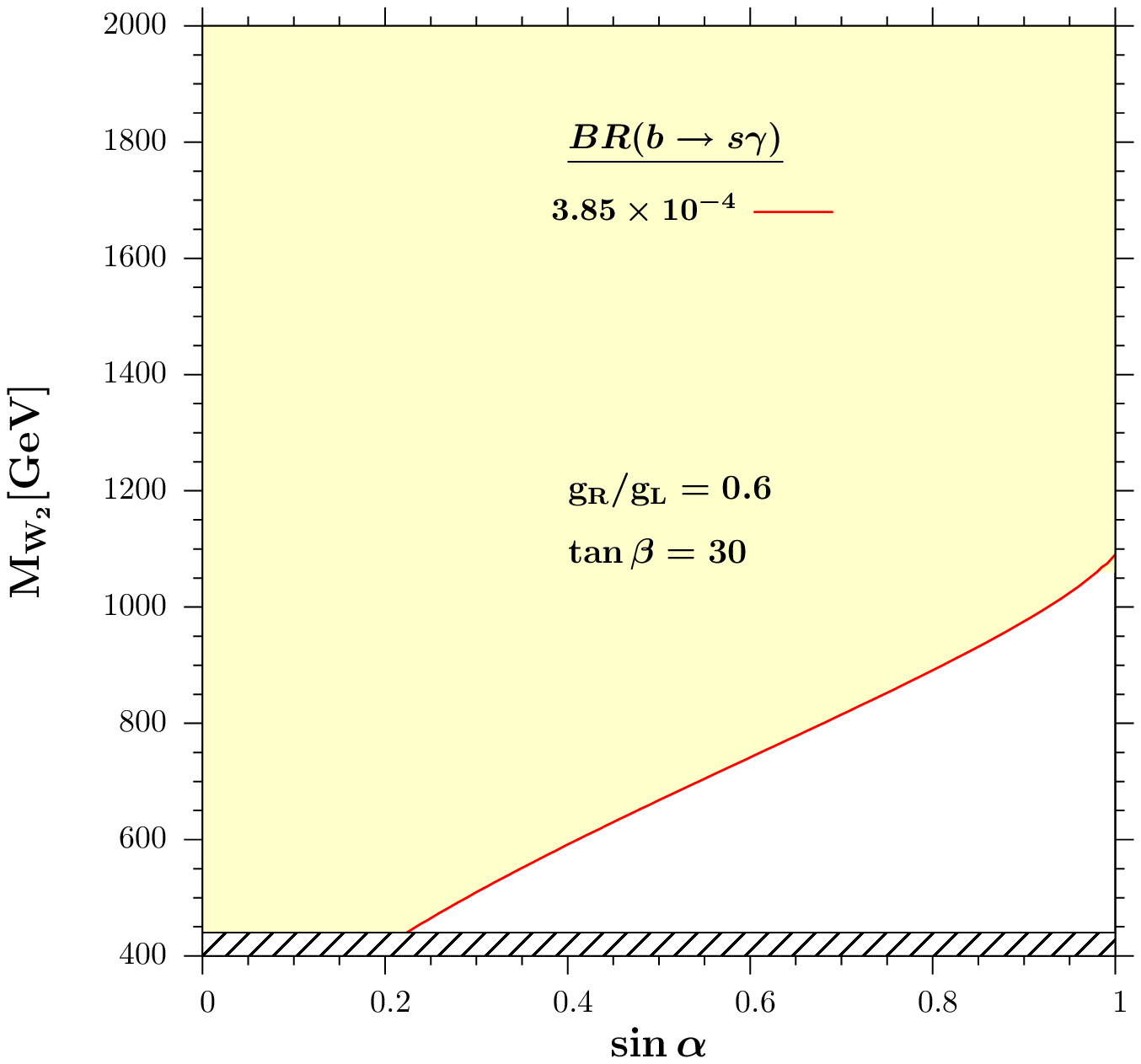} &\hspace*{-0.2cm}
        \includegraphics[width=2.2in,height=2.2in]{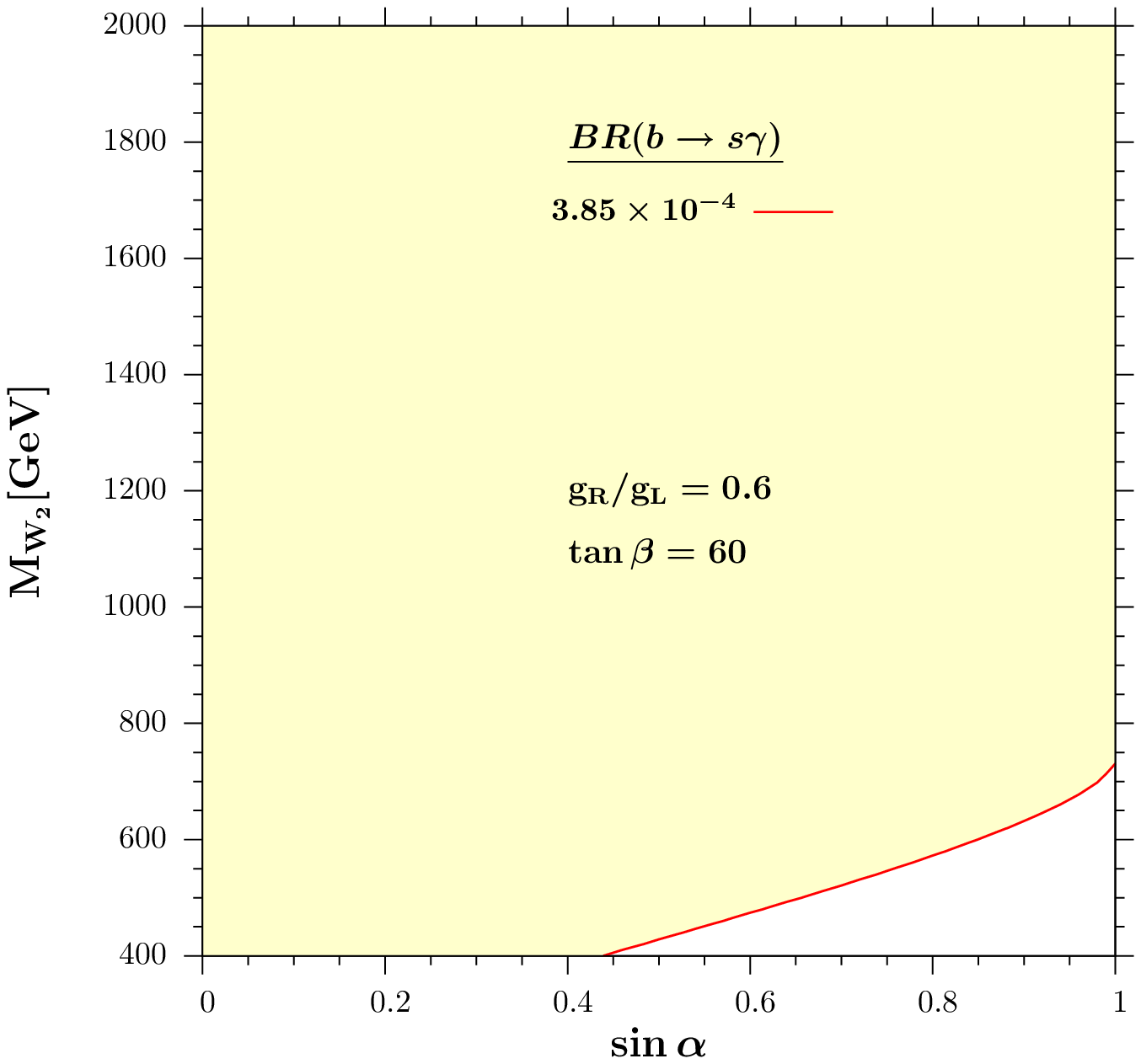} \\
\hspace*{-0.7cm}
	\includegraphics[width=2.2in,height=2.2in]{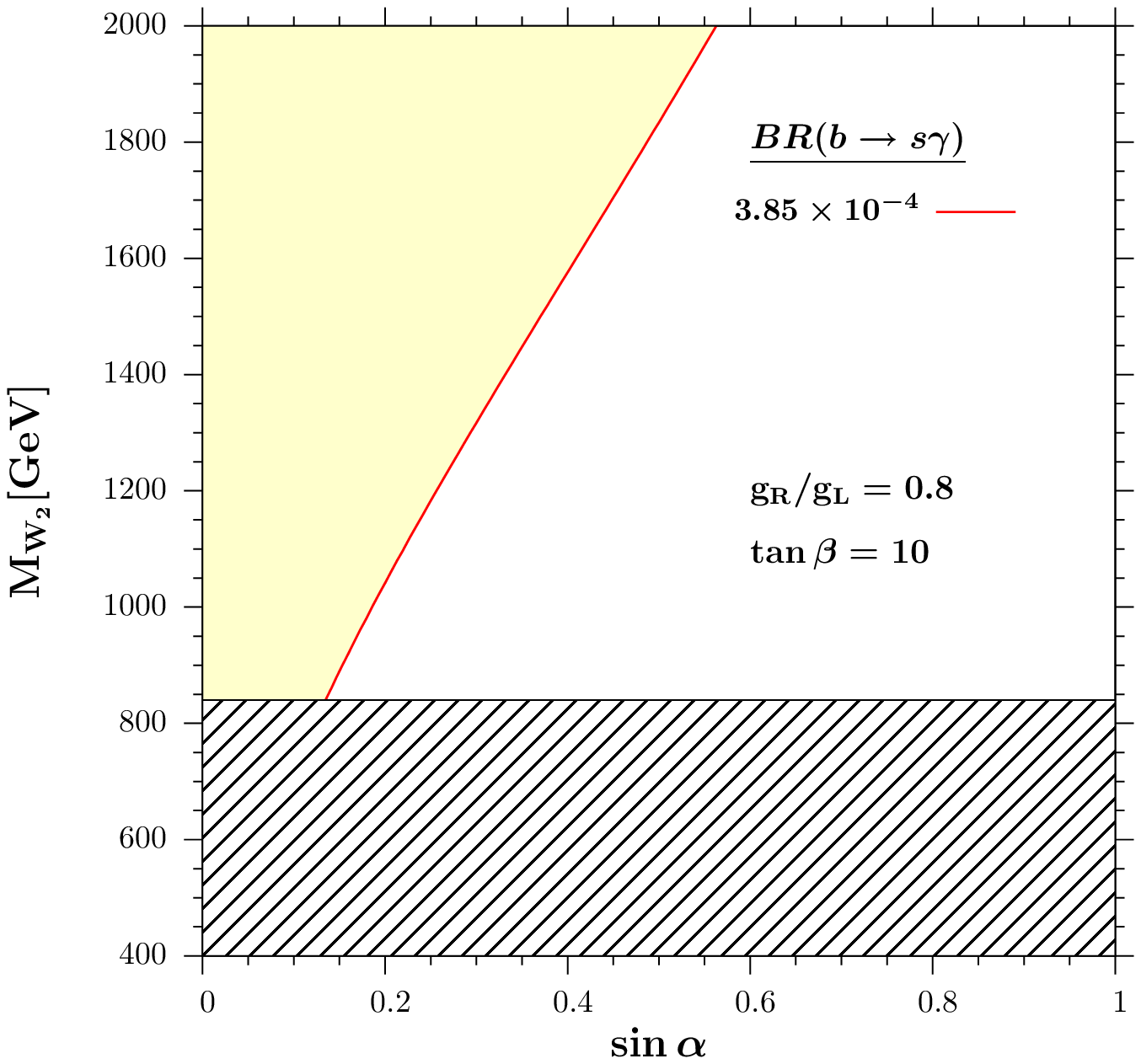} &\hspace*{-0.2cm}
	\includegraphics[width=2.2in,height=2.2in]{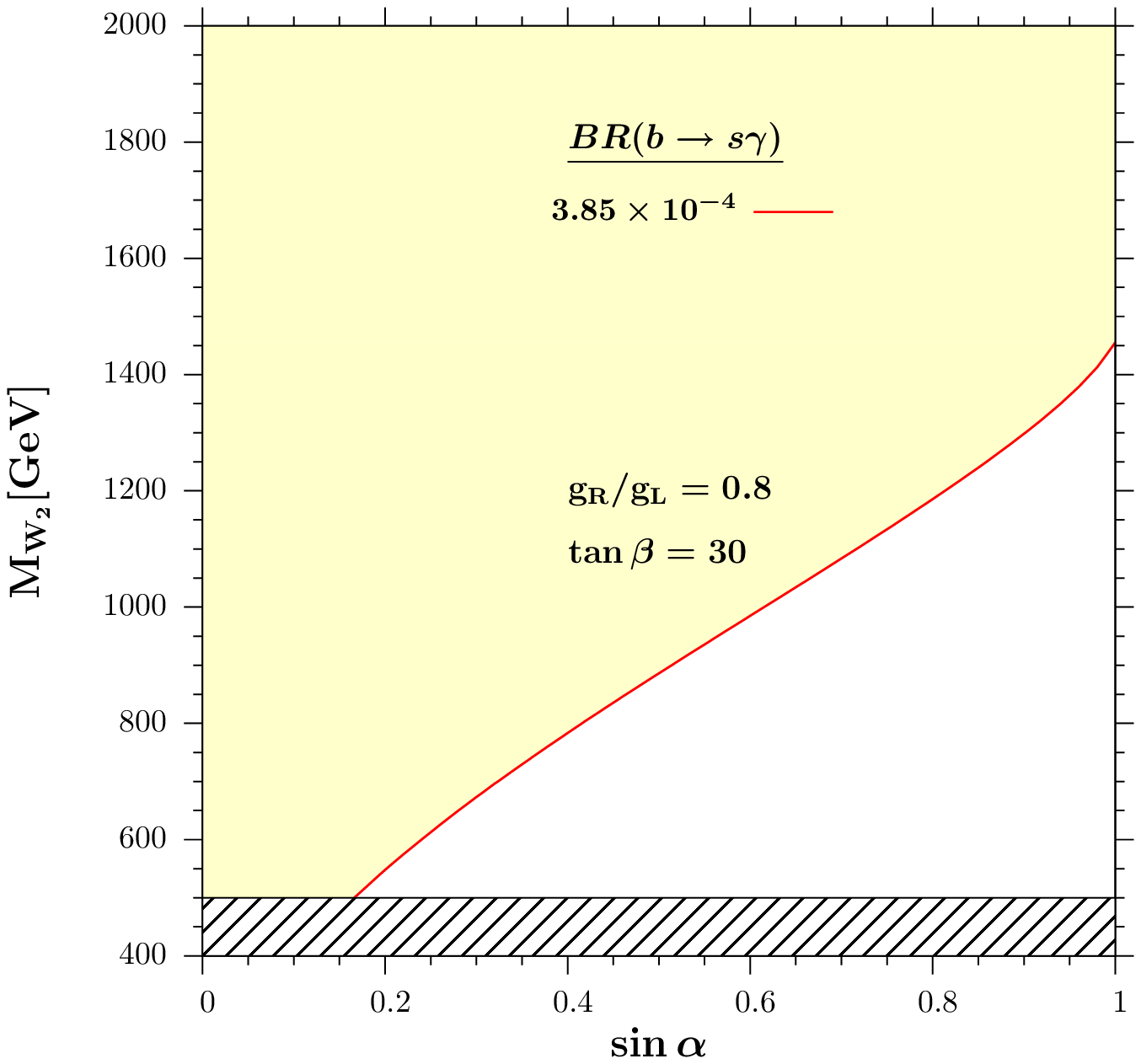} &\hspace*{-0.2cm}
        \includegraphics[width=2.2in,height=2.2in]{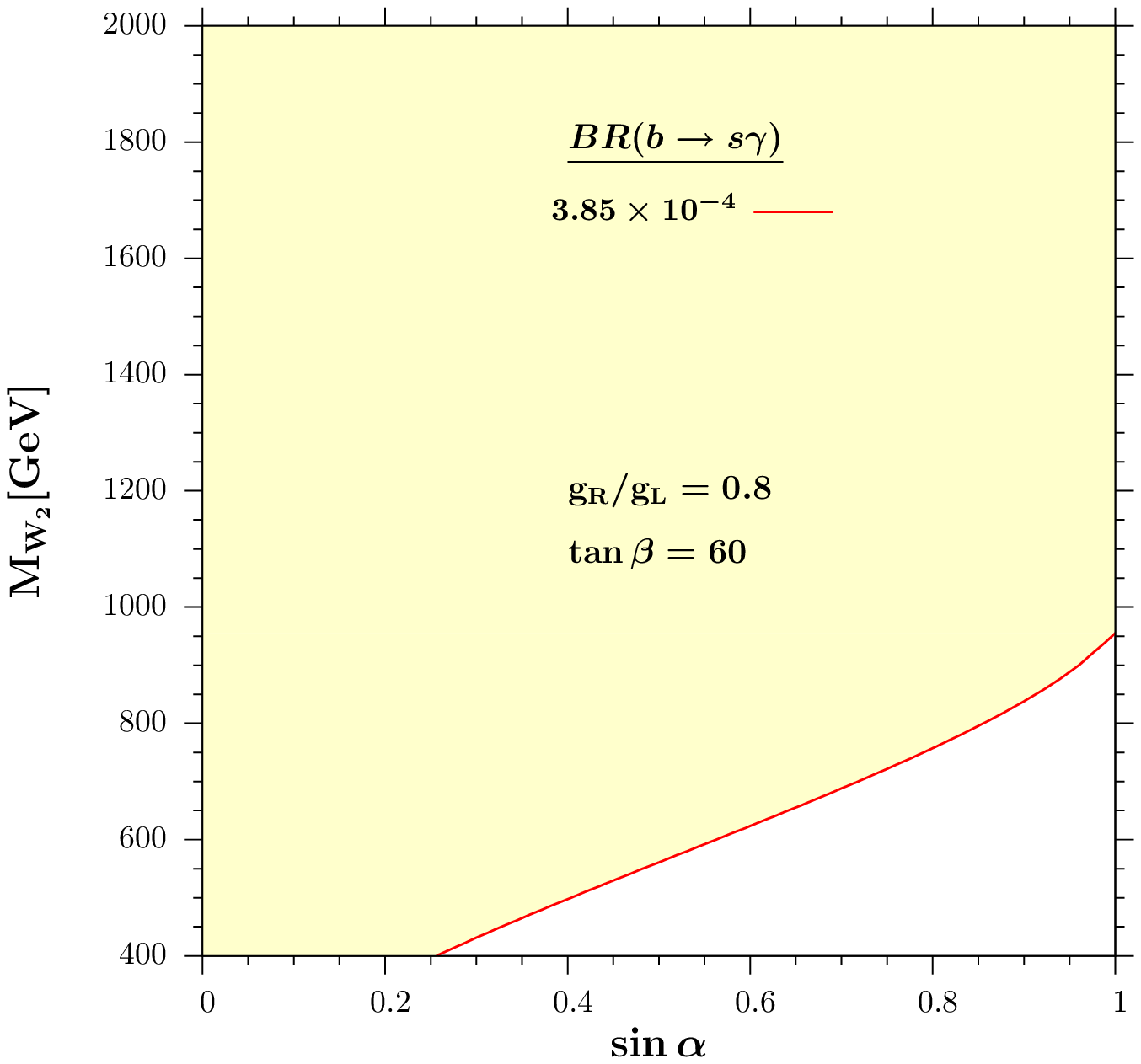} \\
\hspace*{-0.7cm}
	\includegraphics[width=2.2in,height=2.2in]{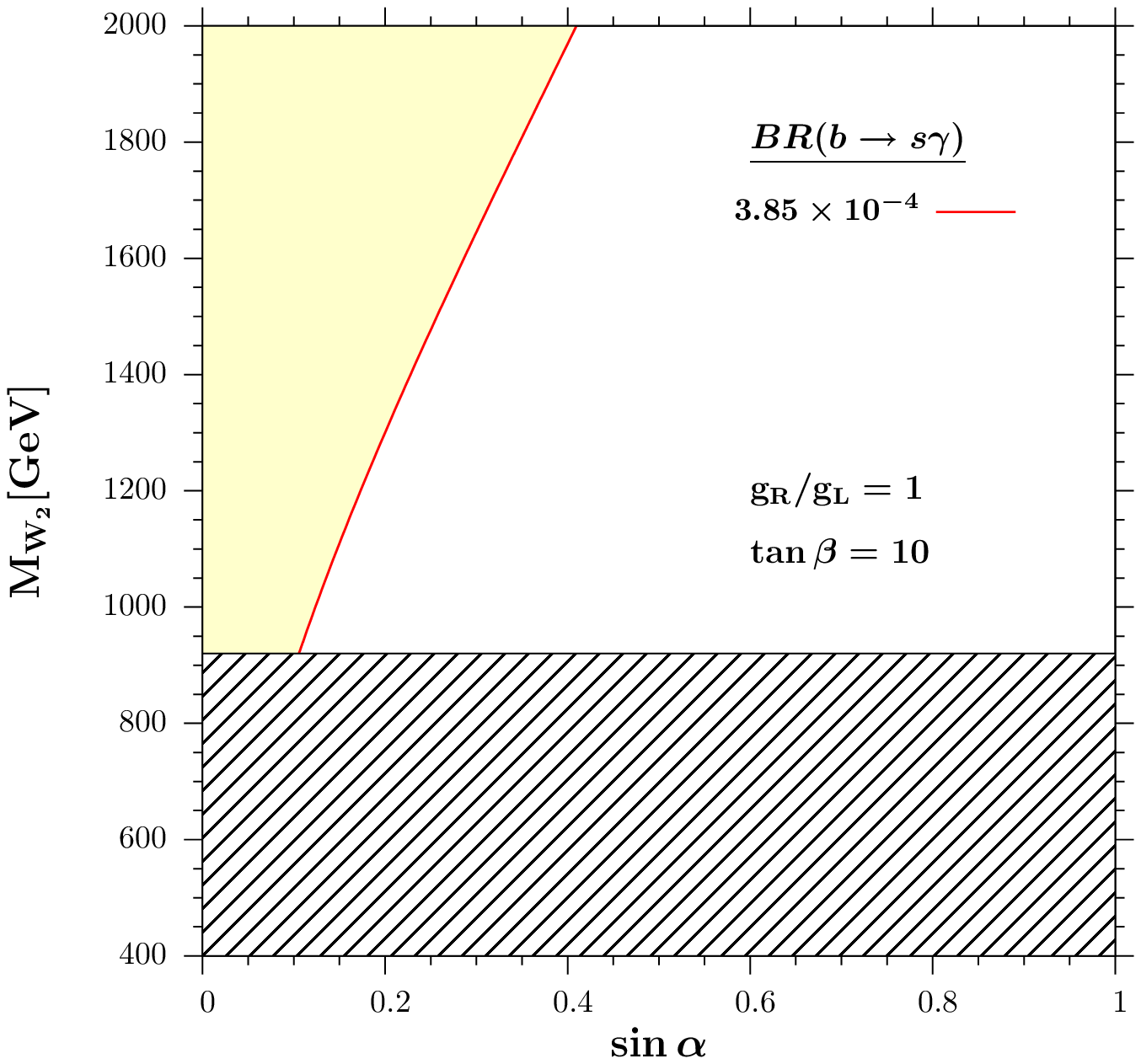} &\hspace*{-0.2cm}
	\includegraphics[width=2.2in,height=2.2in]{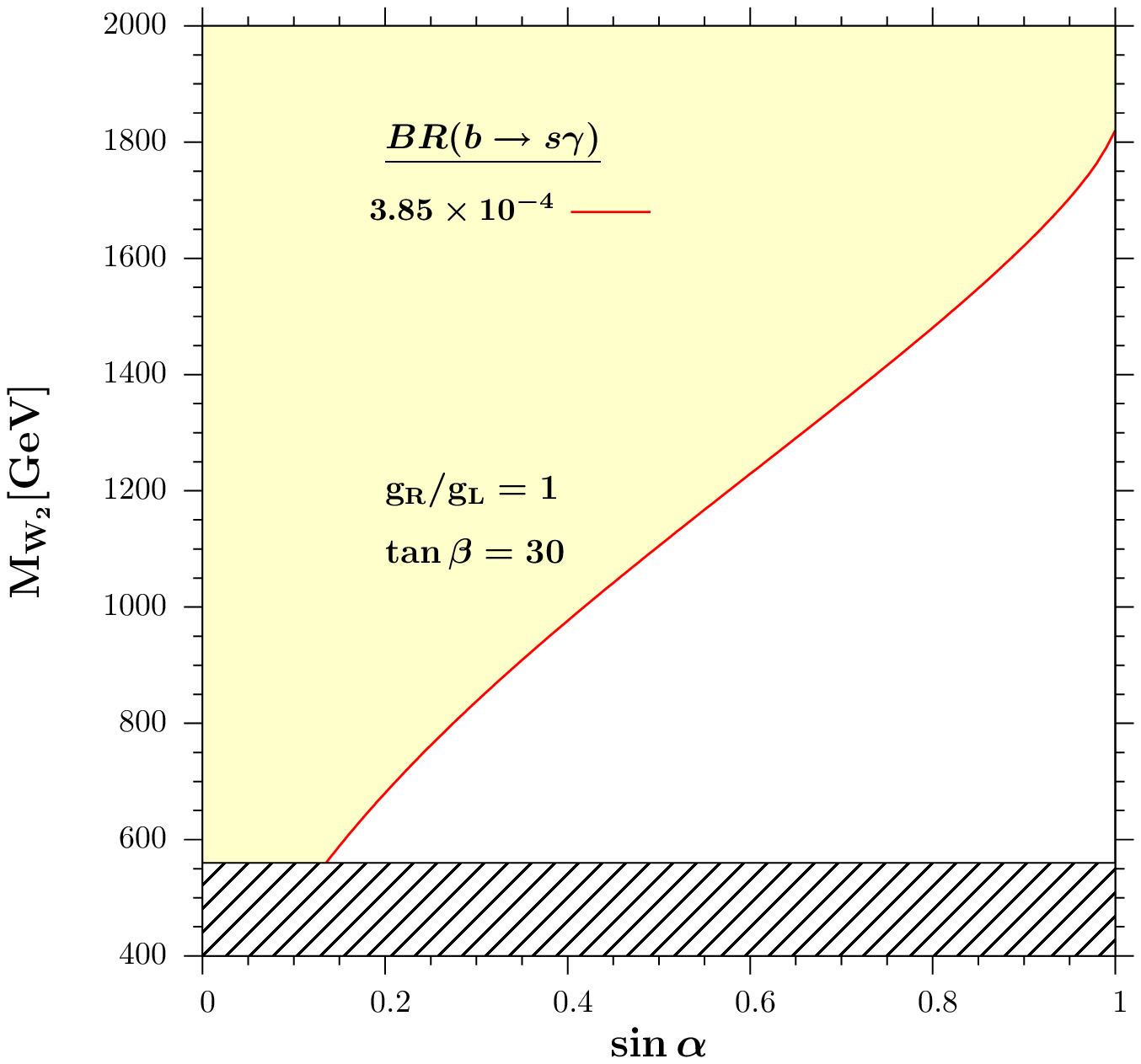} &\hspace*{-0.2cm}
        \includegraphics[width=2.2in,height=2.2in]{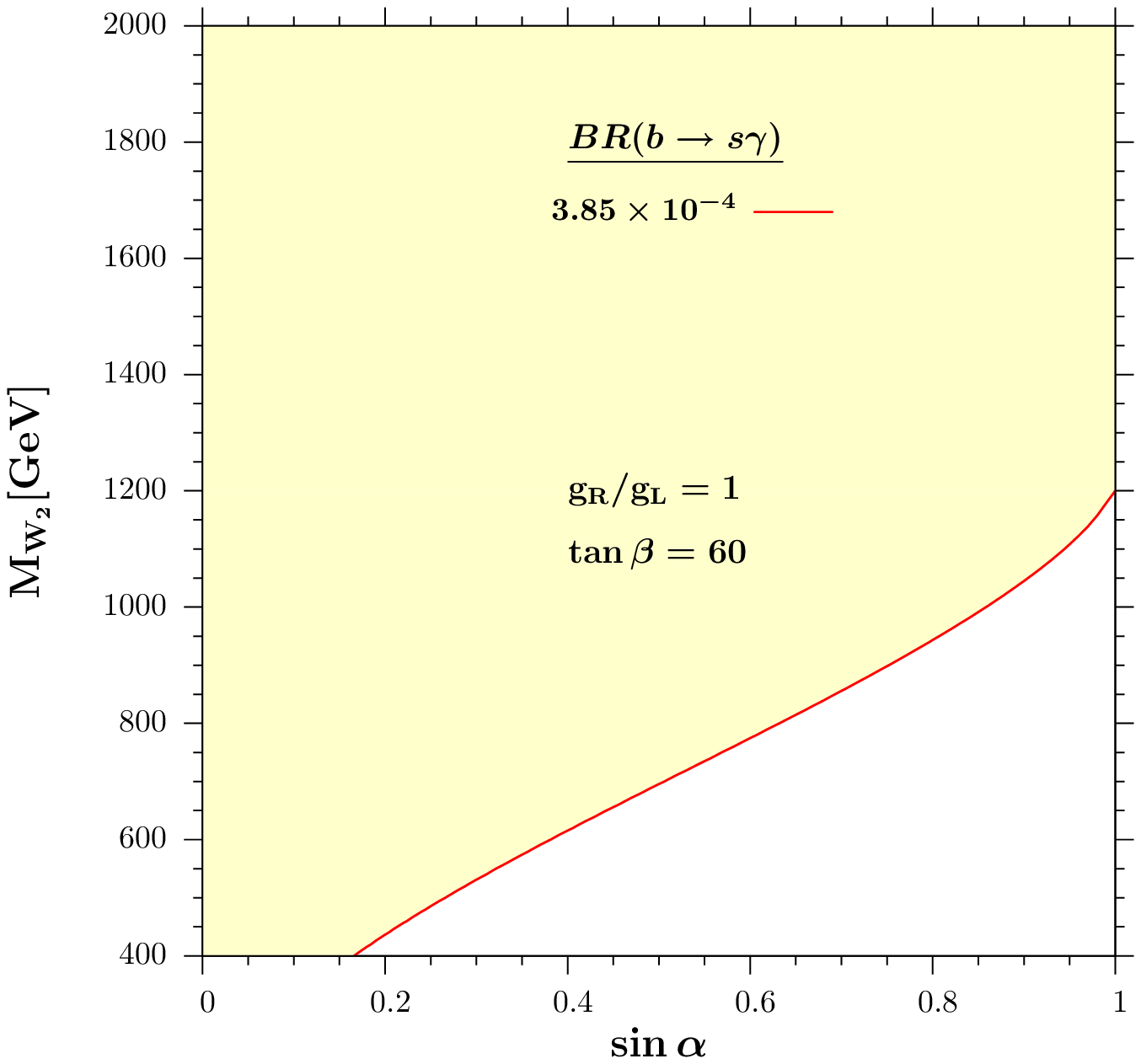} 
	\end{array}$
\end{center}
\vskip -0.3in
      \caption{Contour plot of the $M_{W_2}$ vs $\sin\alpha$ constraint in the $V^R=V_{(A)}$ parametrization,  from $b \to s \gamma$. We fix the BR$(b \to s \gamma)$ to be in the interval $(3.20-3.85) \times 10^{-4}$, and vary $g_R/g_L$ and $\tan \beta$, as indicated in the panels. We take $M_{H^\pm}=10$ TeV. Black-shaded regions represent areas excluded by the $W_R-W_L$ mixing angle, $\xi \le 3 \times 10^{-3}$. Regions highlighted in yellow represent allowed parameter spaces.}
\label{fig:Br2}
\end{figure}

In Fig.~\ref{fig:Br2}  we present the dependence of the branching ratio of
$b\to s \gamma$ in a contour plot in $M_{W_2}-\sin \alpha$ plane, with $V^R_{ts}= \sin \alpha$ in the $V^R=V^R_{(A)}$ parametrization. (Note that in $V^R=V_{(B)}$ the contribution to the right-handed quark mixings to $ b \to s$ processes is zero). Fixing the mass of the charged Higgs boson to 
$M_{H^\pm}=10$ TeV{\footnote {As required by the $B^0-{\bar B}^0$ mixing, see discussion in the next subsection.}}, we consider various $\tan\beta$ and $g_R/g_L$ values.  While we allow the ratio of $g_R/g_L$ to vary, it is not allowed to have arbitrary values. As $SU(2)_R \times U(1)_{B-L}$ breaks to $U(1)_Y$, the coupling constants of the three groups $g_R, g_{B-L}$ and $g_Y$ are related, requiring
$g_R/g_L > \tan \theta_W$. For coupling ratios outside this interval, the $Z_R f \bar f$ coupling becomes non-perturbative. 
We restrict the branching ratio to be within the experimentally 
allowed values in the $1 \sigma$ range, 
and  the allowed regions are shaded in yellow, with  upper values in  red. The lower bound value is always allowed by the parameter space chosen. As the SM value in our calculation is BR$(b \to s \gamma)=3.2 \times 10^{-4}$, the region in which $\sin \alpha=0$, which corresponds to no contribution from the right-handed side, is always included in the allowed parameter space.
The  $g_R/g_L$ value is kept constant along the rows of the graphs in Fig.~\ref{fig:Br2}. The values are   $g_R/g_L= 0.6,~ 0.8$ and $1$ for the first, second and third row, respectively. We vary $\tan \beta$ between $10$ and $60$ among the panels. Increasing $\tan \beta$ for a fixed $g_R/g_L$ value widens the allowed 
parameter space for $V^R_{ts}=\sin \alpha$. The reason is that,  for $\tan \beta\ge 5$, the dominant Higgs contribution  is proportional to $1/\cos^2 2 \beta$.  This contribution increases with $\tan \beta$ and thus requires a larger compensating $W_2$ contribution, thus enlarging the parameter space allowed to satisfy the experimental bounds.
 Taking $\tan \beta \to 0$ and $M_{W_2} \to \infty$ does not reduce the model to the SM for 
the chosen Higgs mass; one would also need to take $M_{H^\pm} \to \infty$ limit to recover the SM. Going down the plots along the columns of Fig.~\ref{fig:Br2}, 
we investigate the effects of varying the ratio $g_R/g_L$.  For low $\tan \beta$, the parameter regions available for $V^R_{ts}= \sin \alpha$  are reduced because one effectively increases the contribution of $W_2$ for a fixed Higgs contribution; 
while increasing $\tan \beta$ increases the Higgs contribution, opening more parameter space for $V^R_{ts}= \sin \alpha$.  The region shaded is excluded by the restriction on the $W_R-W_L$ mixing 
angle, $\xi < 3 \times 10^{-3}$. In conclusion, Fig.~\ref{fig:Br2} shows that large values of $\tan \beta$ insure that a large parameter space for 
$V^R_{ts}= \sin \alpha$ is allowed as $M_{W_2}$ gets larger; while smaller values of $g_R/g_L$ allow larger flavor violation in the right-handed sector, even for low $W_2$ masses.
\begin{figure}[htb]
\begin{center}$
\begin{array}{cc}
\includegraphics[width=2.2in,height=2.2in]{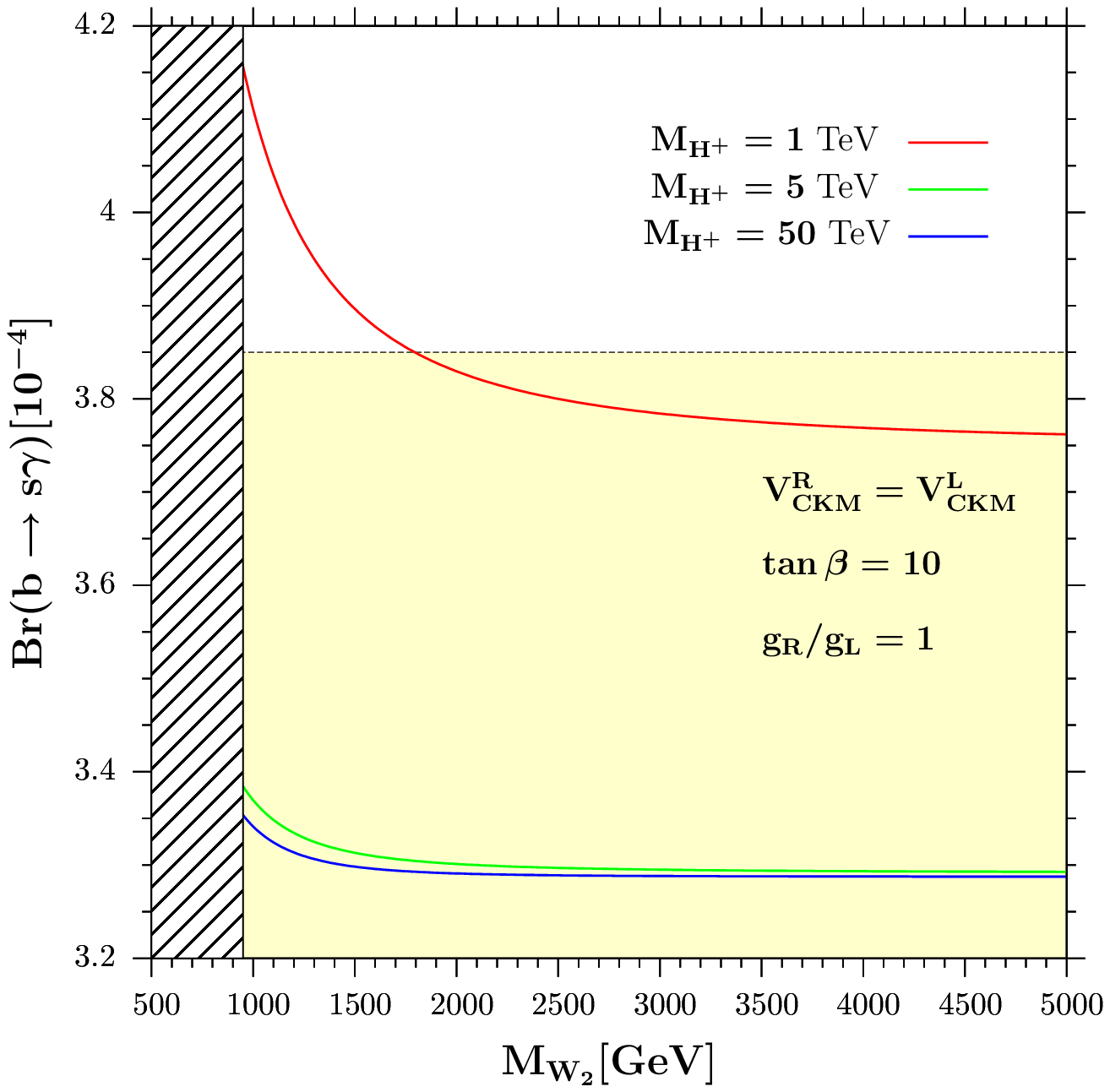} ~~&~~ 
\includegraphics[width=2.2in,height=2.2in]{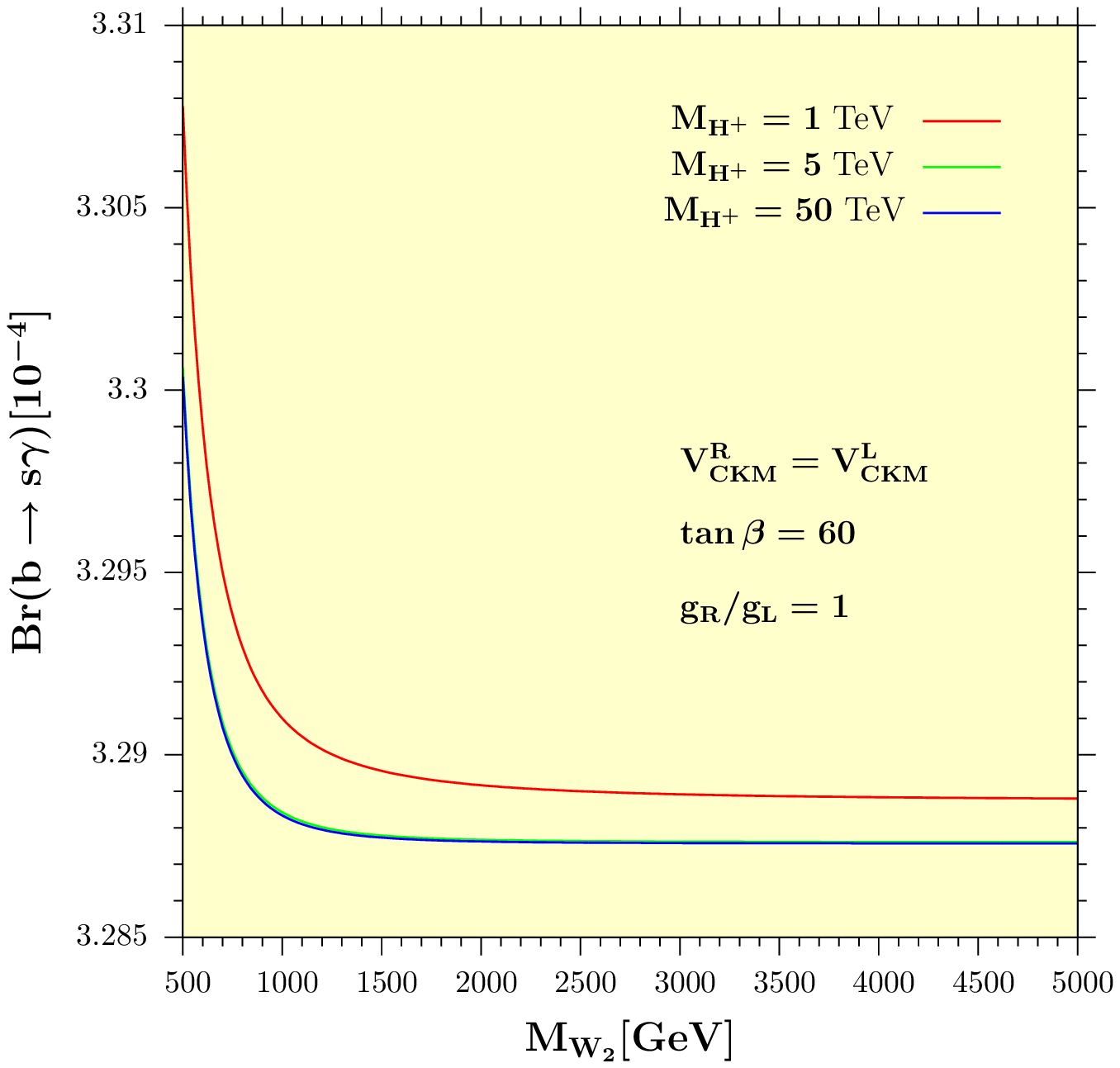} \\
\end{array}$
\end{center}
\vskip -0.3in
\caption{BR($b \to s \gamma$) as a function of the $W_2$ mass in the manifest LRSM, $V_{CKM}^R=V_{CKM}^L$. We take $\tan \beta=10$ in the left panel and $\tan \beta=60$ in the right panel. The curves in red, green and blue correspond to, respectively $M_{H^\pm}=1, ~5$ and $50$ TeV. Yellow highlighted regions represent allowed spaces; the black shaded region is excluded by the $W_L-W_R$ mixing angle. }
\label{fig:Br3}
\end{figure}
For comparison, we investigate the same dependence in the manifest left-right symmetric model in Fig.~2. There is not $\sin \alpha$ dependence there, as the flavor violation in the right-handed sector is fixed; and so is $g_R=g_L$. As in our model, large $\tan \beta$ allows for a larger parameter space. The main difference lies in the fact that in manifest left-right symmetry $V_{ts}^R \sim {\cal O}(10^{-2})$ while in our model, $V_{ts}^R=\sin\alpha$ is allowed to vary and be large. Thus in the manifest left-right model the contribution for $W_2$ is relatively smaller, allowing for contributions from lighter charged Higgs. The $W_2$ mass is required to be at least $1$ TeV for $\tan \beta=10$, while for $\tan \beta=60$, the $W_2$ mass is allowed to be as light as $500$ GeV. Higgs masses of $1$ TeV are ruled out for $M_{W_2} < 2$ TeV for $\tan \beta=10$, but not for $\tan \beta=60$. In both cases, the Higgs contribution decouples for $M_{H^\pm} \ge 5$ TeV, while no such statement can be made in our model, where both $V_{ts}^R$ and $g_R/g_L$ are allowed to vary.

\begin{figure}[htb]
\vskip -0.3in 
\begin{center}$
	\begin{array}{ccc}
\hspace*{-0.7cm}
	\includegraphics[width=2.2in,height=2.2in]{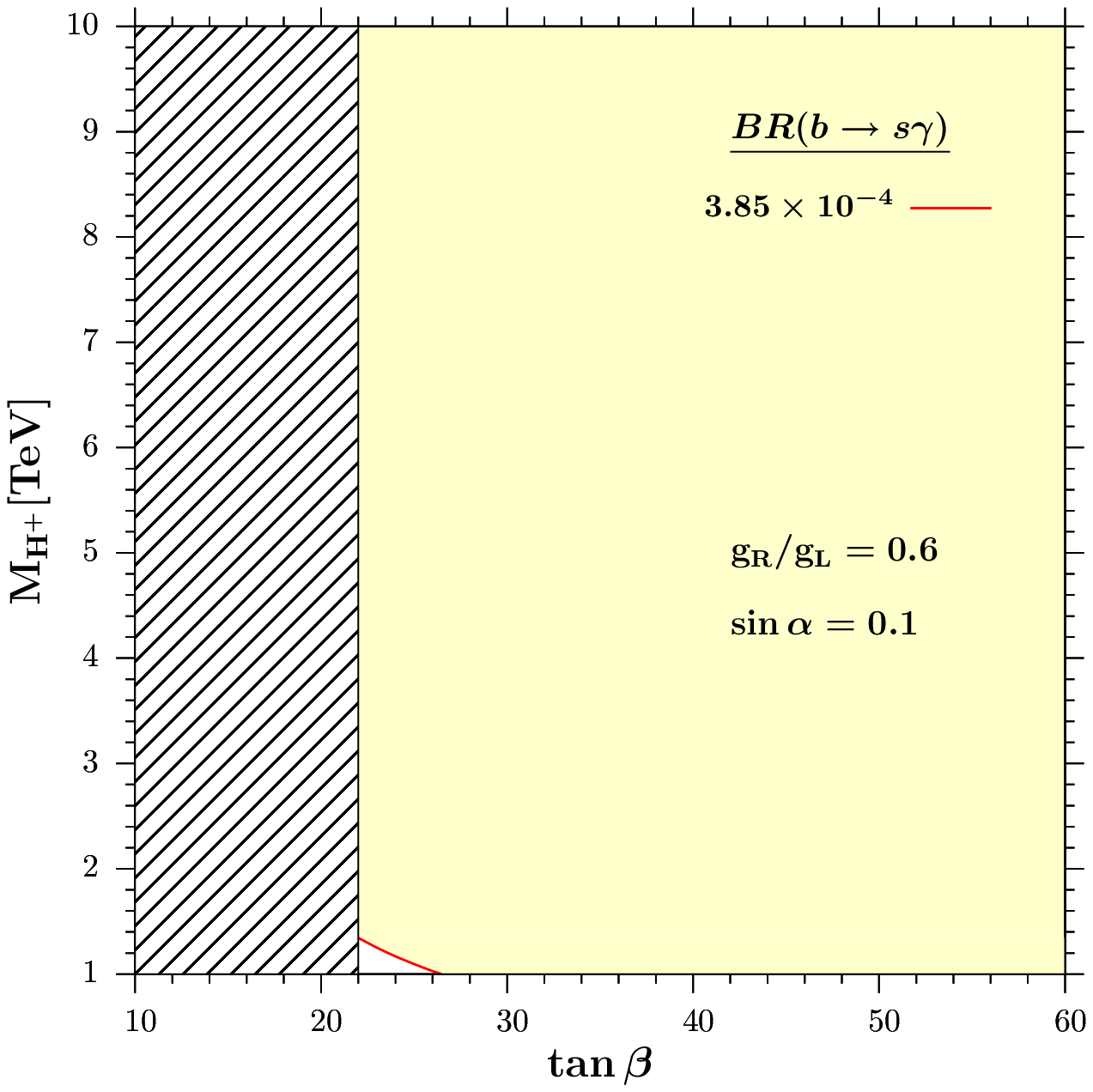} &\hspace*{-0.2cm}
	\includegraphics[width=2.2in,height=2.2in]{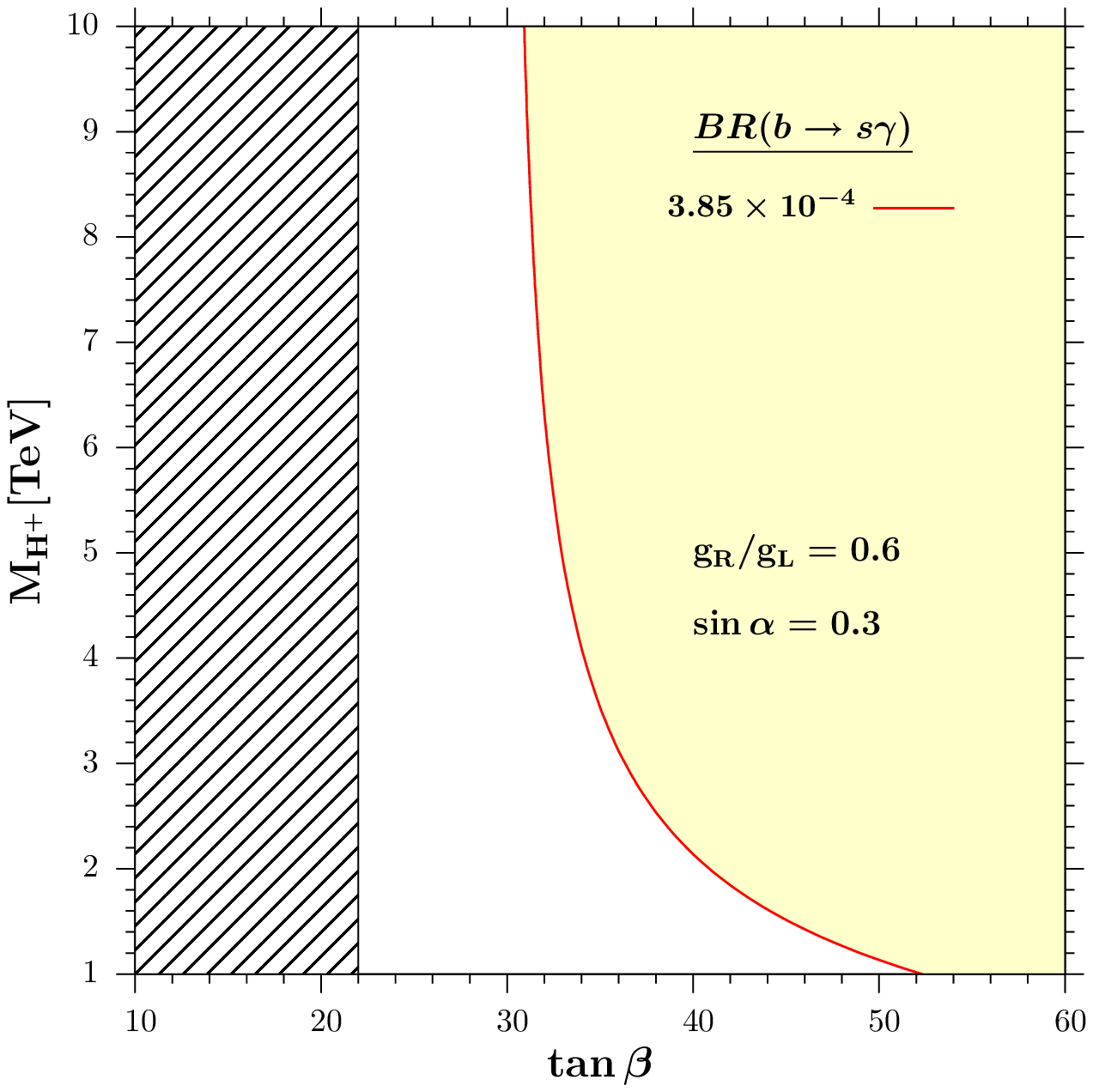} &\hspace*{-0.2cm}
        \includegraphics[width=2.2in,height=2.2in]{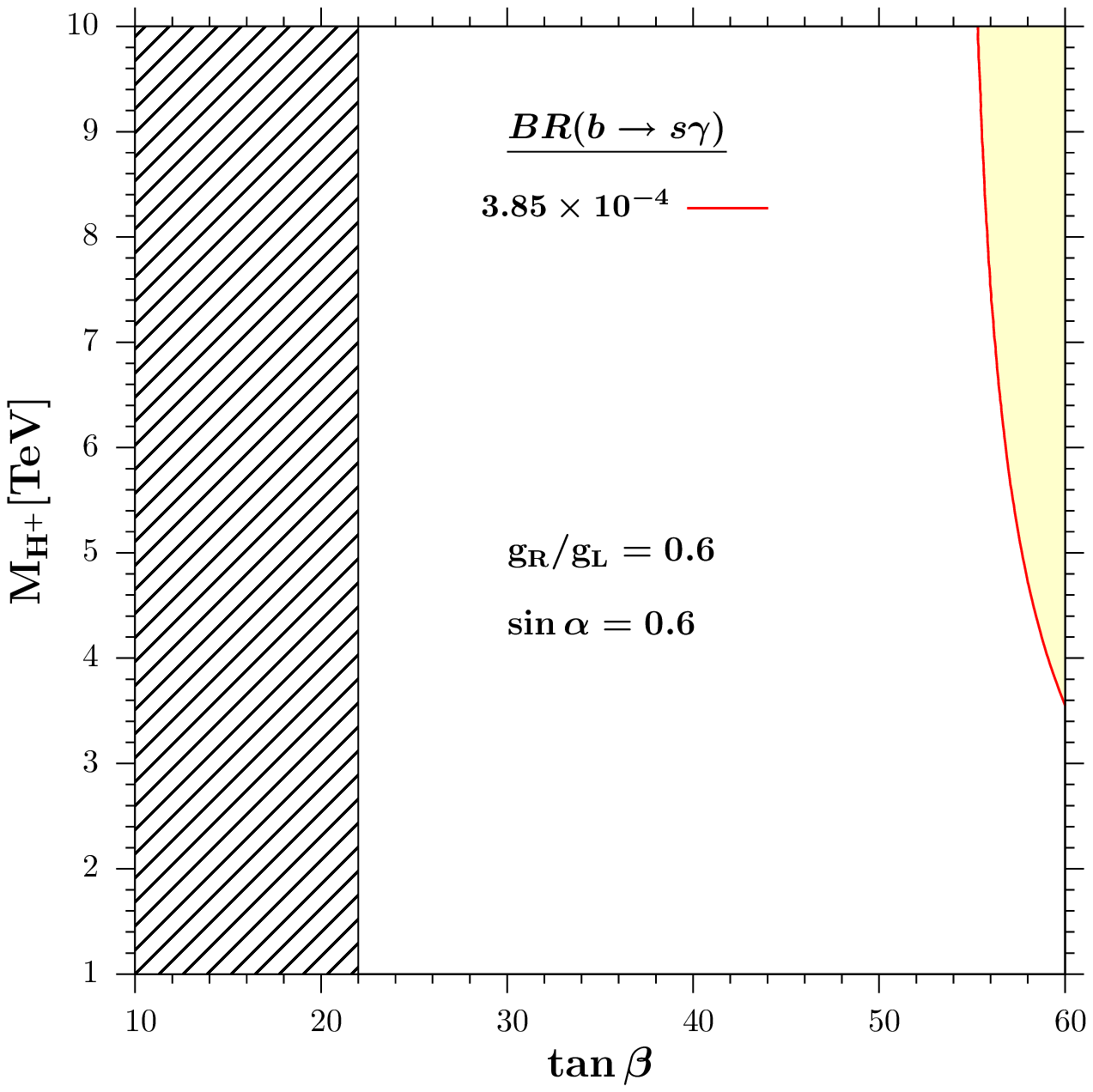} \\
\hspace*{-0.7cm}
	\includegraphics[width=2.2in,height=2.2in]{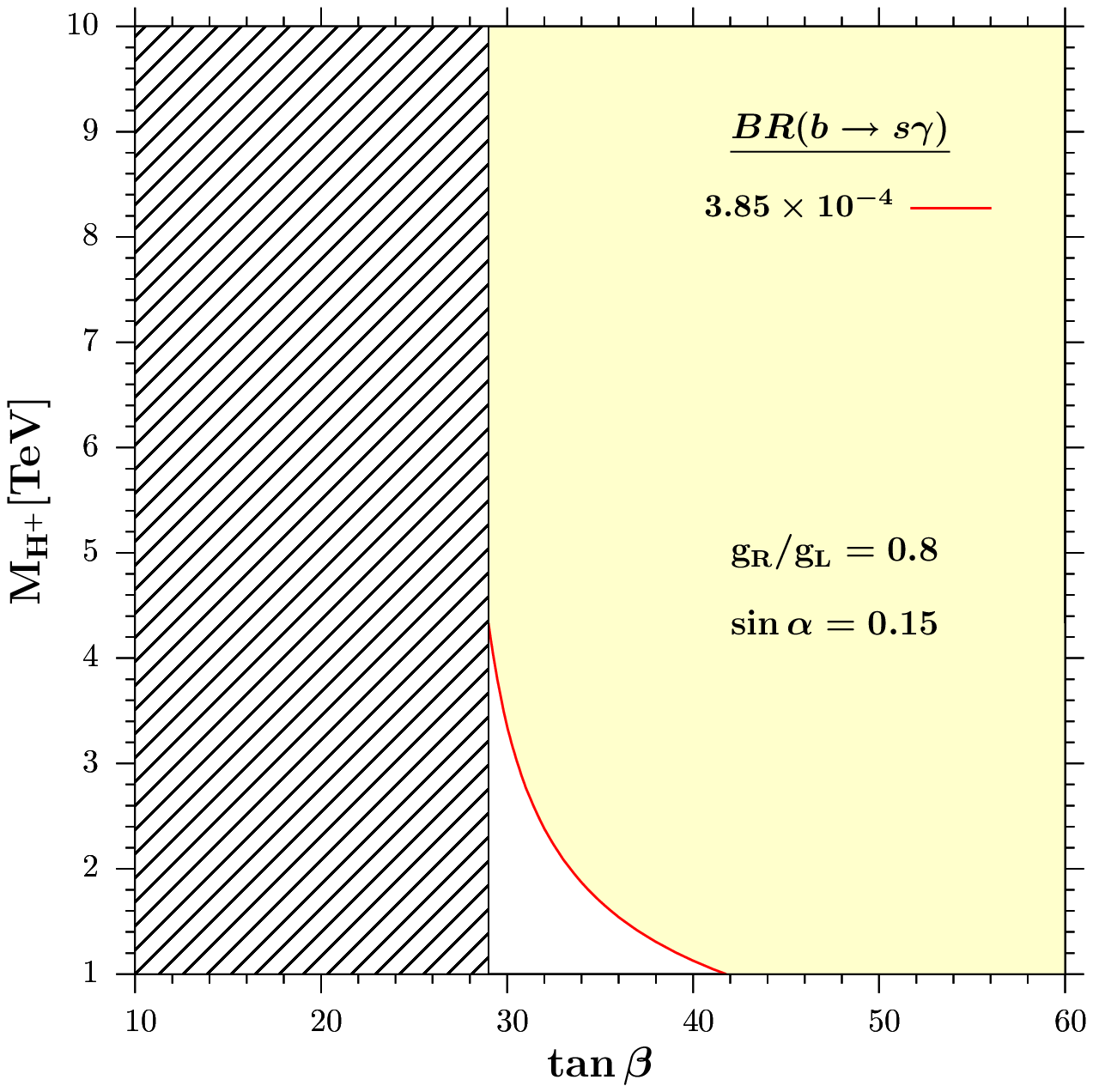} &\hspace*{-0.2cm}
	\includegraphics[width=2.2in,height=2.2in]{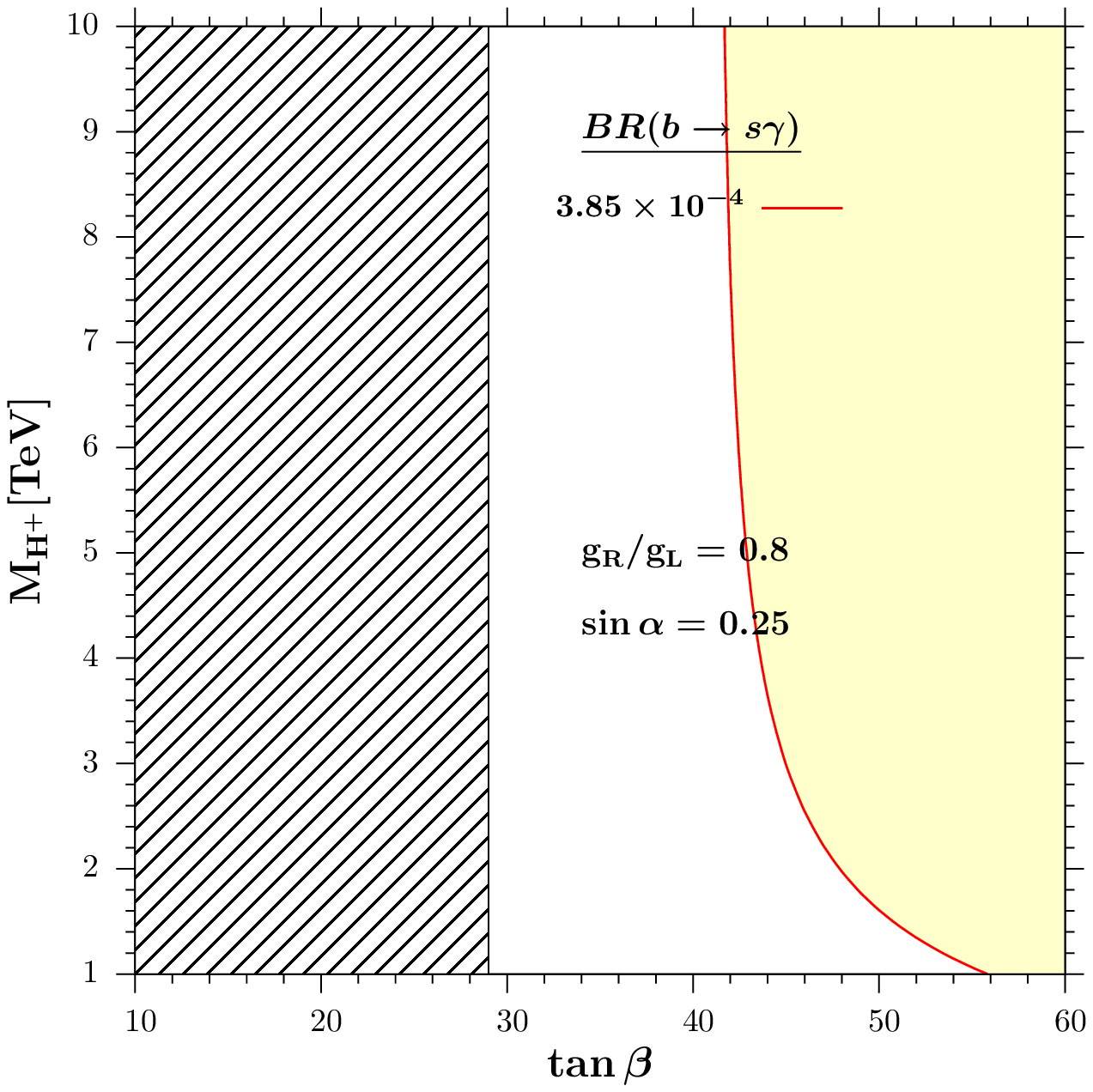} &\hspace*{-0.2cm}
        \includegraphics[width=2.2in,height=2.2in]{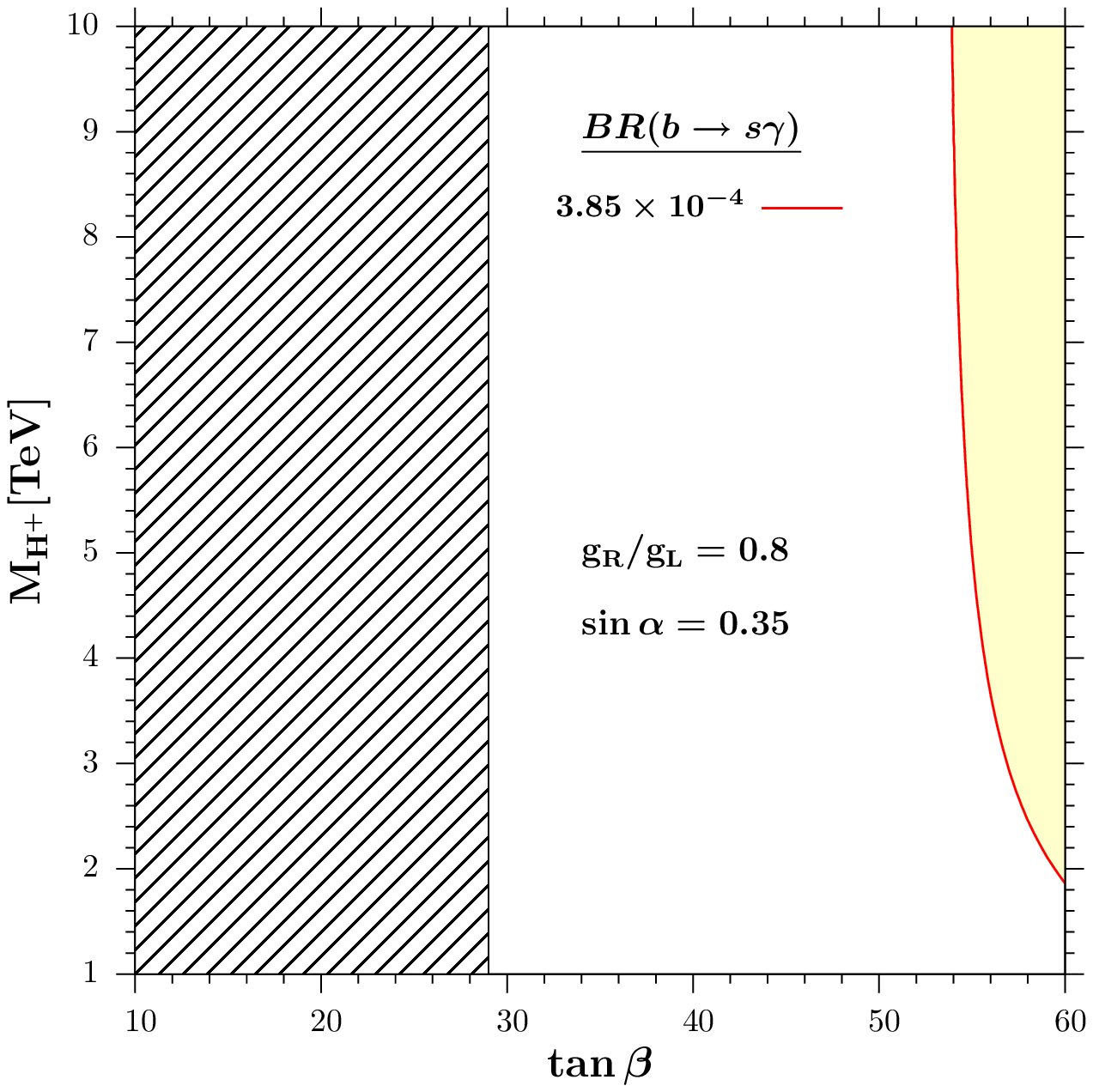} \\
\hspace*{-0.7cm}
	\includegraphics[width=2.2in,height=2.2in]{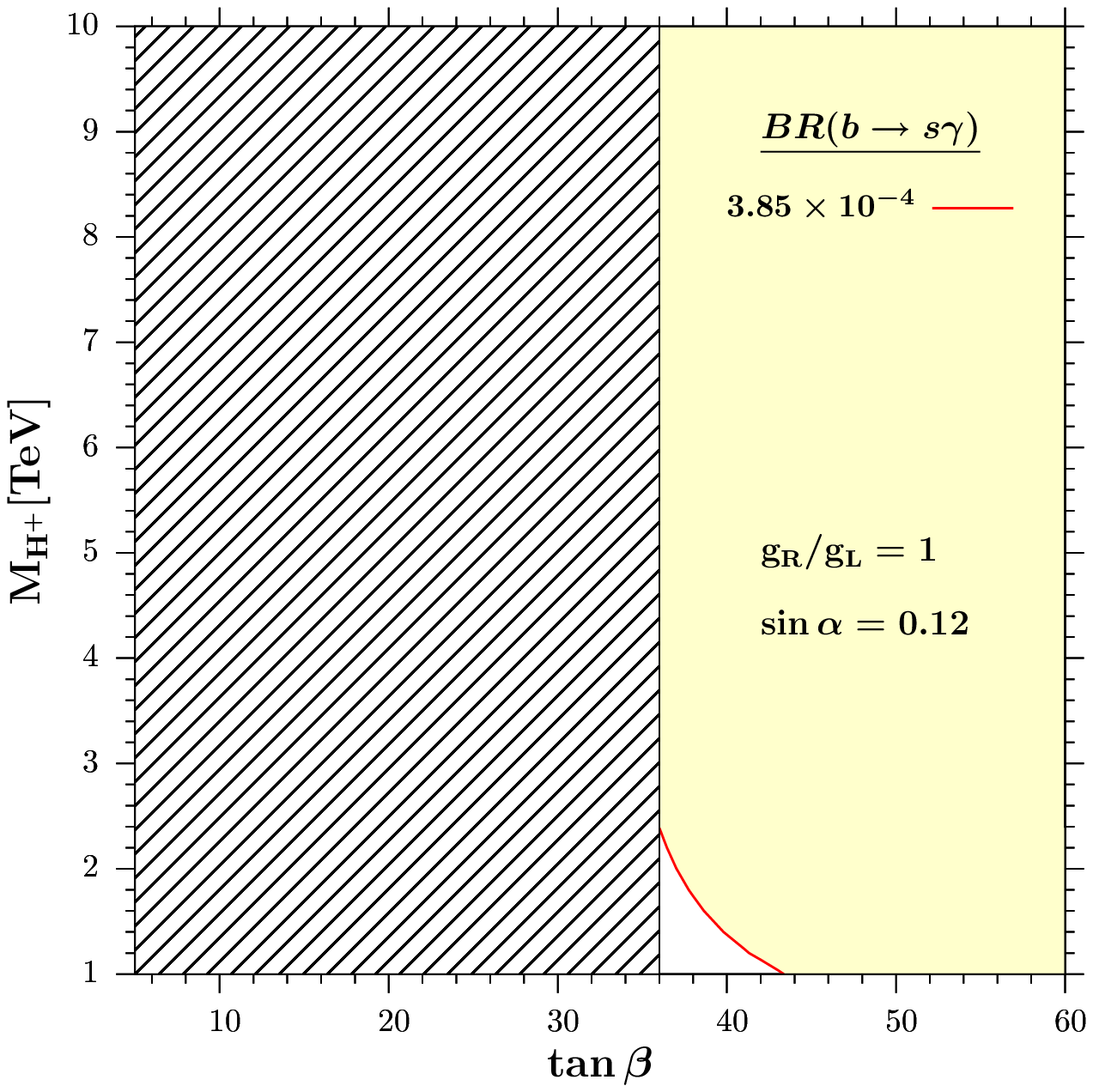} &\hspace*{-0.2cm}
	\includegraphics[width=2.2in,height=2.2in]{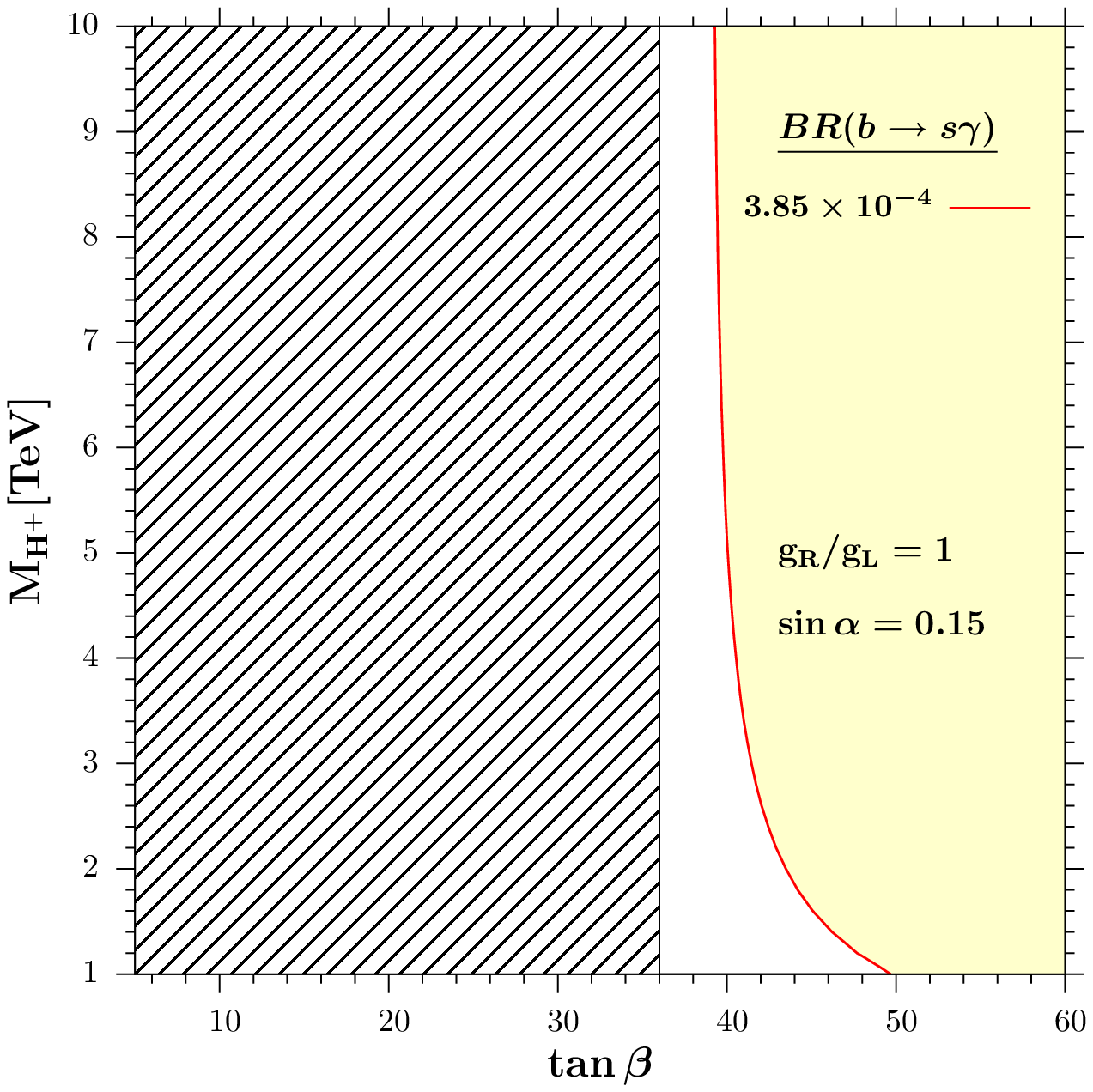} &\hspace*{-0.2cm}
        \includegraphics[width=2.2in,height=2.2in]{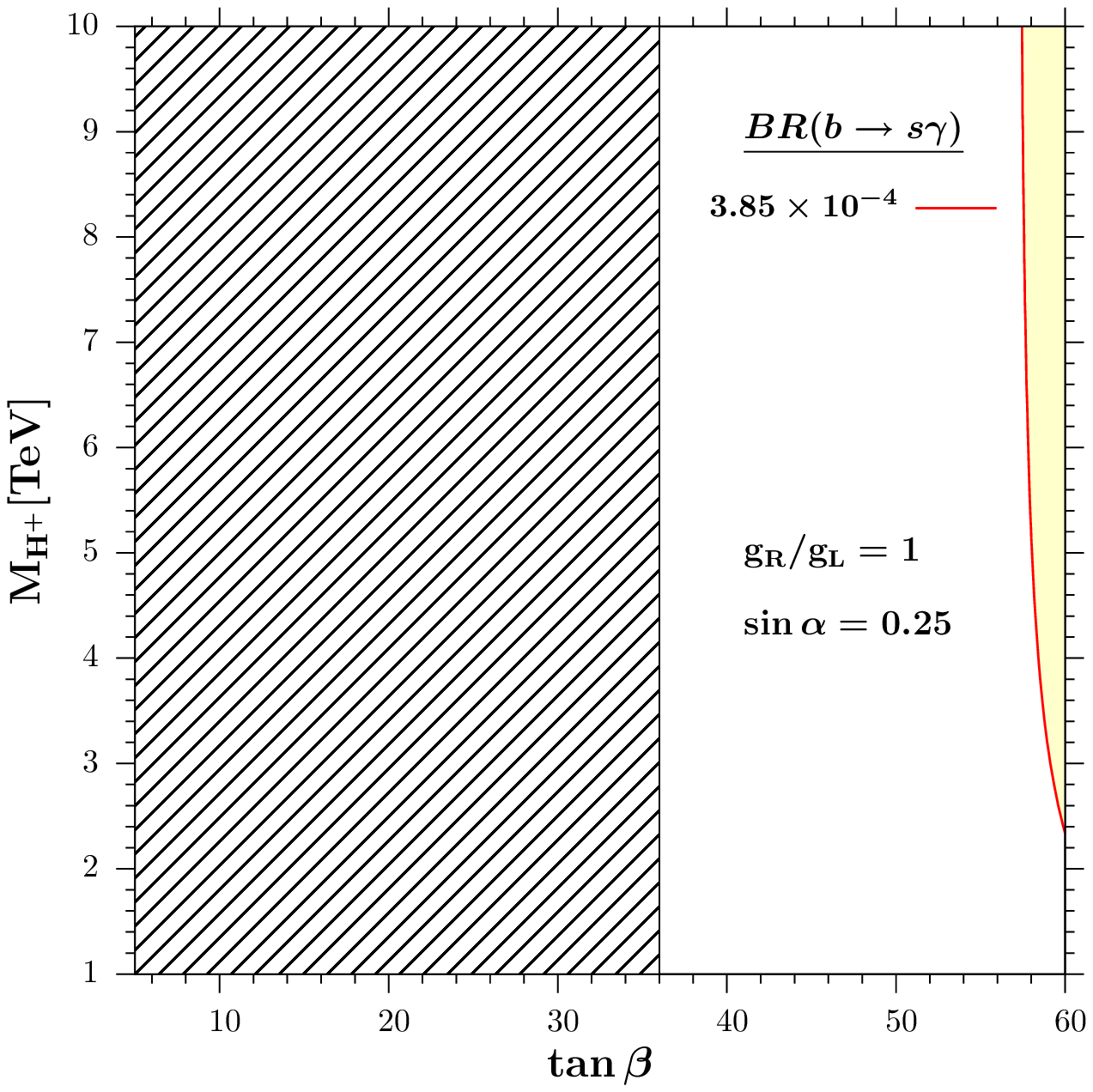} 
	\end{array}$
\end{center}
\vskip -0.3in
      \caption{Contour plot of the $M_{H^\pm}$ vs $\tan\beta$ constraint in the $V^R=V_{(A)}$ parametrization,  from $b \to s \gamma$. We fix the BR$(b \to s \gamma)$ to be in the interval $(3.20-3.85) \times 10^{-4}$, and vary $g_R/g_L$ and $\sin \alpha$, as indicated in the panels. We take $M_{W_2}=500$ GeV. Shaded regions represent areas excluded by the $W_R-W_L$ mixing angle, $\xi \le 3 \times 10^{-3}$. Regions highlighted in yellow represent allowed parameter spaces.}
\label{fig:Br4}
\end{figure}

In Fig.~\ref{fig:Br4} we investigate the dependence of the branching ratio of $b\to s \gamma$ on the $H^\pm$ mass and $\tan\beta$ 
in the $V^R=V_{(A)}^R$ parametrization. We fix  the mass  $M_{W_2}=500$ GeV (as we are interested in 
the consequences of a light gauge boson) and vary $V^R_{ts}= \sin \alpha$ and $g_R/g_L$. We again restrict the branching ratio 
to be within $1 \sigma$ range and give contour plots for the allowed regions (highlighted in yellow, with  upper values 
in  red; as before, lower values are always allowed in the chosen parameter space). For each of the rows of plots in  Fig.~\ref{fig:Br4} we keep $g_R/g_L$ constant and choose values for $V^R_{ts}= \sin \alpha$. 
For fixed ratios $g_R/g_L$, increasing $\sin \alpha$ shifts  the allowed parameter space to higher values of $\tan \beta$, and this result is independent of $M_{H^\pm}$. 
The result is in complete agreement with our observations on the $\tan \beta$ influence in Fig.~\ref{fig:Br2},   where the Higgs contribution was  
needed to compensate for a large flavor mixing in the right-handed sector.  Going down the plots along the columns of Fig.~\ref{fig:Br4}, we 
analyze the effects of varying $g_R/g_L$.   The second row shows that for larger $g_R/g_L$ ratio,  allowed parameter  regions are moving towards larger $\tan \beta$.   For the last row, where $g_R/g_L=1$, the allowed region of the parameter space is extremely sensitive to $\sin \alpha$, and consistent with the data only for very 
small values for $V^R_{ts}= \sin \alpha$.  Even for relatively small right-handed flavor violation, $\sin \alpha =0.25$, most of the region of 
the parameter space is ruled out. Here the contribution from the right-handed gauge boson is large,  large flavor violation requires a 
very large Higgs term contribution, and even large values of $\tan \beta$ are insufficient to generate compensating terms. Here again, the region 
shaded is excluded by the restriction of the $W_R-W_L$ mixing angle $\xi < 3 \times 10^{-3}$; this region depends only on the ratio $g_R/g_L$. In conclusion, we see from Fig.~\ref{fig:Br4} that  larger 
values of $\tan \beta$ and smaller values of $g_R/g_L$ satisfy the $b \to s \gamma$ branching ratio constraints  for a wide parameter space for 
$M_{H^\pm}$, while  low values for $V^R_{ts}= \sin \alpha$ are required  for low $W_2$ masses.

 Following previous studies, we do not analyze $b \to d \gamma$ transitions. Finding new physics effects in the $b \to d$ transition may be easier than in $b \to s$ because the SM amplitude is suppressed in $b \to d$. In the SM, $b \to  s \gamma$ and $b \to  d\gamma$ are both described by a common Wilson coefficient, $C^7_L$. This is also true in any model within a minimal flavor  violating framework in which the flavor changing interactions are determined by the left-CKM angles. However, the experimental measurement for $b \to d \gamma$ is not very precise \cite{btodgamma}
 $$ BR_{\rm Exp}(b \to d \gamma)=\left (1.63 ^{+0.30}_{-0.24} \pm0.16 \right) \times 10^{-6}.$$
Since SM predictions for exclusive modes such as $B \to \rho \gamma$ or $B \to \omega \gamma$ \cite{btodgamma} suffer from large model-dependent uncertainties, it is necessary to measure the inclusive rate for $B \to  X_d \gamma$. The largest experimental challenge is the huge background due to $b \to s \gamma$. The only possible way is probably to sum up exclusive $b \to d \gamma$ modes, perhaps from Belle and KEKB.

\subsection{$B_{d,s}^0- {\bar B}_{d,s}^0$ Mixing} 
\label{sec:B0B0bar}
The  $\Delta B=2$ flavor changing decays have been studied in the context of minimal left-right symmetric models  \cite{Altarelli:1987zf, Gilman:1983ce}.
The mass difference between $B_q^0$ and $ {\bar B}_q^0$ is defined as:
\begin{equation}
\Delta m_q = \frac {\left | \langle B^0_q | H_{eff}^{\Delta B=2} | {\bar B}^0_q \rangle \right |}{m_{B_q}}.
\end{equation}
The effective Hamiltonian $H_{eff}^{(\Delta B=2)}$ for $B^0-\bar{B^0}$ transition is obtained by integrating out the internal loop in the box diagrams responsible for this process.
\begin{equation}
 H_{eff}^{(\Delta B=2)}=\sum_i^6 C_i Q_i + \sum_i^3 \tilde{C}_i \tilde{Q}_i,
\end{equation}
with the following four-quark operators
\begin{eqnarray}
Q_1 &=& (\bar{q}^\alpha \gamma^\mu P_L b^\alpha)\otimes(\bar{q}^\beta \gamma_\mu P_L b^\beta)~~,~~
\tilde{Q}_1 = (\bar{q}^\alpha \gamma^\mu P_R b^\alpha)\otimes(\bar{q}^\beta \gamma_\mu P_R b^\beta), \nonumber \\
Q_2 &=& (\bar{q}^\alpha P_L b^\alpha)\otimes(\bar{q}^\beta P_L b^\beta)~~~~~~~~,~~\,
\tilde{Q}_2 =(\bar{q}^\alpha P_R b^\alpha)\otimes(\bar{q}^\beta P_R b^\beta), \nonumber \\
Q_3 &=& (\bar{q}^\alpha P_L b^\beta)\otimes(\bar{q}^\beta P_L b^\alpha)~~~~~~~~,~~\, 
\tilde{Q}_3 = (\bar{q}^\alpha P_R b^\beta)\otimes(\bar{q}^\beta P_R b^\alpha), \nonumber \\
Q_4 &=& (\bar{q}^\alpha P_L b^\alpha)\otimes(\bar{q}^\beta P_R b^\beta),  \nonumber \\
Q_5 &=& (\bar{q}^\alpha P_L b^\beta)\otimes(\bar{q}^\beta P_R b^\alpha), \nonumber \\
Q_6 &=& (\bar{q}^\alpha \gamma^\mu P_L b^\alpha)\otimes(\bar{q}^\beta \gamma_\mu P_R b^\beta),
\end{eqnarray}
where the superscripts $\alpha$, $\beta$ denote color indices, and $q$ stands for either $d$ or $s$ quark. We used the parametrization of the matrix elements of the operators in terms of the bag parameters in Vacuum Insertion Approximation
\begin{eqnarray}
\langle B^0 | Q_1(\mu) | \bar{B^0}\rangle ~&=& ~~~\frac{1}{3} m_{B_q}^2 f_{B_q}^2 B^q_1(\mu), \nonumber \\ 
\langle B^0 | Q_2(\mu) | \bar{B^0}\rangle ~&=& -\frac{5}{24} \left(\frac{m_{B_q}}{m_b + m_q}\right)^2m_{B_q}^2 f_{B_q}^2 B^q_2(\mu), \nonumber \\
\langle B^0 | Q_3(\mu) | \bar{B^0}\rangle ~&=& ~~\frac{1}{24} \left(\frac{m_{B_q}}{m_b + m_q}\right)^2m_{B_q}^2 f_{B_q}^2 B^q_3(\mu), \nonumber \\
\langle B^0 | Q_4(\mu) | \bar{B^0}\rangle ~&=& ~~~\,\frac{1}{4} \left(\frac{m_{B_q}}{m_b + m_q}\right)^2m_{B_q}^2 f_{B_q}^2 B^q_4(\mu), \nonumber \\
\langle B^0 | Q_5(\mu) | \bar{B^0}\rangle ~&=& ~~\frac{1}{12} \left(\frac{m_{B_q}}{m_b + m_q}\right)^2m_{B_q}^2 f_{B_q}^2 B^q_5(\mu), \nonumber \\
\langle B^0 | Q_6(\mu) | \bar{B^0}\rangle ~&=& -\frac{1}{6} \left(\frac{m_{B_q}}{m_b + m_q}\right)^2m_{B_q}^2 f_{B_q}^2 B^q_6(\mu), 
\label{VIA}
\end{eqnarray}
where $m_{B_q}$ is the mass of the $B_q$ meson, $m_b$ and $m_q$ are the masses of $b$ quark and $d$ or $s$ quark respectively. And the same expressions for the operators $Q_{1,2,3}$ in (\ref{VIA}) are valid for the operators $\tilde{Q}_{1,2,3}$. Performing the renormalization group (RG) evolution down to $m_b$ scale, the associated Wilson coefficients $C_i$'s acquire next-to-leading (NLO) QCD correcting factors
\begin{eqnarray}
 C_i(m_b)= \eta_i(m_b) C_i(m_t),
\end{eqnarray}
where $\eta_i(m_b)$ are the QCD correction factors at NLO \cite{Buras:2001ra},
\begin{eqnarray}
\eta_i(m_b)=\eta_i^{(0)}(m_b) + \frac{\alpha_s(m_b)}{4\pi}\,\eta_i^{(1)}(m_b).
\label{eta}
\end{eqnarray}
We took $\alpha_s(m_b) = 0.22$ and listed the QCD correction parameters $\eta_i(m_b)$ at NLO for all the operators in the Appendix \ref{QCD}. For the meson masses and decay constants, we used the following values
\begin{eqnarray}
 m_{B_d}=5.28 ~{\rm GeV}~~~&,&~~~m_{B_s}=5.37 ~{\rm GeV}, \nonumber \\
 f_{B_d}=0.21 ~{\rm GeV}~~~&,&~~~f_{B_s}=0.25 ~{\rm GeV},
\end{eqnarray}
and the bag-parameters at $\mu=m_b$ scale are given in Table I.
\begin{table}[htb]
 \begin{center}
  \begin{tabular}{|c|c||c|c|}\hline
 ~~$B^d_1(m_b)$~~ & ~~0.87~~ & ~~$B^s_1(m_b)$~~ & ~~0.86~~ \\
 $B^d_2(m_b)$ & 0.82 & $B^s_2(m_b)$ & 0.83 \\
 $B^d_3(m_b)$ & 1.02 & $B^s_3(m_b)$ & 1.03 \\
 $B^d_4(m_b)$ & 1.16 & $B^s_4(m_b)$ & 1.17 \\
 $B^d_5(m_b)$ & 1.91 & $B^s_5(m_b)$ & 1.94 \\ 
 $B^d_6(m_b)$ & 1.00 & $B^s_6(m_b)$ & 1.00 \\ \hline
\end{tabular}
 \end{center}
\caption{Bag-parameter values taken from lattice improved calculations in the RI-MOM renormalization scheme \cite{Becirevic:2001xt}, with the running quark masses $m_b(m_b)=4.5$ GeV and $m_d(m_b)=5.4$ MeV. Notice that we took $B_6=1$ for both cases since the bag parameters for the relevant operator is not known yet.}
\end{table}


All the contributions from $W_{1,2},G_{1,2}$ and charged Higgs bosons are encoded in Wilson coefficients ($C_i$ and $\tilde{C}_i$) in terms of reduced Passarino-Veltman functions. We do not give explicit expressions for the different contributions, in the interest of brevity, as some have been presented before. For the analytical evaluation of the diagrams we again used the {\tt FeynArts} to generate the amplitudes in 
't Hooft-Feynman gauge with the approximation of neglecting external momenta.  However, in the limit of vanishing external momenta, all four-point functions in {\tt LoopTools} are known to be ill-defined, so when using them in  numerical calculations we introduced analytical expressions for all the relevant four-point functions, and we listed them in the Appendix \ref{PVint}.

Experimentally,  the mass differences are known with high
precision \cite{hfag,CDFD0} 
\begin{equation}
\Delta m_d = (0.508\pm 0.004)/\text{ps}\,,\qquad
\Delta m_s = (17.77\pm0.10\pm 0.07)/{\rm ps}.
\end{equation}
However, evaluation of the SM contributions is less precise   \cite{etaB}. The measured value can be explained by the SM within 20\% theoretical uncertainty $\Delta m_d$ is $(0.53 \pm 0.08)$ ps$^{-1}$, the error arising from uncertainties in $\overline{MS}$ mass values, bag parameters and the decay constant \cite{Mahmoudi:2009zx}. This  is consistent with our results. If we were to strictly impose the experimental constraints, we might incorrectly omit an important part of the parameter space. Estimating the theoretical errors conservatively at 15\%{\footnote{This is the same as assuming a Gaussian distribution and calculating the total error from the experimental and theoretical ones.}}, we restrict the parameter space for $\Delta m_d=(0.43-0.58)$ ps$^{-1}$ and $\Delta m_s=(15-20)$ ps$^{-1}$. We evaluate the SM contributions as: $\Delta m_d=0.48$ ps$^{-1}$ and $\Delta m_s=17.66$ ps$^{-1}$. 
 The parameters are, as before $M_{W_2}, \,M_{H^\pm}, \,\tan\beta, g_R/g_L$ and $\sin \alpha$, the measure of flavor violation in the right-handed 
quark sector. 

\begin{figure}[htb]
\begin{center}$
	\begin{array}{ccc}
\hspace*{-0.7cm}
	\includegraphics[width=2.2in,height=2.2in]{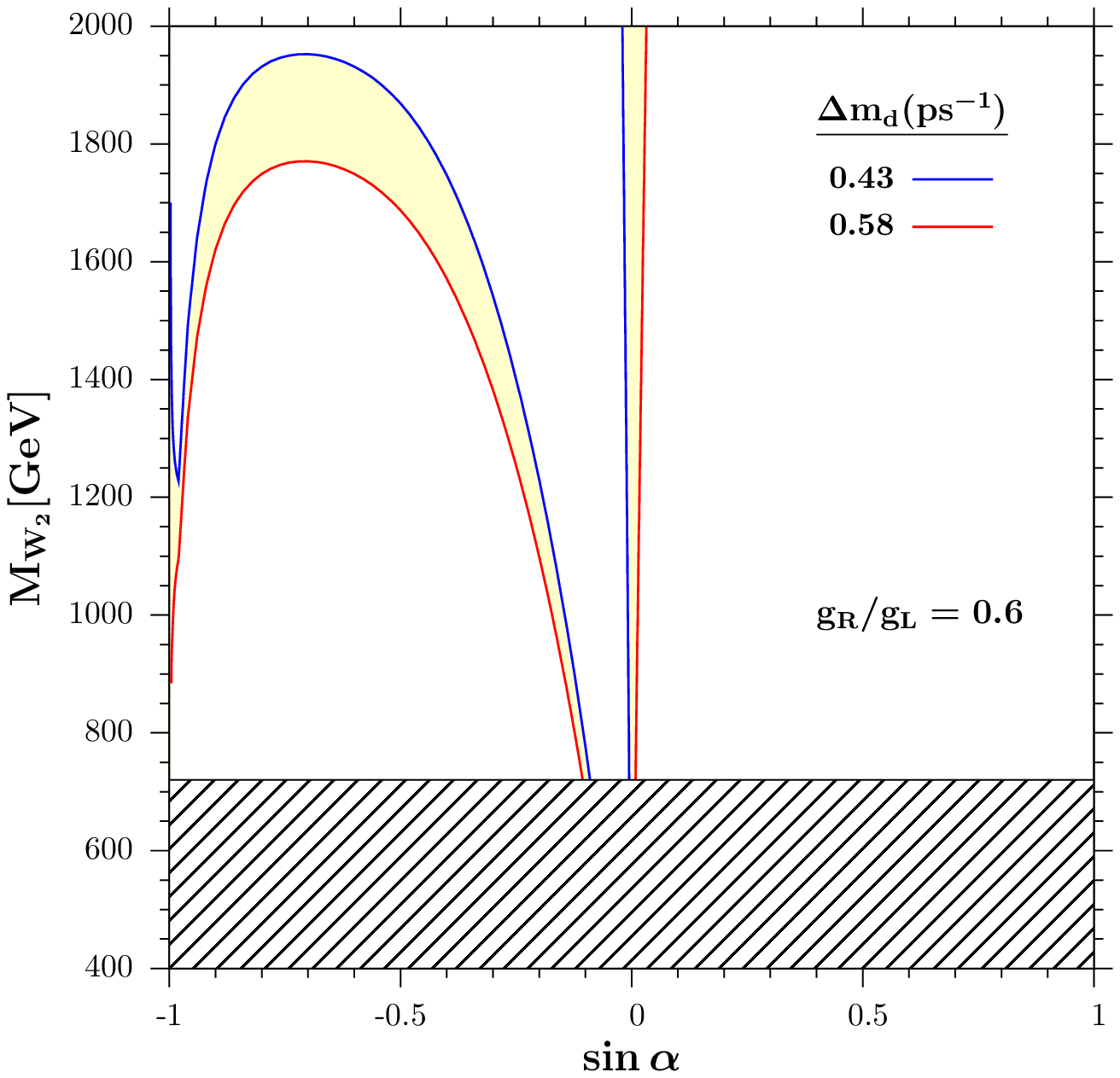} &\hspace*{-0.2cm}
	\includegraphics[width=2.2in,height=2.2in]{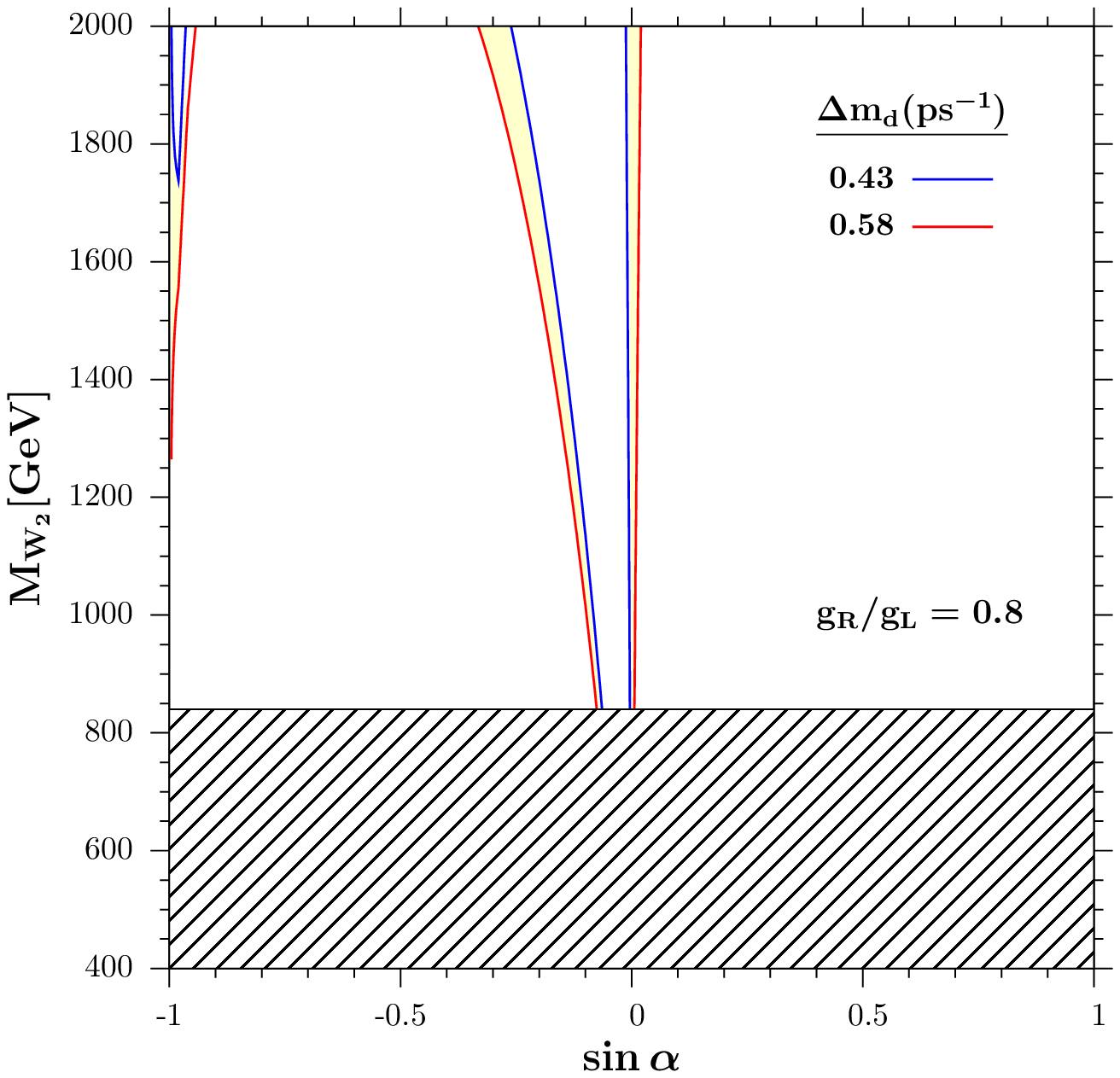} &\hspace*{-0.2cm}
        \includegraphics[width=2.2in,height=2.2in]{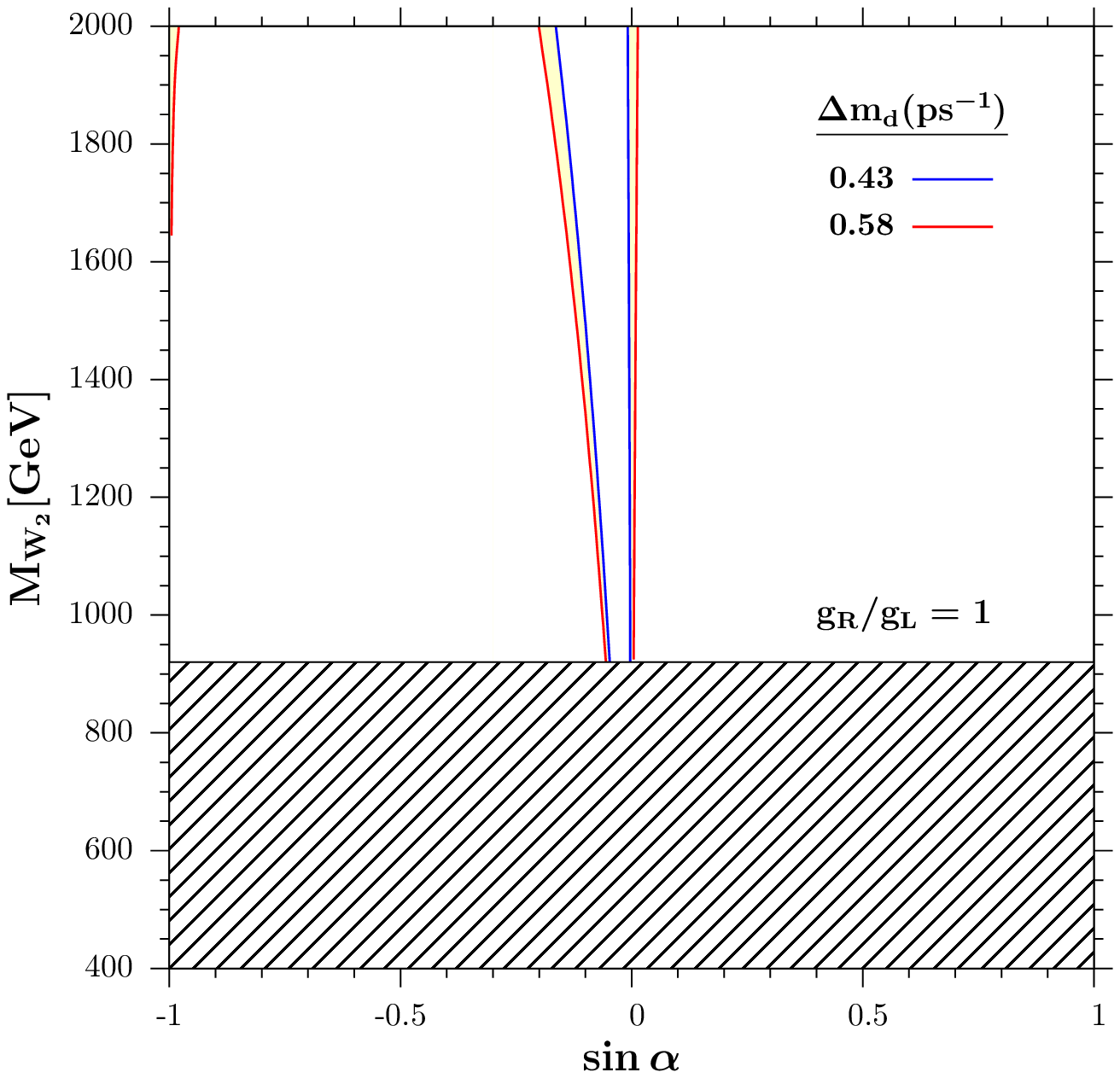} \\ 
\hspace*{-0.7cm}
	\includegraphics[width=2.2in,height=2.2in]{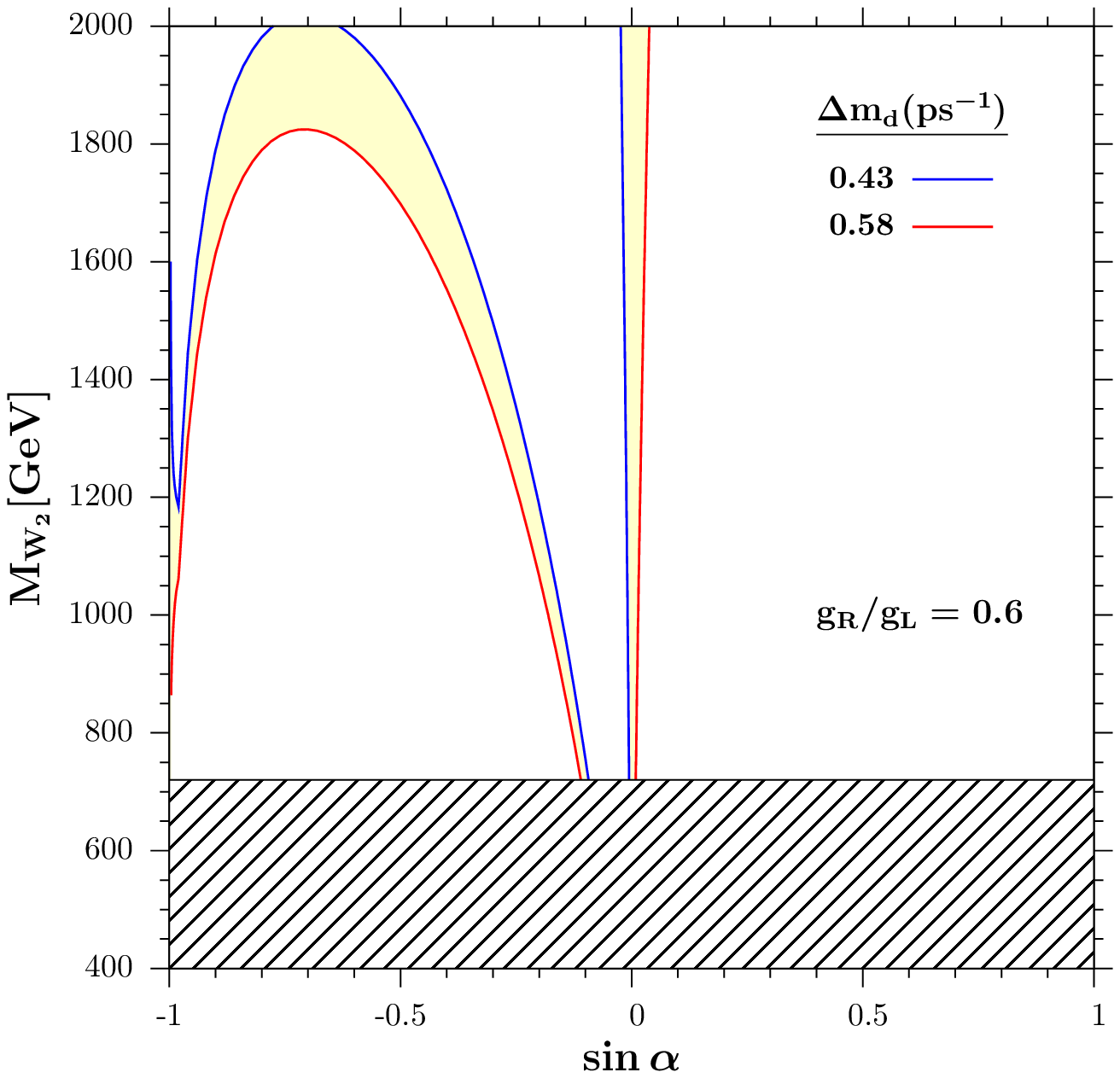} &\hspace*{-0.2cm}
	\includegraphics[width=2.2in,height=2.2in]{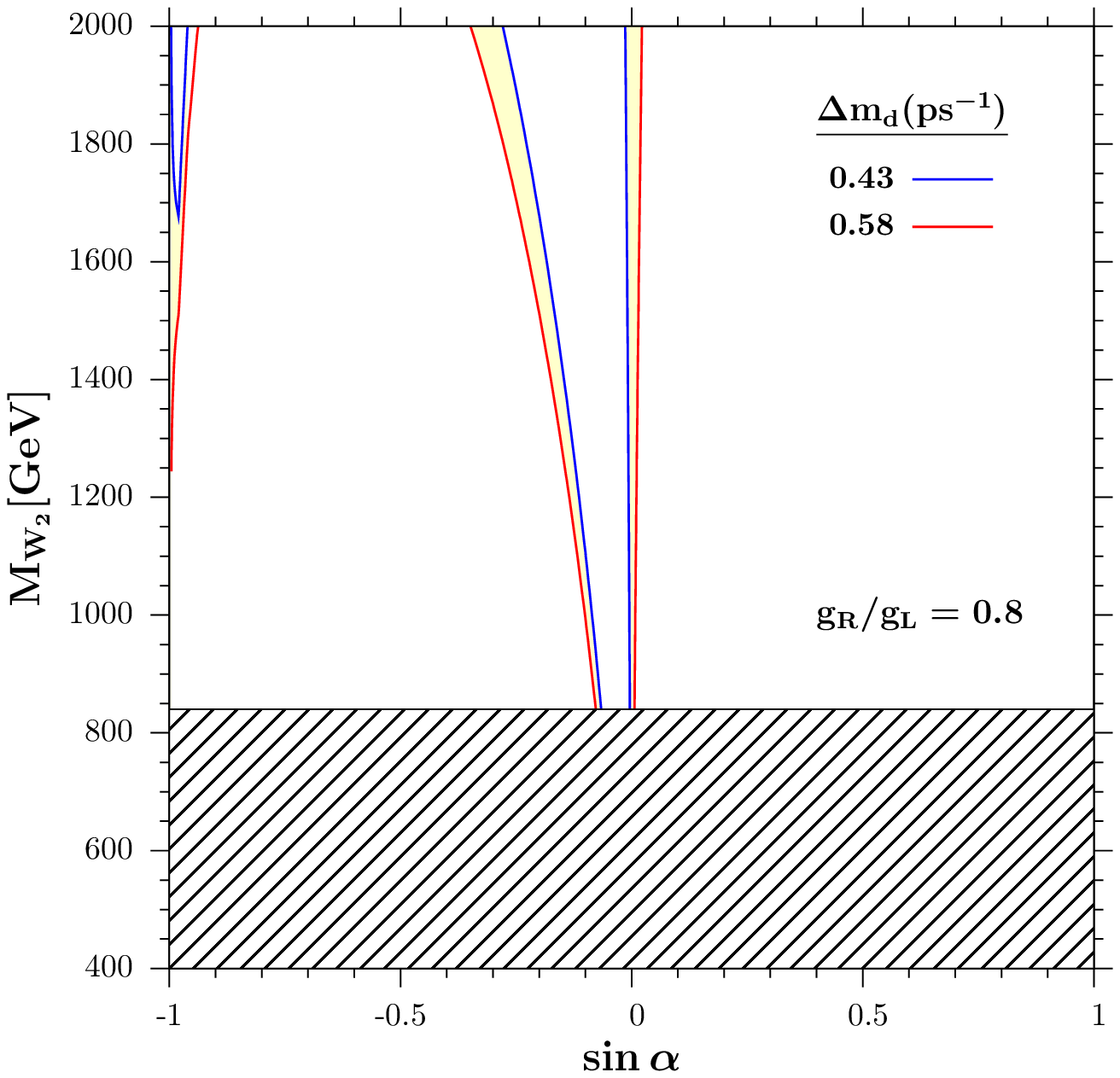} &\hspace*{-0.2cm}
	\includegraphics[width=2.2in,height=2.2in]{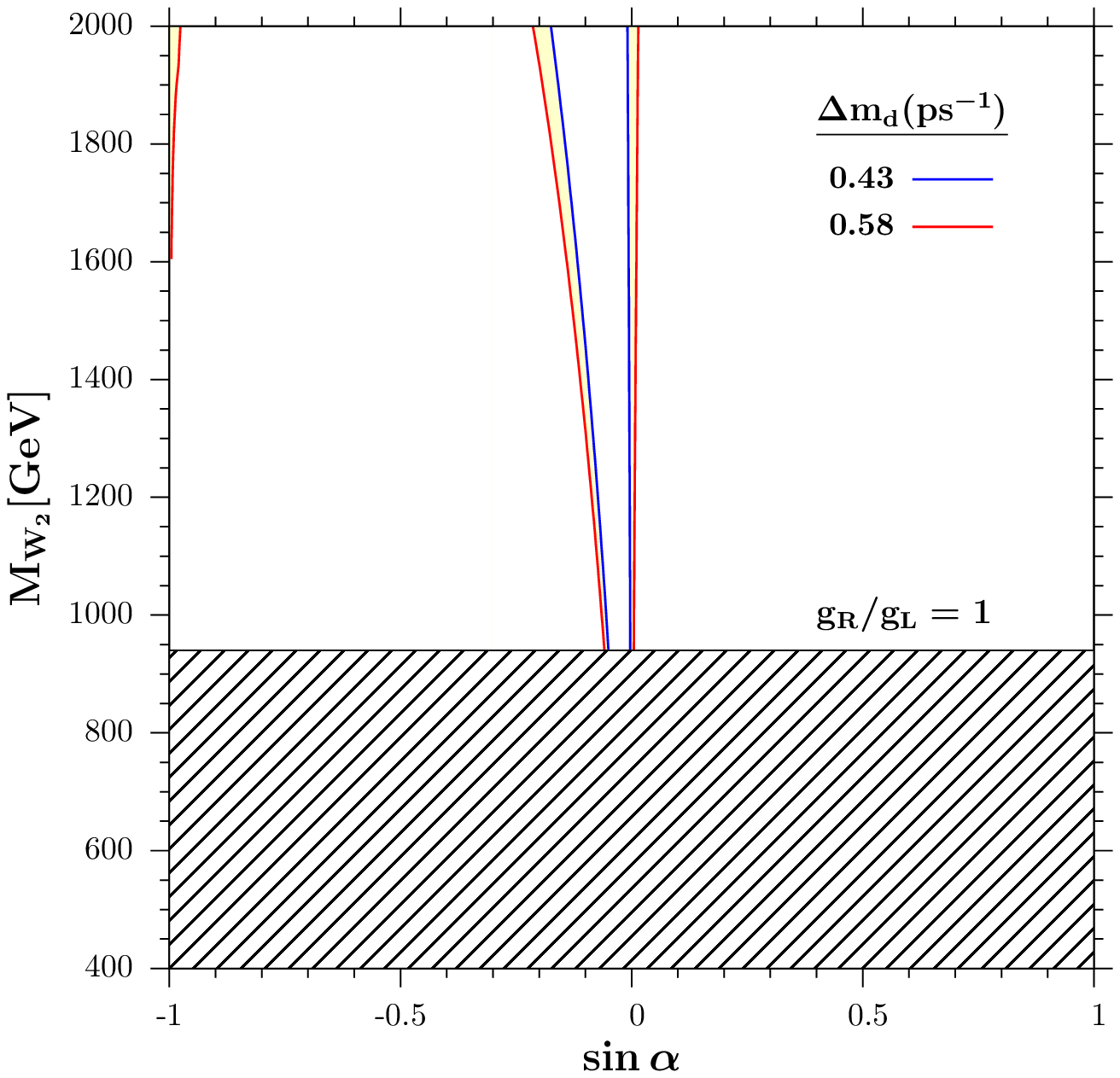}
	\end{array}$
\end{center}
\vskip -0.3in
      \caption{Contour plot of the $M_{W_2}$ vs $\sin\alpha$ constraint in the $V^R=V_{(B)}$ parametrization,  for the $B^0_d-{\bar B}^0_d$ mass difference. We fix $\Delta m_d$ mass difference to be in the interval $(0.43-0.58)$ $ps^{-1}$  (represented by blue and red curves, respectively) and vary $g_R/g_L$, as indicated in the panels. We take $M_{H^\pm}=10$ TeV in upper panels and $M_{H^\pm}=20$ TeV in lower panels and $\tan \beta=10$.  Regions shaded are restricted by the $W_1-W_2$ mixing angle, $\xi \le 3 \times 10^{-3}$. Regions highlighted in yellow represent the allowed parameter spaces.}
\label{fig:B03}
\end{figure}
In Fig.~\ref{fig:B03} we show contour plots of the $M_{W_2}$ versus $V_{td}^R= \sin \alpha$ in the $V_{(B)}^R$ parametrization for the $B_d^0-{\bar B}^0_d$ mass difference, for several values of $g_R/g_L$. The results are very sensitive to this ratio, and we can satisfy the mass difference for any  $W_2$ mass in the $500$ GeV to $2$ TeV range consistently only for small $\sin \alpha$. Increasing $g_R/g_L$ restricts the parameter space further from $W_L-W_R$ mixing. While the Higgs contribution compensates for some of the contributions from $W_2$, the  $W_2$ contribution to the mass difference  appears dominant for the chosen values  $M_{H^\pm}=10$  TeV and $M_{H^\pm}=20$ TeV for $g_R/g_L=0.6,\, 0.8$ and $1$. The interplay between the $W_2$ and $H^{\pm}$ contributions is responsible for allowed regions of parameter space away from $\sin \alpha=0$, for regions around $M_{W_2} \sim 1.8$ TeV. Note that, as the SM value is within the range considered, the region around $\sin \alpha = 0$ is always allowed, and in fact, increasing the ratio $g_R/g_L$, this is the parameter region that consistently survives, corresponding to a very small flavor violation in the right quark system. The sign of $\sin \alpha $ is relevant, with more parameter regions available for $\sin \alpha <0$.  As before, the shaded regions are restricted by the $W_1-W_2$ mixing angle, $\xi \le 3 \times 10^{-3}$.

\begin{figure}[htb]
\begin{center}$
	\begin{array}{ccc}
\hspace*{-0.7cm}
	\includegraphics[width=2.2in,height=2.2in]{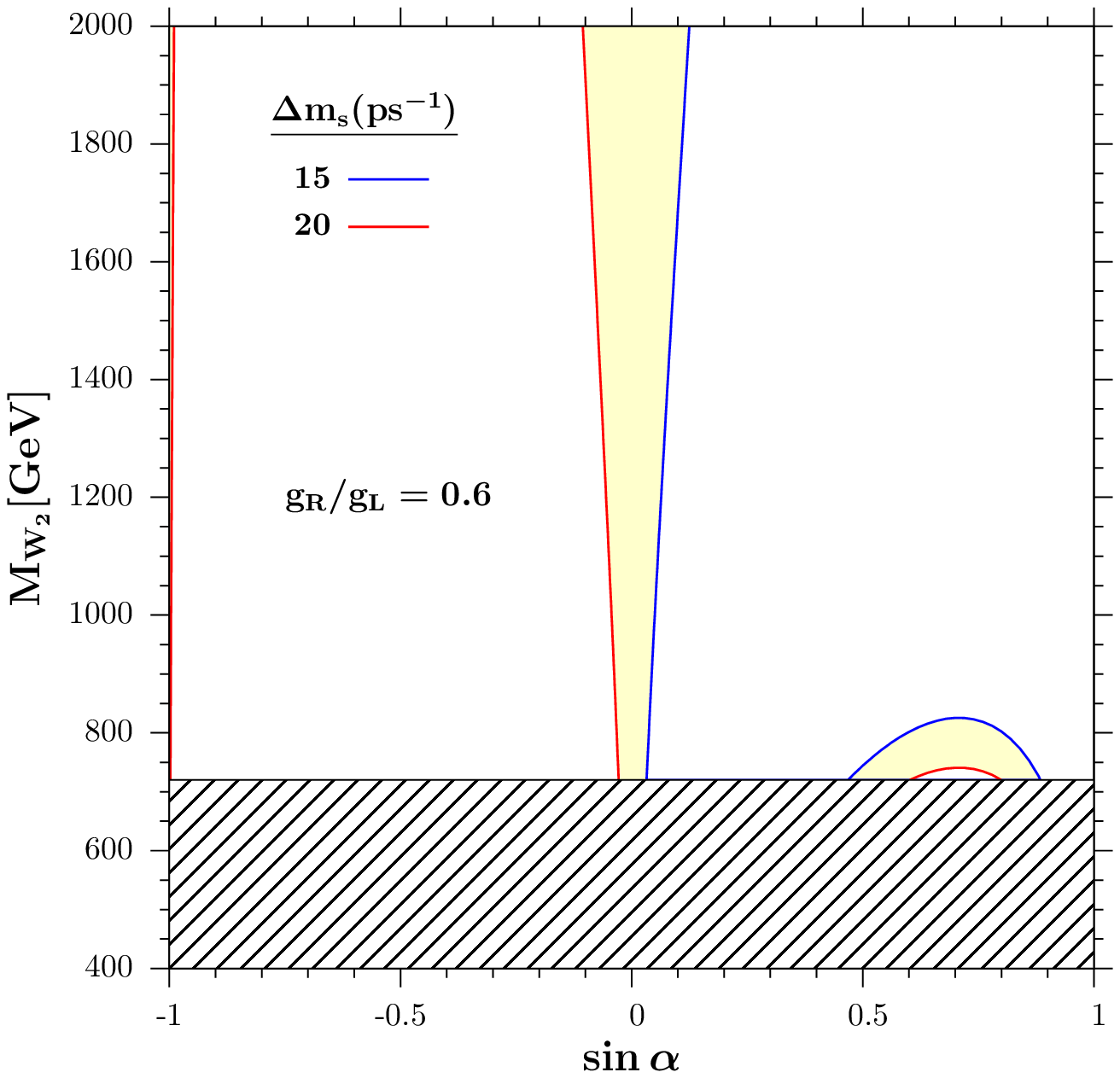} &\hspace*{-0.2cm}
	\includegraphics[width=2.2in,height=2.2in]{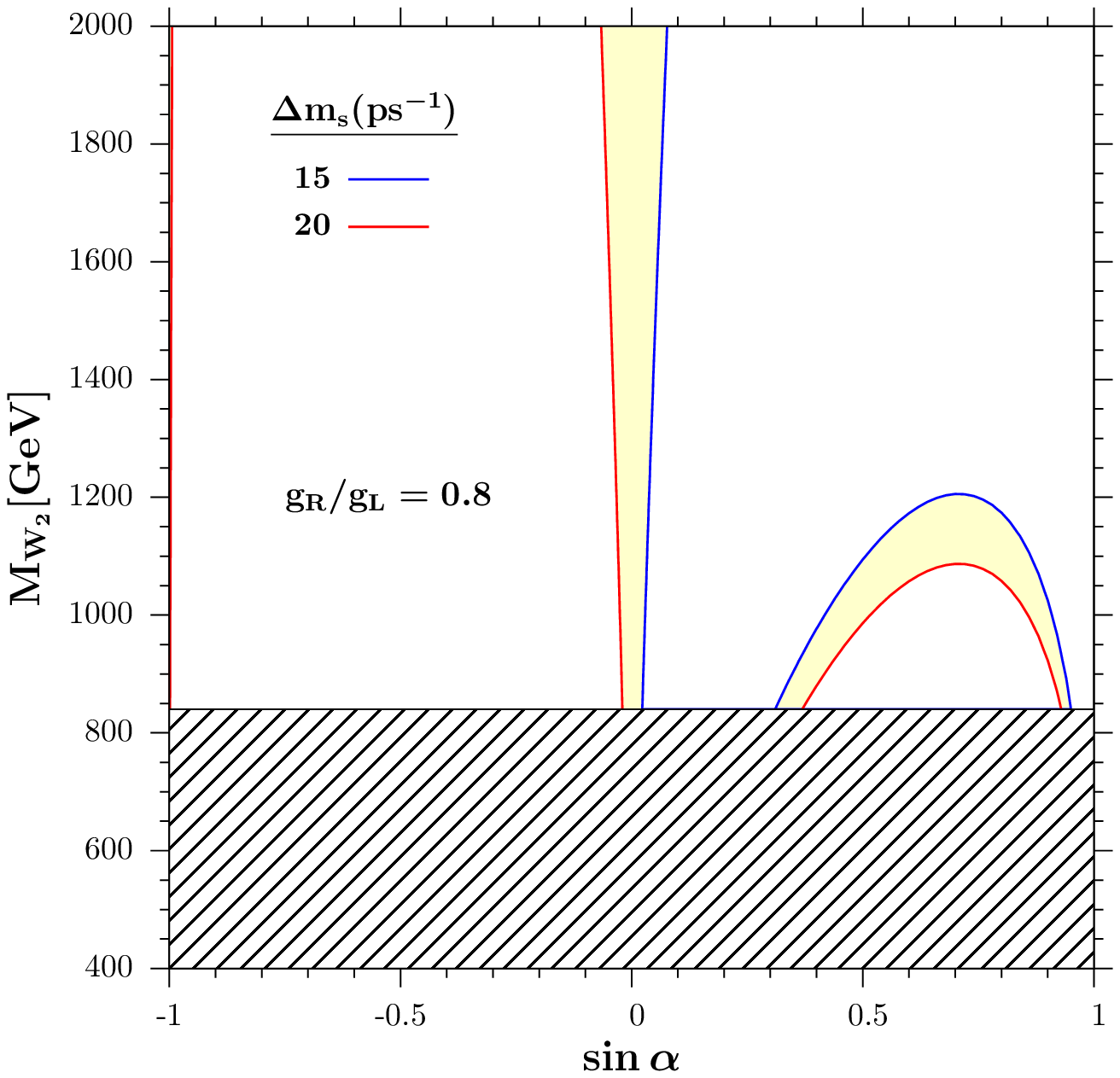} &\hspace*{-0.2cm}
        \includegraphics[width=2.2in,height=2.2in]{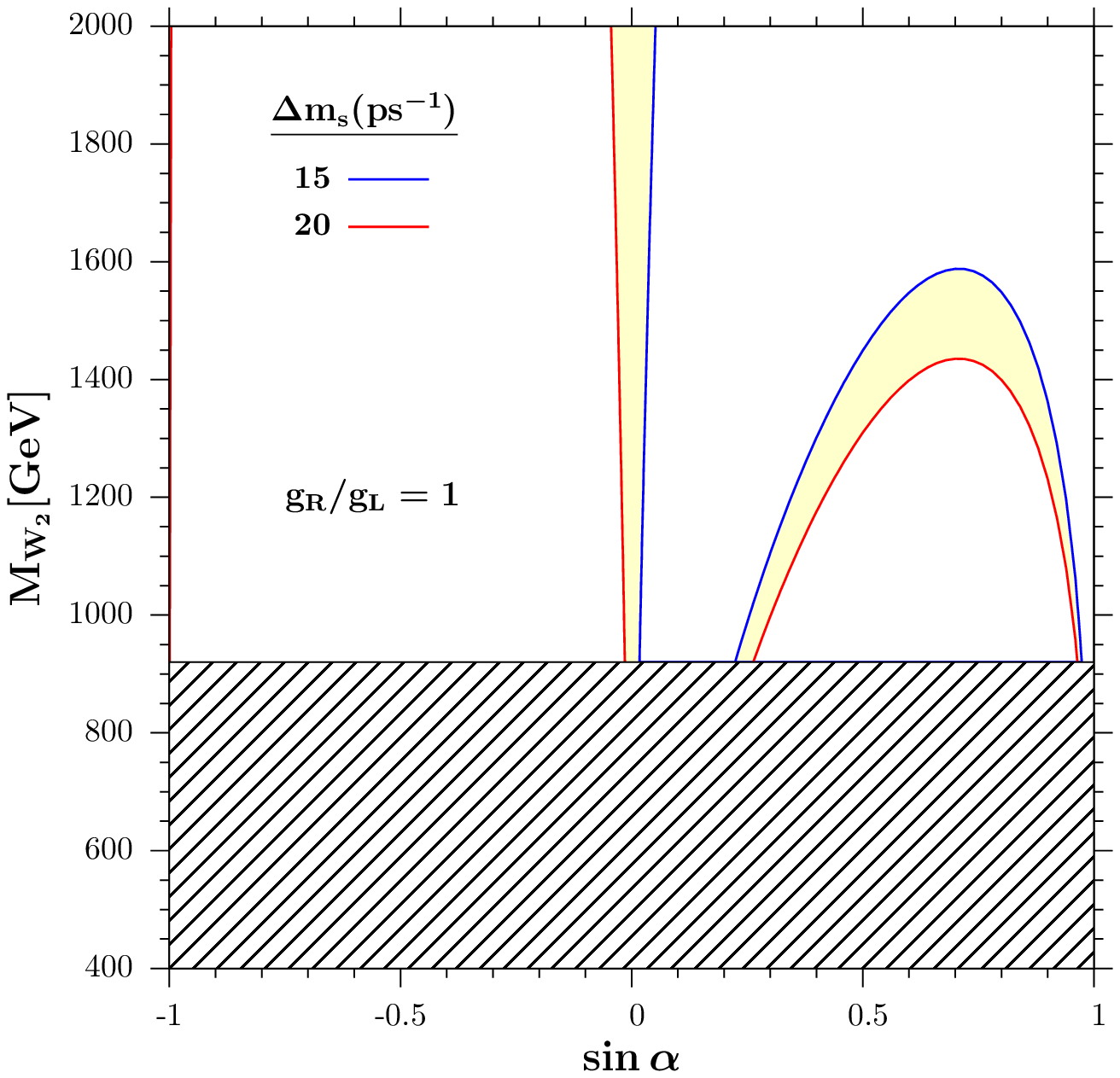}\\ 
\hspace*{-0.7cm}
	\includegraphics[width=2.2in,height=2.2in]{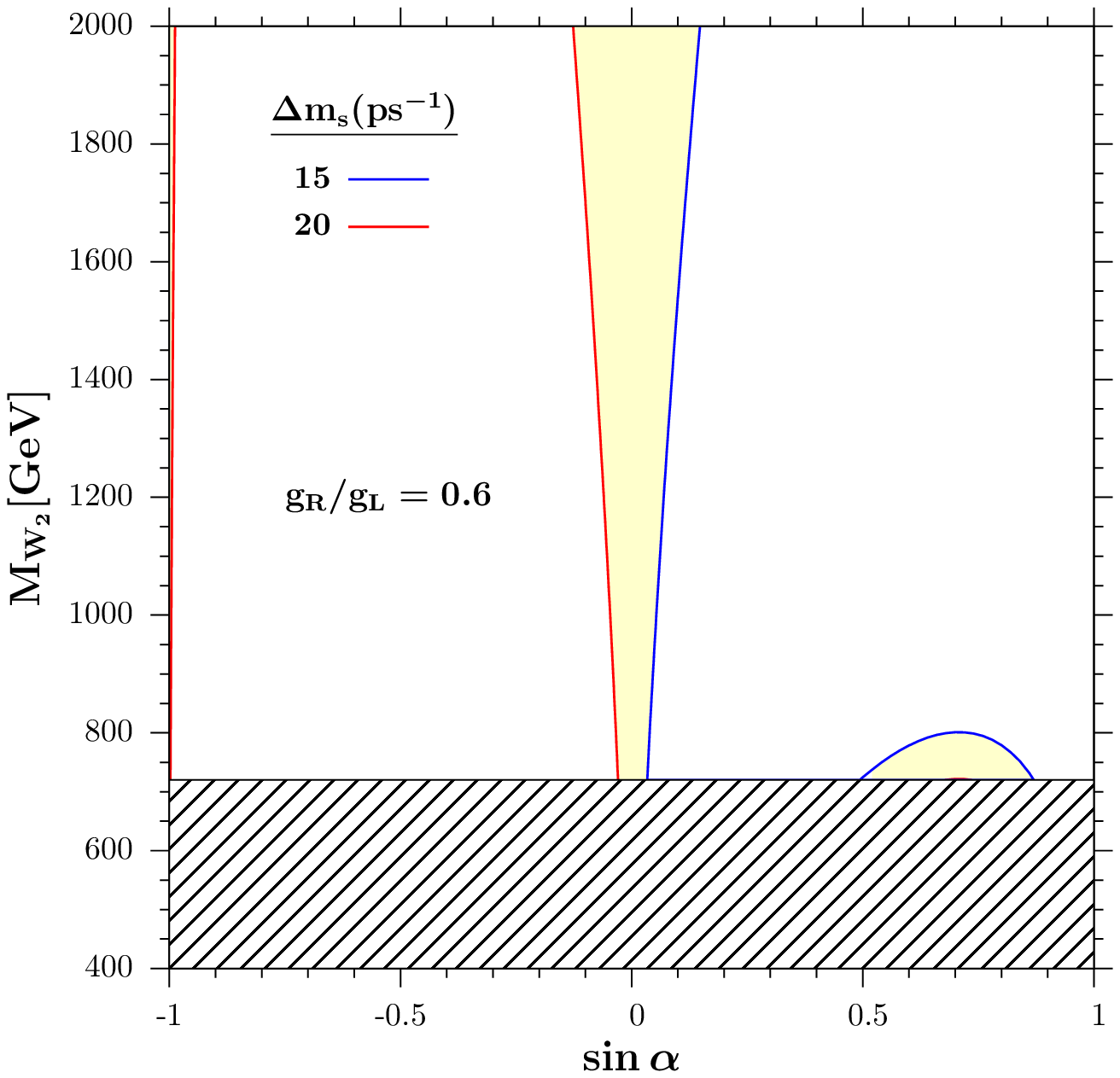} &\hspace*{-0.2cm}
	\includegraphics[width=2.2in,height=2.2in]{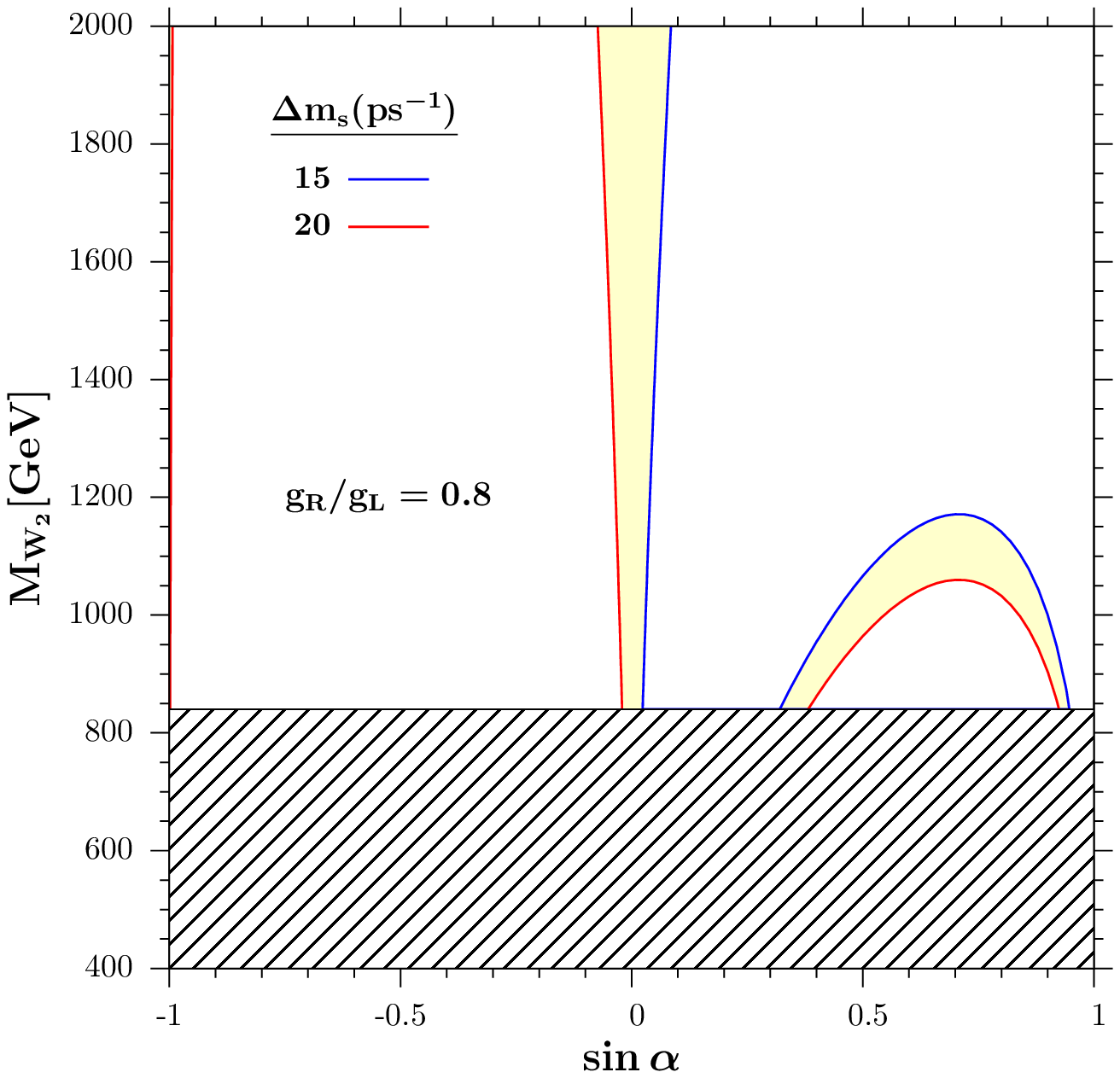} &\hspace*{-0.2cm}
	\includegraphics[width=2.2in,height=2.2in]{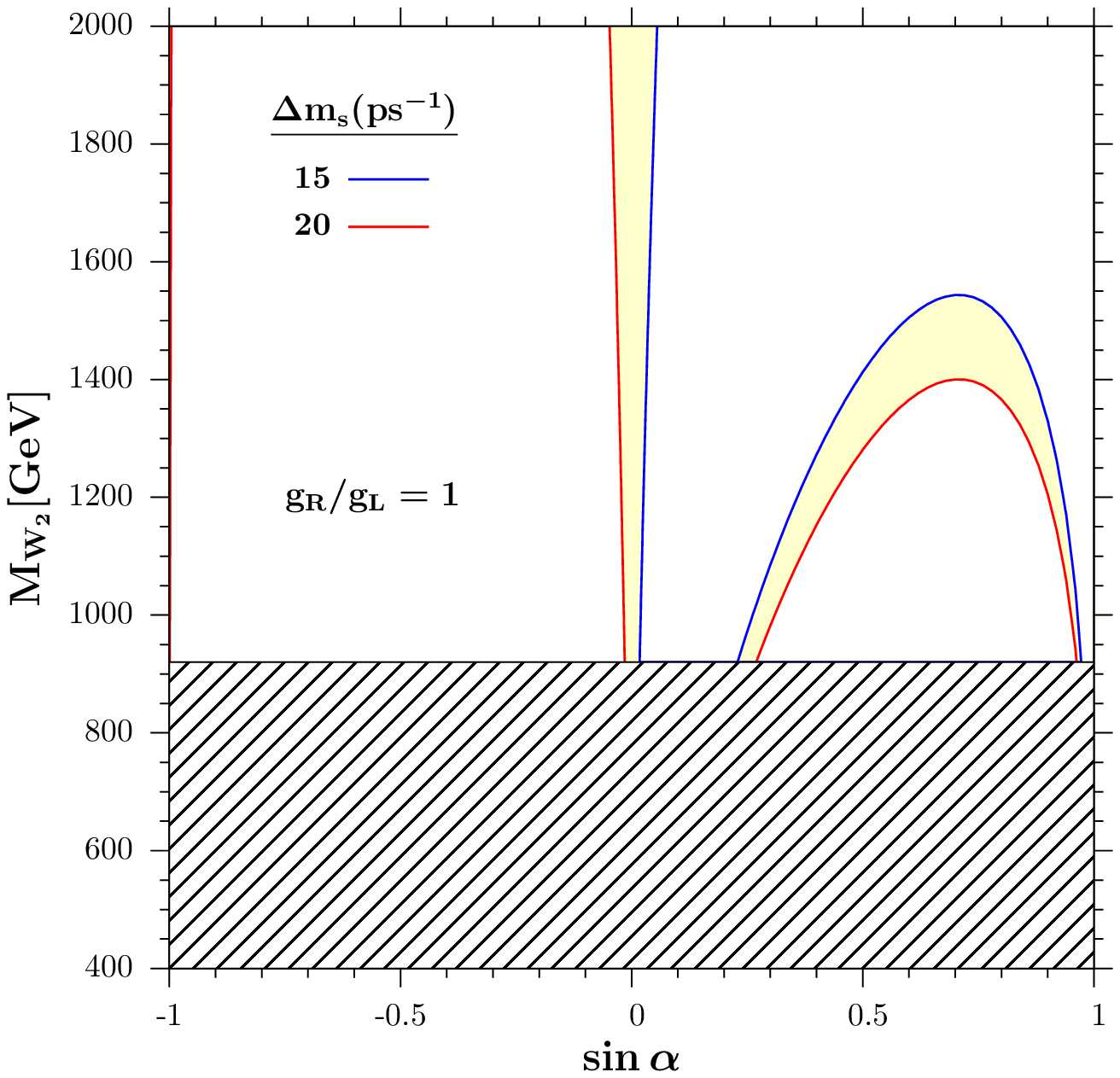}
	\end{array}$
\end{center}
\vskip -0.3in
      \caption{Contour plot of the $M_{W_2}$ vs $\sin\alpha$ constraint in the $V^R=V_{(A)}$ parametrization,  for the $B^0_s-{\bar B}^0_s$ mass difference. We fix $\Delta m_s$ mass difference to be in the interval $(15-20)$ $ps^{-1}$  (represented by blue and red curves, respectively) and vary $g_R/g_L$, as indicated in the panels. We take $M_{H^\pm}=10$ TeV in upper panels and $M_{H^\pm}=20$ TeV in lower panels, and $\tan \beta=10$ throughout.  Regions shaded are restricted by the $W_1-W_2$ mixing angle, $\xi \le 3 \times 10^{-3}$. Regions highlighted in yellow represent the allowed parameter spaces.}
\label{fig:B05}
\end{figure}

Similarly, in Fig.~\ref{fig:B05} we show the contour plot for the $B_s^0-{\bar B}^0_s$ mass difference, with restrictions on $M_{W_2}- \sin \alpha$ plane in the $V_{(A)}^R$ parametrization. 
 The difference is that in this case, the constraints on the parameter space are slightly less stringent and  a larger region of ($M_{W_2}, \sin \alpha$) is allowed than in the $\Delta m_d$ case.  
 In the allowed range, 
the experimental bounds allow a significant region of the parameter space   around $\sin \alpha \in (-0.1, 0.1)$ even for $g_R/g_L =1$, and increasing for  $g_R/g_L=0.6$ and $0.8$. The interplay between the charged Higgs and $W_2$ contributions are more pronounced for $g_R/g_L=1$, where a region of the parameter space opens for $M_{W_2} \sim 1.2-1.6$ TeV. (This region is present, to a lesser extent, for $g_R/g_L=0.8$ in the $M_{W_2} \sim 1-1.2$ TeV region.)

\begin{figure}[htb]
\begin{center}$
	\begin{array}{cc}
	\includegraphics[width=2.2in,height=2.2in]{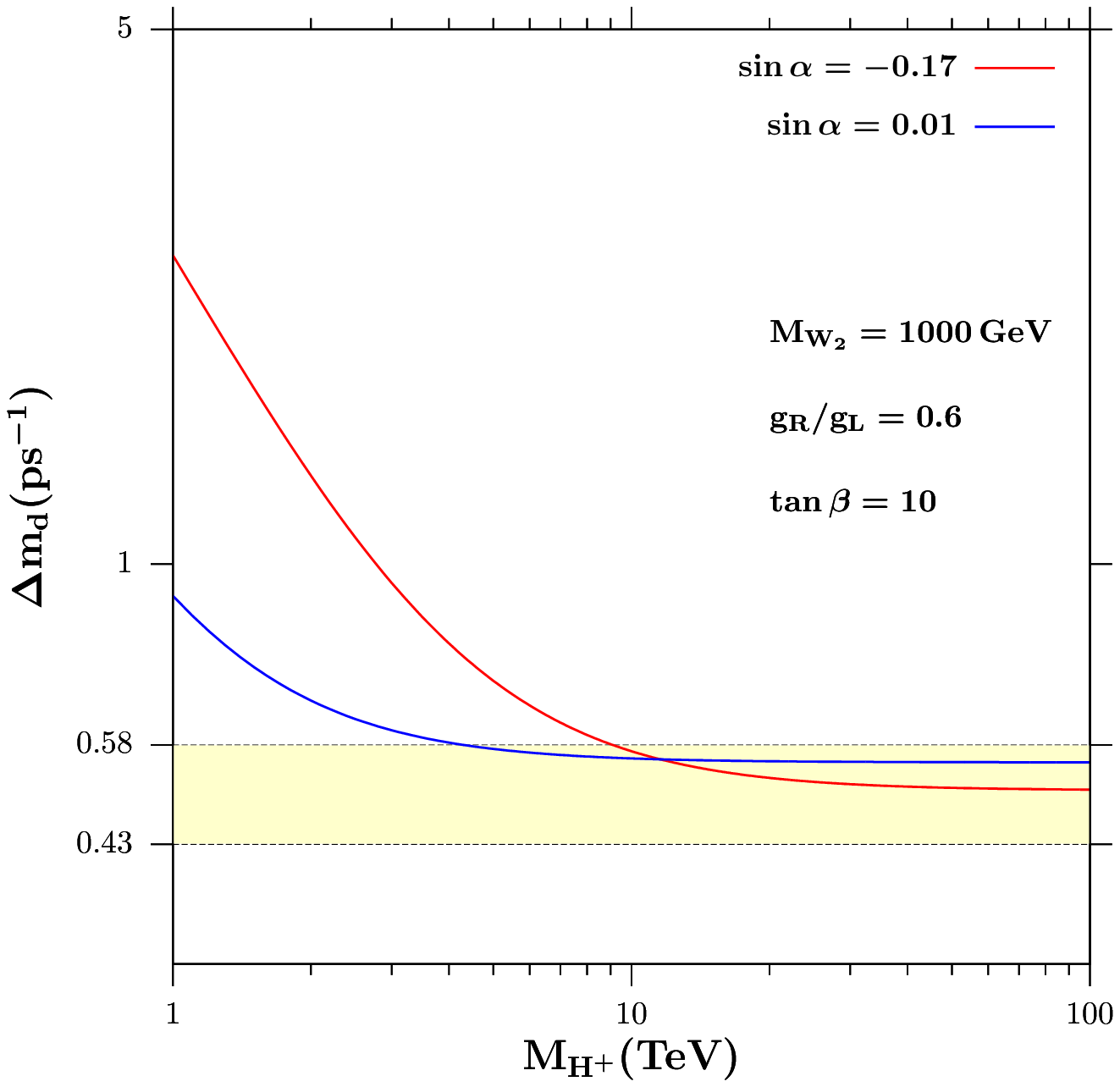} ~~&~~
	\includegraphics[width=2.2in,height=2.2in]{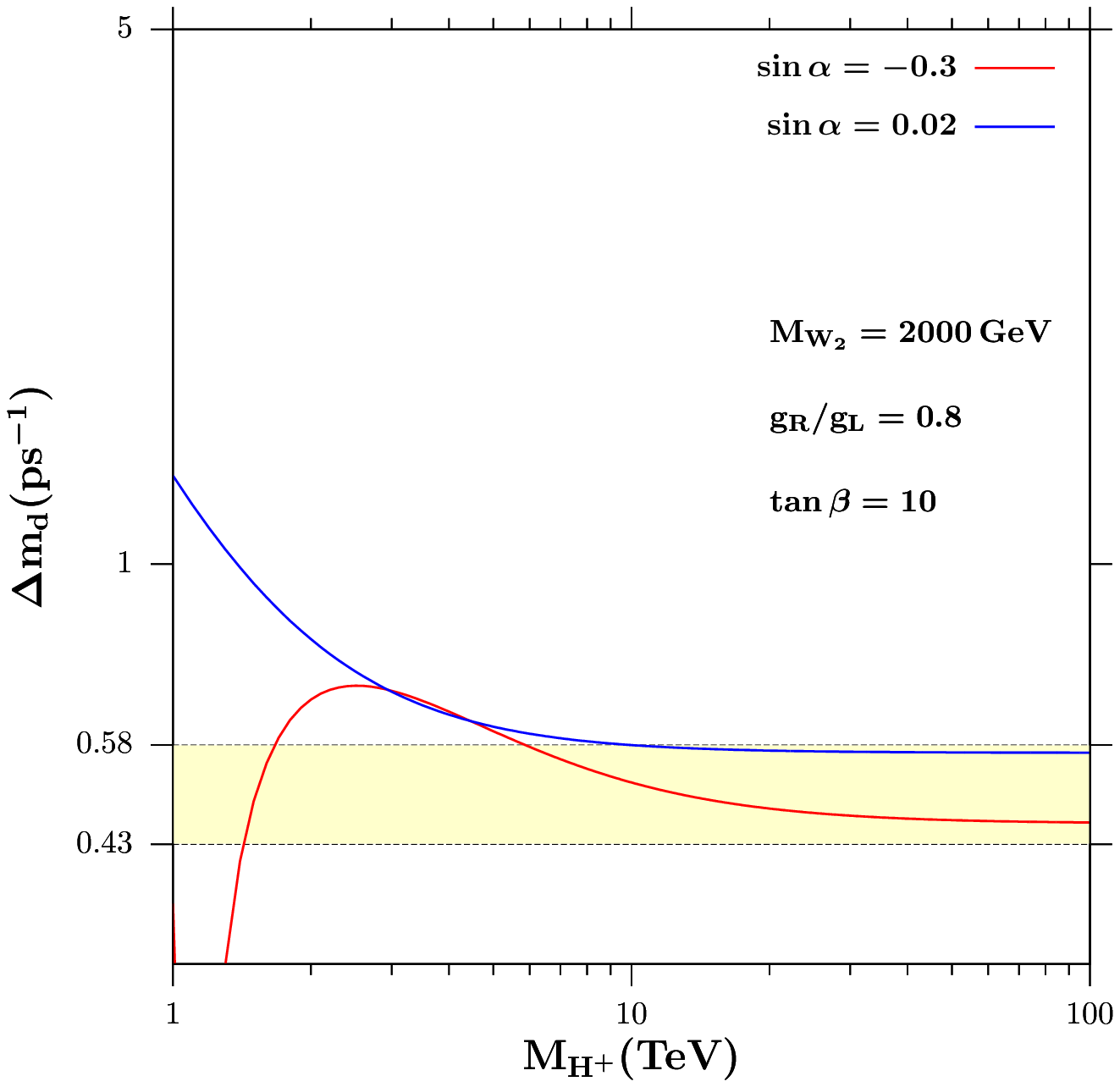} \\       
	\includegraphics[width=2.2in,height=2.2in]{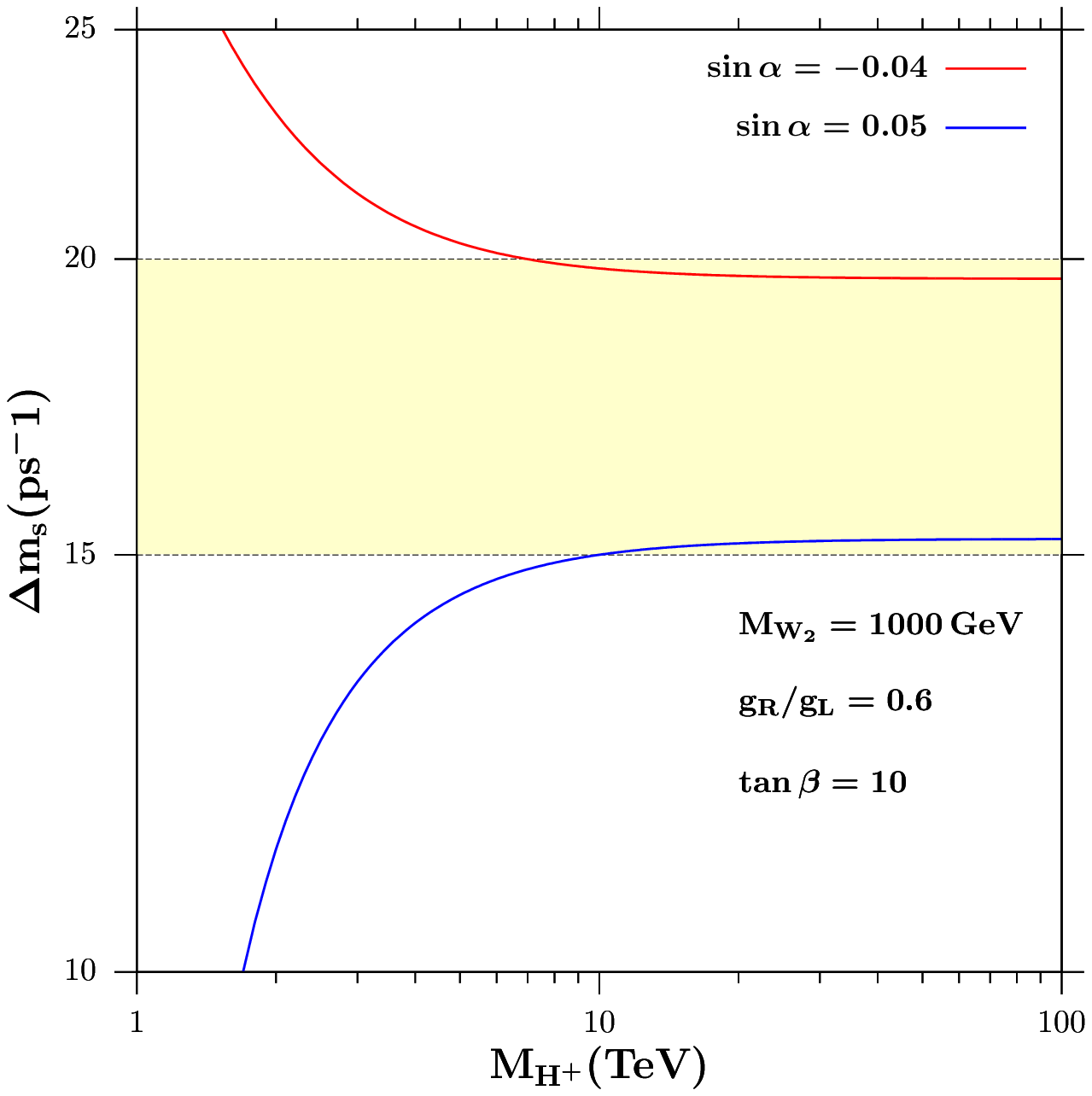} ~~&~~
	\includegraphics[width=2.2in,height=2.2in]{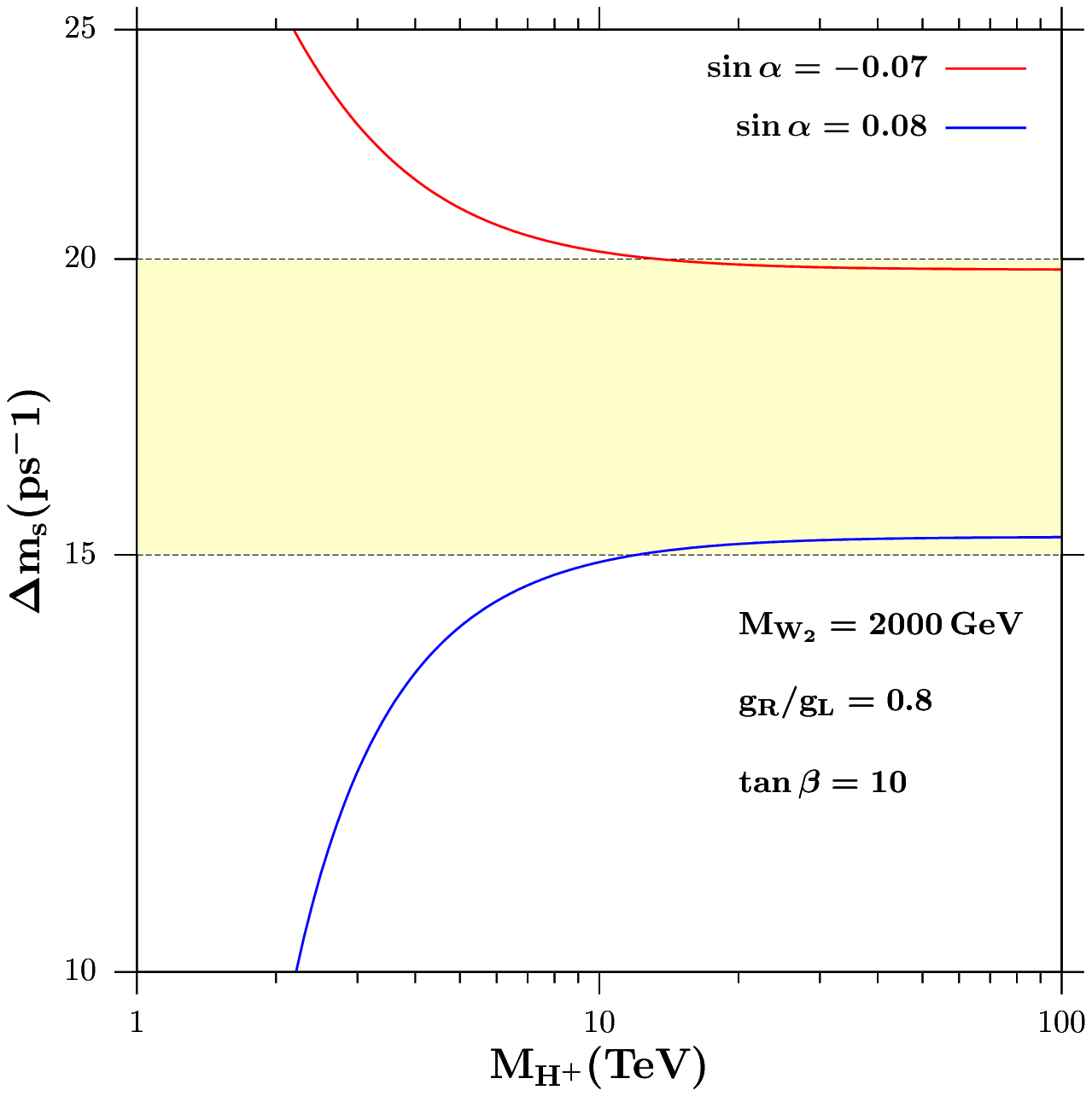} \\
		\end{array}$
\end{center}
\vskip -0.3in
      \caption{$\Delta m_{d,s}$ dependence on the charged Higgs mass $M_{H^{\pm}}$. We fix $\tan \beta=10$ and show curves for  negative and positive values of $\sin \alpha$, in red and blue respectively, chosen in each panel to fit within the experimental range.  The upper row represents $\Delta m_d$ dependence, the lower is for the $\Delta m_s$. The upper left panel corresponds to $M_{W_2}=1$ TeV, $g_R/g_L=0.6$, the right one to $M_{W_2}=2$ TeV, $g_R/g_L=0.8$.   The yellow highlighted regions represent allowed parameter regions between $\Delta m_d= (0.43-0.58)$ ps$^{-1}$ and $\Delta m_s= (15-20)$ ps$^{-1}$.}
\label{fig:B02}
\end{figure}

In  Fig.~\ref{fig:B02}, we show the dependence of $\Delta m_{d}$ (upper row) and $\Delta m_{s}$ (lower row) on the charged Higgs mass, for two values of $g_R/g_L: ~0.6$ and $0.8$.  We include a sample of significant plots, for two values of $M_{W_2}$, $M_{W_2}=1$ and $2$ TeV, for values $\sin \alpha$ chosen to fit within the allowed experimental range.  One can see, comparing the top panels, that the $B_d^0-\bar{B}_d^0$ mass difference is  sensitive to both the $M_{W_2}$ mass and to the measure of CKM flavor violation in the right-handed quark sector, $\sin\alpha$. For $g_R/g_L=0.6$ and $M_{W_2}=1$ TeV, the charged Higgs mass must be $M_{H^\pm} \ge 10$ TeV for $\sin \alpha \in (-0.17, 0.01)$ interval. This constraint is relaxed for $g_R/g_L=0.8$ and $M_{W_2}=2$ TeV, when $\sin \alpha \in (-0.3,0.02)$ for $M_{H^\pm} \ge 7$ TeV; while  outside this $\sin \alpha$ interval, the bounds are not satisfied for any charged Higgs masses, and one would need to increase the $W_2$ mass to reproduce the data. In the bottom row, we perform the same analysis for $\Delta m_s$.  The constraints   for $M_{W_2}=1$ TeV, $g_R/g_L=0.6$ (left panel) are satisfied for $M_{H^\pm} \ge 7$ TeV, but in a smaller region, for $\sin \alpha \in (-0.04, 0.05)$, than those for $\Delta m_d$. For $M_{W_2}=2$ TeV, to remain within the bounds for $g_R/g_L=0.8$ (right panel) requires $M_{H^\pm} \ge 10$ TeV for $\sin \alpha \in (-0.07, 0.08)$. The horizontal region highlighted in yellow corresponds to the allowed region between the  bounds, $\Delta m_d= (0.43-0.58)$ ps$^{-1}$, and $\Delta m_s= (15-20)$ ps$^{-1}$.
   As in the $b \to s \gamma$, our model requires heavier Higgs bosons especially  for larger flavor violation in the right-handed quark sector.

\begin{figure}[htb]
\begin{center}$
\begin{array}{cc}
\includegraphics[width=2.2in,height=2.0in]{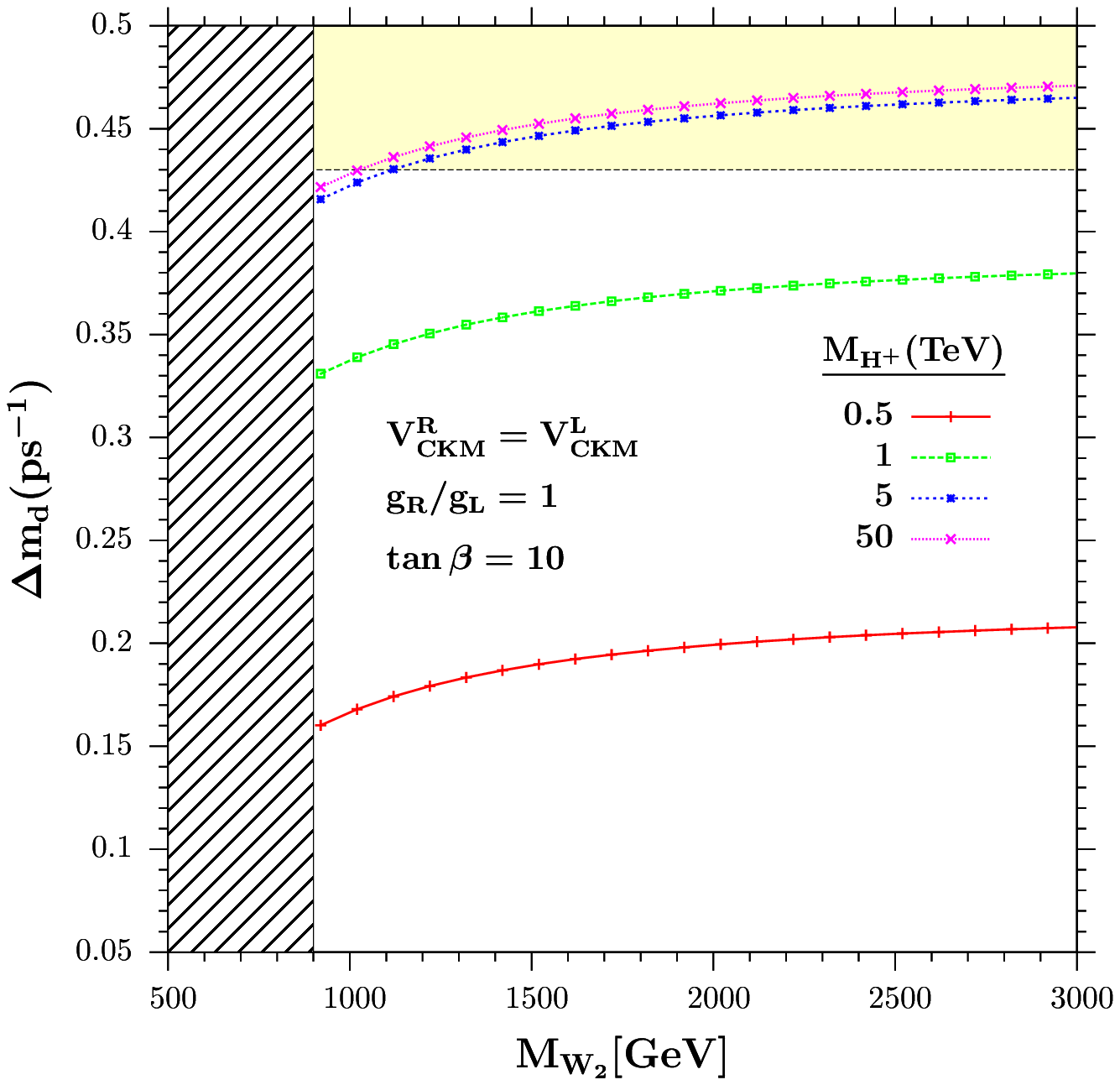} ~~&~~
\includegraphics[width=2.2in,height=2.0in]{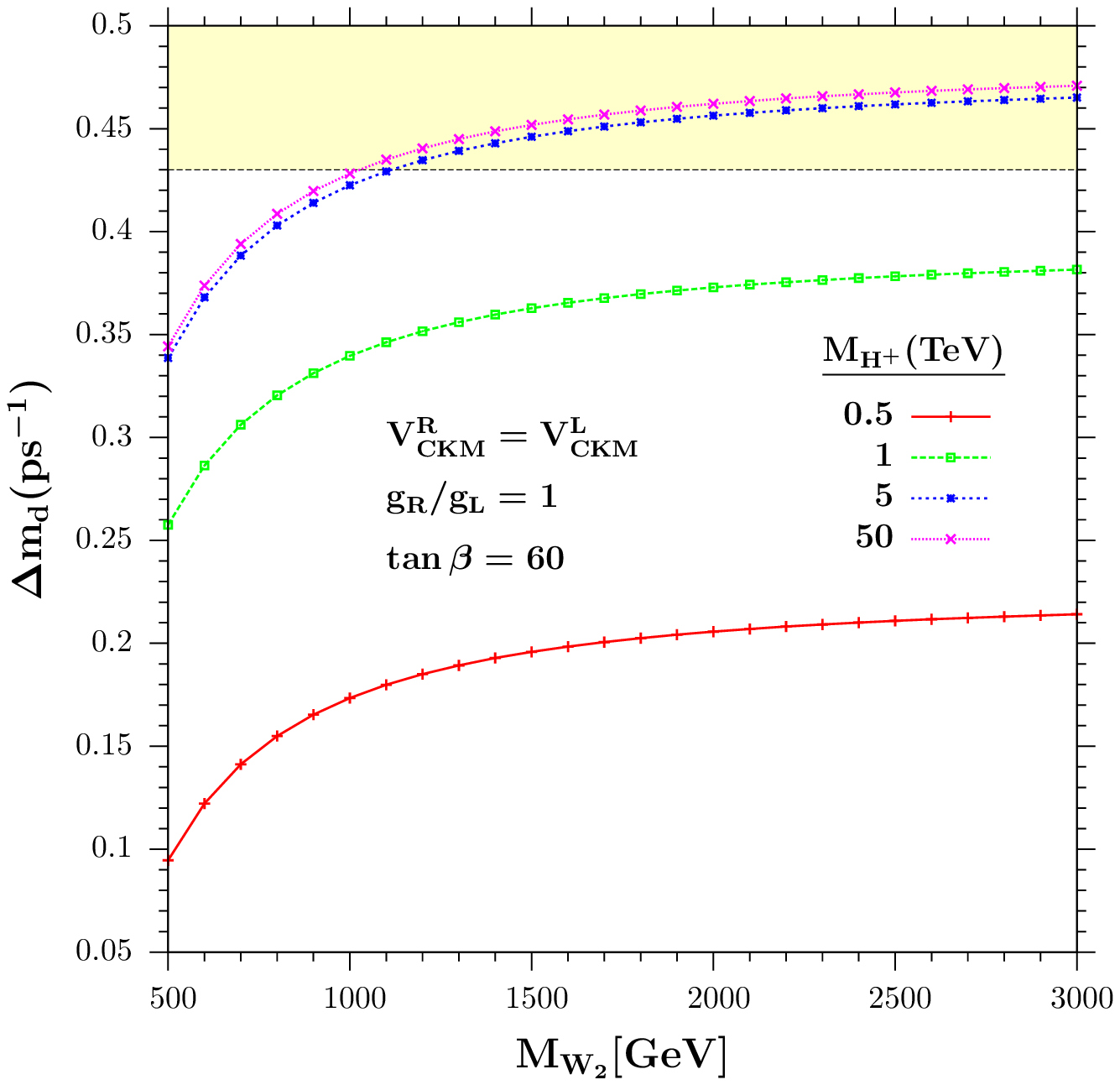} \\
\includegraphics[width=2.2in,height=2.0in]{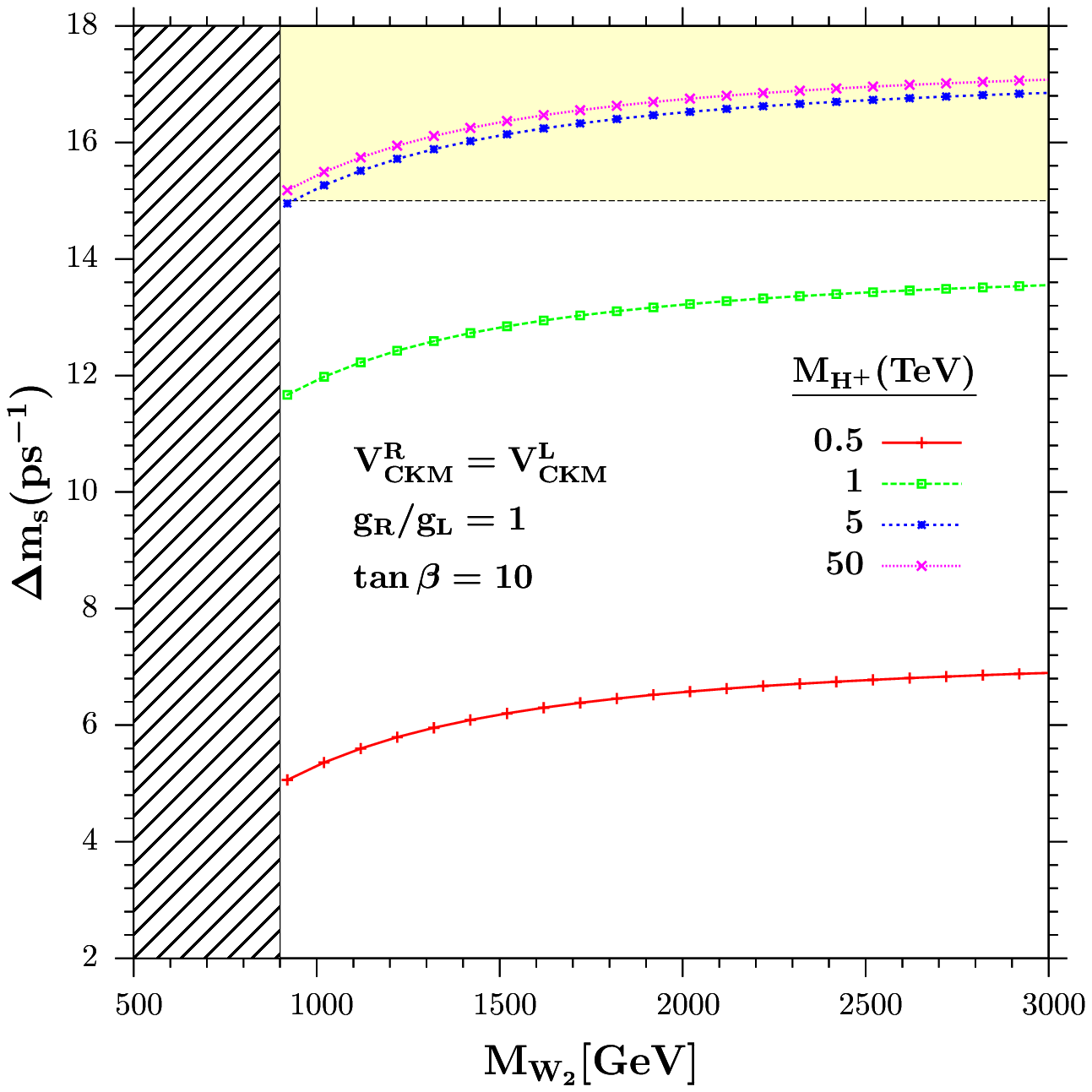} ~~&~~
\includegraphics[width=2.2in,height=2.0in]{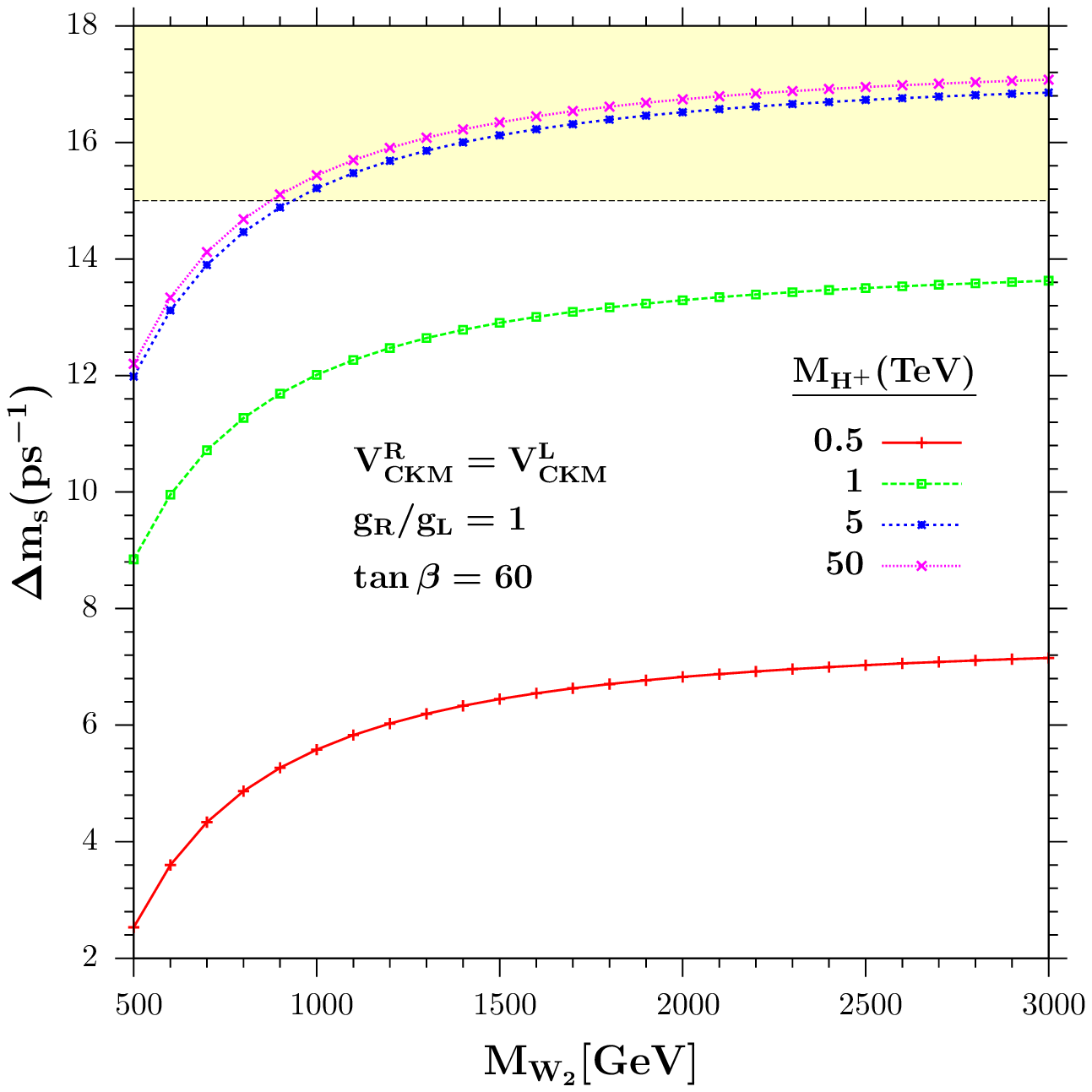}
\end{array}$
\end{center}
\vskip -0.3in
\caption{$\Delta m_{d,s}$ dependence on ${W_2}$ mass in manifest left-right symmetric model for $M_{H^{\pm}}=0.5, 1, 5$ and $50$ TeV. We show $\Delta m_d$ in the upper panels, and $\Delta m_s$ in the lower ones. The left row corresponds to $\tan \beta=10$, the right one to $\tan \beta=60$. Regions shaded are restricted by the $W_1-W_2$ mixing angle, $\xi \le3 \times 10^{-3}$. Regions highlighted in yellow represent the allowed parameter spaces.}
\label{fig:B04}
\end{figure}

In the manifest left-right case, with $V^R_{CKM}$ and $g_R=g_L$ fixed, Higgs masses are required to be $5$ TeV or larger for both $\tan \beta=10$ and $60 $, while $M_{W_2} >1 $ TeV, as shown in Fig.~{\ref{fig:B04}, where we study the dependence of $\Delta m_{d,s}$ with $W_2$ mass for four values of the charged Higgs mass, $0.5, 1, 5$ and $50$ TeV. 
Note that there is no new information provided by $\Delta m_s$ data, and that the manifest left-right contribution is also largely insensitive to $\tan \beta$.

 \section{Conclusions}
\label {sec:conclude}
With the advent of the data from LHC, we expect to observe physics beyond the SM. The left-right  model  is perhaps the simplest such scenario, with  the right handed quarks belonging to doublets and participating in charged flavor violating interactions. Models in which the right-handed sector mimics exactly the left-handed one, such as the manifest or the pseudo-manifest left-right model, have been explored thoroughly and are very restrictive. Motivated by the possibility of additional gauge bosons that may be observed at LHC, as well as some shortcomings  of a left-right symmetric quark flavor sector, we investigated here an asymmetric left-right parametrization for the  quark mixing matrix (Langacker and Sankar)  in the context of B physics. This parametrization has several attractive feature:  while respecting family unitarity, it is general. It allows for variations in the right-handed coupling constant. And it is simple, thus predictive   (the right handed quark mixing matrix depends on one additional parameter only). 

Note that our results are quite general, if we restrict ourselves to parametrizing two family mixings only, in the CP conserving case, as setting $V^R_{ts}= \sin \alpha$ in the $V^R_{(A)}$ parametrization, and $V^R_{td}= \sin \alpha$ in the $V^R_{(B)}$ parametrization, satisfy  general unitarity constraints.

We include existing restrictions on the $W_L-W_R$ mixing angle $\xi$  coming from $K^0-\bar {K^0}$ mixing, while not restricting ourselves to any particular scenario for  the nature or masses of the neutrinos. 
 We provide additional constraints from BR($b \to s \gamma$) and $B^0_{d,s}-\bar {B}^0_{d,s}$ mixing. 
 Defining the parametrizations as $V^R_{(A)}~(V^R_{ts} \ne 0$, $V_{td}^R=0$) and $V^R_{(B)}~(V^R_{td} \ne 0$,  $V_{ts}^R=0$), we set constraints on $\sin \alpha,\, M_{W_2}, \,g_R/g_L, \tan \beta$ and $M_{H^\pm}$. We have used exact numerical evaluations and the existing packages {\tt FeynArts}  for generating  the amplitudes, then {\tt FormCalc} and 
{\tt LoopTools} packages to evaluate the loop contributions, adding modifications as needed.

For the branching ratio $b \to s \gamma$, all parameters play an important role. Smaller values for the ratio $g_R/g_L$ allow for more flavor violation in the right quark sector (larger $\sin\alpha$, smaller $W_2$ masses, wider range for $M_{H^\pm}$). BR($b \to s \gamma$) also depends on $\tan \beta$. Increasing $\tan \beta$ opens larger parameter spaces for both $M_{H^\pm}$ and $M_{W_2}$.
In $\Delta m_{d,s}$ splitting, we find the results be sensitive to the $W_2$ mass, $\sin \alpha$ and the ratio $g_R/g_L$. In the regions allowed 
by the experimental constraints, the results are practically independent of $\tan \beta$.

While a lot of restrictions are interconnected, they share a few general characteristics.  First, the restrictions on  $V^R_{(B)}$, coming from $B^0_d-\bar{B}^0_d$ are more stringent than the combined bounds on $V^R_{(A)}$ coming from $ b \to s \gamma$ and $B^0_s-\bar{B}^0_s$. As these two parametrizations are independent, the larger parameter space available for $V^R_{(A)}$ indicates that in that scenario,  lighter gauge bosons are more likely produced. Second, for any significant regions of parameter space, $g_R/g_L <1$. While decreasing $g_R$ decreases the strength and cross section for right-handed particles, it allows for larger flavor violation in the right-handed sector. It's a delicate balance, as decreasing the amount of right-handed flavor violation makes the model more like the manifest left-right model, and decreasing it even further takes the model to the SM. We restrict $g_R/g_L > \tan \theta_W$ to reproduce correctly the $U(1)_Y$ coupling constant. On the other hand, $g_R/g_L <1$ allows for more flavor violation and smaller $W_2$ masses, while requiring heavy charged Higgs boson masses, $M_{H^\pm} \ge 10$ TeV. The results obtained are consistent with manifest or pseudo-manifest left-right symmetric models, while allowing more flexibility in the parameter space and opening the possibility of observing light gauge bosons at the LHC \cite{FHT}. However, even allowing for more variations of model parameters, the allowed parameter space in $M_{W_2}, \sin \alpha, M_{H^\pm}$ is quite constrained, making the asymmetric left-right  model very predictive.

\section{Acknowledgements}
We are grateful for partial financial support from NSERC of Canada. We would also like to thank Heather Logan for useful discussions and insights.

\section{Appendix}
\label{sec:appendix}

\appendix
\section{QCD correction factors for $B^0_{d,s}-\bar{B}^0_{d,s}$ mixing}
\label{QCD}

We list here the coefficients used to calculate the NLO QCD corrections to $B^0_{d,s}-\bar{B}^0_{d,s}$ mixing in the left-right model, in  eq. (\ref{eta}). The operators $Q_4$ and
$Q_6$ mix under renormalization with an evolution matrix, and the respective Wilson coefficients are
calculated in the following way,
\begin{eqnarray}
\left(
\begin{array}{c}
C_4(m_b) \\ C_6(m_b)
\end{array}
\right)=
\left(
\begin{array}{cc}
\eta_{LR}^{11} & \eta_{LR}^{12} \\
\eta_{LR}^{21} & \eta_{LR}^{22} 
\end{array}
\right)
\left(
\begin{array}{c}
C_4(m_t) \\ C_6(m_t),  
\end{array}
\right)
\end{eqnarray}

and the NLO QCD coefficients $\eta_i(m_b)$ appear in Table \ref{params}.
\begin{table}[htb]
 \begin{center}
  \begin{tabular}{|c|c|c|c|c|c|c|c|c|}\hline
 & $\eta_1$ & $\eta_2$ & $\eta_3$ & $\eta_5$ & $\eta_{LR}^{11}$ & $\eta_{LR}^{12}$ & $\eta_{LR}^{21}$ & $\eta_{LR}^{22}$ \\ \hline
 ~~NLO~~ & ~0.842~ & ~1.648~ & ~1.648~ & ~2.242~ & ~0.920~ & ~-0.039~ & ~-0.877~ & ~2.242~ \\ \hline 
\end{tabular}
 \end{center}
\caption{The QCD correction parameters $\eta_i(m_b)$ used in (\ref{eta}).}
\label{params}
\end{table}

For a detailed analysis of QCD corrections we refer to \cite{Buras:2001ra}.

\section{4-point Passarino-Veltman Integrals at the vanishing external momenta limits}
\label{PVint}
The generic form of 4-point one-loop tensor integrals in 4d is
\begin{equation}
T^{\mu\nu\rho\sigma}=\frac{1}{i\pi^2}\int d^4k\,k^{\mu}\,k^{\nu}\,k^{\rho}\,k^{\sigma}\prod_{i=1}^4\frac{1}{(k+r_i)^2-m_i^2} ,
\end{equation}
where we define the denominators with the conventions of Fig. \ref{fig:generic}.
\begin{figure}[htb]
\begin{center}
\includegraphics[scale=0.8]{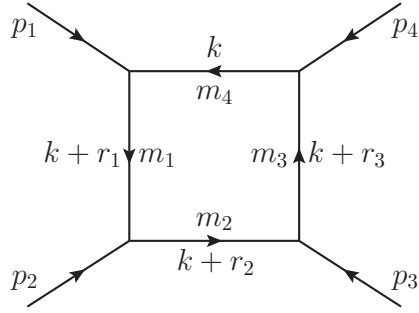} 
\end{center}
\vskip -0.3in
\caption{Momentum and mass conventions used in the Passarino-Veltman for evaluating the box diagrams.}
\label{fig:generic}
\end{figure}
The internal momenta $r_i$ are related to the external momenta through the relations,
\begin{eqnarray}
r_i&=&\sum_{j=1}^i p_j ~~~~,~~~~ i=1,2,3 \nonumber \\
r_4&=&\sum_{j=1}^4 p_j = 0. 
\end{eqnarray}

For $B^0-\bar{B^0}$ mixing we only needed the following scalar and tensor integrals
\begin{eqnarray}
D_0&=&\frac{1}{i\pi^2}\int d^4k \prod_{i=1}^4 \frac{1}{(k+r_i)^2-m_i^2}~, \\
D^{\mu}&=&\frac{1}{i\pi^2}\int d^4k\,k^{\mu} \prod_{i=1}^4 \frac{1}{(k+r_i)^2-m_i^2}~, \\ 
D^{\mu\nu}&=&\frac{1}{i\pi^2}\int d^4k\,k^{\mu}\,k^{\nu} \prod_{i=1}^4 \frac{1}{(k+r_i)^2-m_i^2}~, 
\end{eqnarray}
and the decomposition of tensor integrals in terms of reducible functions are;
\begin{eqnarray}
D^{\mu}&=&\sum_{i=1}^3\,r_i^{\mu}\,D_i~, \\
D^{\mu\nu}&=&g^{\mu\nu}\,D_{00}+\sum_{i,j=1}^3\,r_i^{\mu}\,r_j^{\nu}\,D_{ij}. 
\end{eqnarray}
In {\tt LoopTools}, these coefficient functions ($D_0,D_i,D_{00},D_{ij}$) are evaluated numerically, however at the vanishing
external momenta limits these functions are not well defined. So at this point bypassing the {\tt LoopTools}, we introduced the analytical expressions for those functions in the vanishing external momenta limit,
\begin{eqnarray}
 D(p_1^2,p_2^2,p_3^2,p_4^2,(p_1+p_2)^2,(p_2+p_3)^2,m_1^2,m_2^2,m_3^2,m_4^2),
\end{eqnarray}
where $p_i$'s are external momenta and $m_i$'s are internal masses. Neglecting the external momenta, the structure of those functions might be represented as
\begin{eqnarray}
 D(0,0,0,0,0,0,m_1^2,m_2^2,m_3^2,m_4^2),
\end{eqnarray}
and we will call them for simplicity $D(m_1^2,m_2^2,m_3^2,m_4^2)$. Since we only consider the top quark contributions in the loop, those functions become $D(m_i^2,m_j^2,m_t^2,m_t^2)$ in which $m_i$ and $m_j$ stand for the  boson masses in the loop.

The relevant integrals for $B^0-\bar{B^0}$ mixing are the following:

\begin{eqnarray}
D_0(m_i=m_j)&=&\frac{2\,(1-r)+(1+r)\,\ln{r}}{m_i^4\,(r-1)^3}~, \\ \nonumber \\
D_0(m_i\not=m_j)&=&-\frac{s\,\ln{s}+[(s-r)(r-1)-(s-r^2)\,\ln{r}]}{m_i^4\,(s-r)^2\,(r-1)^2}~, \\ \nonumber \\
D_1(m_i=m_j)&=&-\frac{1+(4-5\,r)\,r+2\,(2+r)\,r\,\ln{r}}{4\,m_i^4\,(r-1)^4}~ ,\\ \nonumber \\
D_1(m_i\not=m_j)&=&\frac{(2\,r-1)\,s^2-(2\,s-1)\,r^2}{2\,m_i^4\,(s-1)\,(s-r)^3\,(r-1)} \nonumber \\ \nonumber \\
&+&\frac{(s-1)^2\,[(r-2)\,s+r^2]\,r\,\ln{r}-(r-1)^2\,[(s-2)\,r+s^2]\,s\,\ln{s}}{2\,m_i^4\,(s-1)^2\,(s-r)^3\,(r-1)^2}~, \\ \nonumber \\
D_2(m_i=m_j)&=&-\frac{r^2+4\,r-5-2\,(2\,r+1)\ln{r}}{4\,m_i^4\,(r-1)^4}~, \\ \nonumber \\
D_2(m_i\not=m_j)&=&-\frac{(r-3)\,s+r+r^2}{4\,m_i^4\,(s-r)^2\,(r-1)^2} \nonumber \\ \nonumber \\
&+&\frac{s^2\,(r-1)^3\ln{s}}{2\,m_i^4\,(s-1)(s-r)^3(r-1)^3}-\frac{[s^2+(r-3)\,s\,r^2+r^3]\,\ln{r}}{2\,m_i^4\,(s-r)^3\,(r-1)^3}~, \\ \nonumber \\
D_3(m_i=m_j)&=&-\frac{r^2+4\,r-5-2\,(2\,r+1)\ln{r}}{4\,m_i^4\,(r-1)^4}~, \\ \nonumber \\
D_3(m_i\not=m_j)&=&-\frac{(r-3)\,s+r+r^2}{4\,m_i^4\,(s-r)^2\,(r-1)^2}  \nonumber \\ \nonumber \\
&+&\frac{s^2\,(r-1)^3\,\ln{s}}{2\,m_i^4\,(s-1)(s-r)^3(r-1)^3}-\frac{[s^2+(r-3)\,s\,r^2+r^3]\,\ln{r}}{2\,m_i^4\,(s-r)^3\,(r-1)^3}~, \\ \nonumber \\
D_{00}(m_i=m_j)&=&-\frac{r^2-1-2\,r\,\ln{r}}{4\,m_i^2\,(r-1)^3}~, \\ \nonumber \\
D_{00}(m_i\not=m_j)&=&-\frac{(s-1)\,s^2\,\ln{s}}{4\,m_i^2\,(s-r)^2\,(r-1)^2} \nonumber \\ \nonumber \\&-& \frac{r\,\lbrace(s-r)(r-1)+[(r-2)\,s+r]\,\ln{r}\rbrace}{4\,m_i^2\,(s-r)^2\,(r-1)^2}~ ,\\ \nonumber \\
D_{11}(m_i=m_j)&=&\frac{-1+r\,[9-(17\,r-9)\,r]+6\,(r+3)\,r^2\,\ln{r}}{18\,m_i^4\,(r-1)^5}~, \\ \nonumber \\
D_{11}(m_i\not=m_j)&=&\frac{[r^2+(2\,r-3)\,s]\,r^2\,\ln{r}}{3\,m_i^4\,(s-r)^4\,(r-1)^2} \nonumber \\ \nonumber \\
&-&\frac{\lbrace s^3+2\,(s-2)\,s^2\,r+[3+(s-3)\,s]\,r^2\rbrace\,s\,\ln{s}}{3\,m_i^4\,(s-1)^3\,(s-r)^4} \nonumber \\ \nonumber \\
&+&\frac{-(s+1)\,s^2+[5+(s-2)\,s]\,s\,r+[2+(5\,s-9)\,s]\,r^2}{6\,m_i^4\,(s-1)^2\,(s-r)^3\,(r-1)}~ ,\\ \nonumber \\
D_{12}(m_i=m_j)&=&\frac{(r-1)\,[1+(r+10)\,r]-6\,(r+1)\,r\,\ln{r}}{12\,m_i^4\,(r-1)^5}~, \\ \nonumber \\
D_{12}(m_i\not=m_j)&=&\frac{2\,(s^2-3\,r+2\,s\,r)\,s^2\,\ln{s}}{12\,m_i^4\,(s-1)^2\,(s-r)^4}\nonumber \\ \nonumber \\
&-&\frac{r\,\lbrace 2\,(r-2)\,s\,r^2+r^3+[3+(r-3)\,r]\,s^2\,\ln{r}\rbrace}{6\,m_i^4\,(s-r)^4\,(r-1)^3}\nonumber \\ \nonumber \\
&-&\frac{-(r+1)\,r^2+[5+(r-2)\,r]\,s\,r+[2+(5\,r-9)\,r]\,s^2}{12\,m_i^4\,(s-1)\,(s-r)^3\,(r-1)^2}~, \\ \nonumber \\
D_{13}(m_i=m_j)&=&\frac{(r-1)\,[1+(r+10)]\,r-6\,(r+1)\,r\,\ln{r}}{12\,m_i^4\,(r-1)^5}~, \\ \nonumber \\
D_{13}(m_i\not=m_j)&=&\frac{2\,(s^2-3\,r+2\,s\,r)\,s^2\,\ln{s}}{12\,m_i^4\,(s-1)^2\,(s-r)^4}\nonumber \\ \nonumber \\
&-&\frac{r\,\lbrace 2\,(r-2)\,s\,r^2+r^3+[3+(r-3)\,r]\,s^2\,\ln{r}\rbrace}{6\,m_i^4\,(s-r)^4\,(r-1)^3}\nonumber \\ \nonumber \\
&-&\frac{-(r+1)\,r^2+[5+(r-2)\,r]\,s\,r+[2+(5\,r-9)\,r]\,s^2}{12\,m_i^4\,(s-1)\,(s-r)^3\,(r-1)^2}~ ,\\ \nonumber \\
D_{22}(m_i=m_j)&=&\frac{17-(r+1)\,9\,r+r^3+6\,(3\,r+1)\,\ln{r}}{18\,m_i^4\,(r-1)^5}~ ,\\ \nonumber \\
D_{22}(m_i\not=m_j)&=&\frac{-s^3\,\ln{s}}{3\,m_i^4\,(s-1)\,(s-r)^4}\nonumber  \\ \nonumber \\
&+&\frac{\lbrace(r-4)\,s\,r^3+[6+(r-4)\,r]\,s^2\,r^2-s^3+r^4\rbrace\,\ln{r}}{3\,m_i^4\,(s-r)^4\,(r-1)^4}\nonumber \\ \nonumber \\
&+&\frac{11\,s^2-7\,(s+1)\,s\,r+2\,[1+(s-5)\,s]\,r^2+5\,(s+1)\,r^3-r^4}{18\,m_i^4\,(s-r)^3\,(r-1)^3}~,\\ \nonumber \\
D_{23}(m_i=m_j)&=&\frac{17-(r+1)\,9\,r+r^3+6\,(3\,r+1)\,\ln{r}}{36\,m_i^4\,(r-1)^5} ,\\ \nonumber \\
D_{23}(m_i\not=m_j)&=&\frac{-s^3\,\ln{s}}{6\,m_i^4\,(s-1)\,(s-r)^4}\nonumber  \\ \nonumber \\
&+&\frac{\lbrace(r-4)\,s\,r^3+[6+(r-4)\,r]\,s^2\,r^2-s^3+r^4\rbrace\,\ln{r}}{6\,m_i^4\,(s-r)^4\,(r-1)^4}\nonumber \\ \nonumber \\
&+&\frac{11\,s^2-7\,(s+1)\,s\,r+2\,[1+(s-5)\,s]\,r^2+5\,(s+1)\,r^3-r^4}{36\,m_i^4\,(s-r)^3\,(r-1)^3}~,\\ \nonumber \\
D_{33}(m_i=m_j)&=&\frac{17-(r+1)\,9\,r+r^3+6\,(3\,r+1)\,\ln{r}}{18\,m_i^4\,(r-1)^5} ~,\\ \nonumber \\
D_{33}(m_i\not=m_j)&=&\frac{[(s-1)\,\ln{r}-(r-1)\,\ln{s}]\,s^3}{3\,m_i^4\,(s-1)\,(s-r)^4\,(r-1)}\nonumber \\ \nonumber \\
&-&\frac{\lbrace (r-3)\,s\,r^2+r^3+[3+(r-3)\,r]\,s^2\,r\rbrace\,\ln{r}}{3\,m_i^4\,(s-r)^3\,(r-1)^4}\nonumber \\ \nonumber \\
&+&\frac{11\,s^2-7\,(s+1)\,s\,r+2\,[1+(s-5)\,s]\,r^2+5\,(s+1)\,r^3-r^4}{18\,m_i^4\,(s-r)^3\,(r-1)^3}~,
\end{eqnarray}
where we define the parameters as 
\begin{eqnarray}
 r=\left(\frac{m_t}{m_i}\right)^2  ~~~~~~\text{and}~~~~~~ s=\left(\frac{m_j}{m_i}\right)^2~.
\end{eqnarray}


\end{document}